\documentclass[aps, twocolumn,prx, showpacs, longbibliography, preprintnumbers,amsmath,amssymb,superscriptaddress]{revtex4-1}
\usepackage{silence}

\usepackage{svg}
\usepackage{amsmath}
\usepackage{amsthm}
\usepackage{amssymb}

\usepackage{amsmath,amsfonts,amssymb,graphics,graphicx,epsfig,color,times,indentfirst,layout,hyperref,multirow,bm}
\usepackage{hyperref, soul}
\usepackage{ulem}

\usepackage[shortlabels]{enumitem}

\hypersetup{linktocpage,,colorlinks=true}
\ActivateWarningFilters[pdftoc]
\usepackage[colorinlistoftodos,prependcaption,textsize=tiny]{todonotes}
\usepackage{regexpatch}

\newcommand{\starop}[3]{
\begin{gathered}
\xymatrix@=1cm{
& #3 
\\
    #3 & #1 \ar@{-}[u]|{\textstyle#2}  \ar@{-}[d]|{\textstyle#2}  \ar@{-}[l]|{\textstyle#2}  \ar@{-}[r]|{\textstyle#2}  & #3
\\
& #3
}
\end{gathered}
}

\newcommand{\plaquetteop}[3]{
\begin{gathered}
\xymatrix@=0.7cm{%
    #3\ar@{-}[r]|{\textstyle#2}\ar@{}[dr]|{\textstyle#1} & #3\ar@{-}[d]|{\textstyle#2} \\
    #3\ar@{-}[u]|{\textstyle#2} & #3\ar@{-}[l]|{\textstyle#2}
}
\end{gathered}
}

\newcommand{\fplaquetteop}[9]{
\begin{gathered}
\xymatrix@=0.7cm{%
    #6\ar@{-}[r]|{\textstyle#2}\ar@{}[dr]|{\textstyle#1} & #7\ar@{-}[d]|{\textstyle#3} \\
    #8\ar@{-}[u]|{\textstyle#4} & #9\ar@{-}[l]|{\textstyle#5}
}
\end{gathered}
}

\newcommand{\linkop}[2]{
\begin{gathered}
\xymatrix@=0.7 cm{
    #2 \ar@{-}[r]|{\textstyle#1} & #2
}
\end{gathered}
}

\newcommand{\flinkop}[3]{
\begin{gathered}
\xymatrix@=2em{
    #2 \ar@{-}[d]|{\textstyle#1} \\
    #3{} \\
}
\end{gathered}
}

\newcommand{\coordop}[4]{
\begin{gathered}
\xymatrix@=0.5cm{
    &&&\\
    &&&\\
    &#1\ar@{-}[uu]|{\textstyle#4}\ar@{-}[rr]|{\textstyle#3}\ar@{-}[dl]|{\textstyle#2}&&\\
    &&&\\
}
\end{gathered}
}

\usepackage{environ}
\NewEnviron{smallgathered}{
\begin{gathered}
    \scalebox{0.7}{$\BODY$}
    \end{gathered}
}

\usepackage[all]{xy}

\DeclareMathOperator\supp{supp}

\theoremstyle{plain}

\theoremstyle{definition}

\begin{document}

\title{Non-Abelian fracton order from gauging {a mixture of subsystem and global symmetries}
}

\author{Yi-Ting Tu}
\email{ricktu256@gmail.com}
\affiliation{Department of Physics, National Tsing Hua University, Hsinchu 30013, Taiwan}

\author{Po-Yao Chang}
\email{pychang@phys.nthu.edu.tw}
\affiliation{Department of Physics, National Tsing Hua University, Hsinchu 30013, Taiwan}

\begin{abstract}
We construct a general gauging procedure of a pure matter theory on a lattice with a mixture of subsystem and global symmetries.
This mixed symmetry can be either a semidirect product of a subsystem symmetry and a global symmetry,
or a non-trivial extension of them. 
We demonstrate this gauging procedure on a cubic lattice in three dimensions with four examples:
$G=\mathbb{Z}_3^{\text{sub}} \rtimes \mathbb{Z}_2^{\text{glo}}$, 
$G=(\mathbb{Z}_2^{\text{sub}} \times \mathbb{Z}_2^{\text{sub}}) \rtimes \mathbb{Z}_2^{\text{glo}}$,
$1\to \mathbb {Z}_2^\text {sub}\to G\to \mathbb {Z}_2^\text {glo}\to 1$, and
$1\to \mathbb {Z}_2^\text {sub}\to G\to K_4^\text {glo}\to 1$. The former two cases and the last one produce the non-Abelian fracton orders. 
Our construction provides an identification of the electric charges of these fracton orders with irreducible representations of the symmetry.
Furthermore, by constraining the local Hilbert space, the magnetic fluxes with different geometry (tube-like and plaquette-like)
satisfy a subalgebra of the quantum double models (QDMs). This algebraic structure leads to an identification of
the magnetic fluxes to the conjugacy classes of the symmetry.
\end{abstract}

\maketitle


\section{Introduction}

The discovery of topological quantum phases of matter revolutionizes our understanding of quantum many-body phases and leads to many theoretical developments and remarkable experimental results. Fractional quantum Hall states~\cite{Tsui1982,Laughlin1983} are the most studied topological quantum phases that host
fractional excitations with non-Abelian braiding statistics, robust gapless edge modes, and ground state degeneracies. 
The long-range entanglement and intrinsic strong correlations in 
these topological quantum phases bring a key challenge of studying these phases.   

The concept of the emergent ``gauge" degrees of freedom (DOF) in the topological quantum phases provides the breakthrough of tackling these systems. For example, the notion of the flux attachment in the composite electrons gives rise to the dynamical gauge fields
in the fractional quantum Hall 
states. The fractional braiding statistics can be understood from the knot structures of these dynamical gauge fields described by the topological quantum field theories (TQFT)~\cite{Witten1989, Witten1992,Lopez1991, Dong2008, Wen2016}. 
The topological properties of these phases at low-energy are governed by the TQFTs. The structure of TQFT allows us to extract their mathematical structure and enable us to apply the language of the modular tensor category~\cite{KITAEV20062, Rowell2009} to describe these systems. On the other hand, various exactly-solvable models of the lattice gauge theories~\cite{PhysRevD.11.395} also describe the fundamental excitations of the topological quantum phases and
provide insights from studying these models. The famous models are the Kitaev's quantum double models (QDMs)~\cite{A.Yu.Kitaev2002}, which are the exactly-solvable limit of the lattice gauge theories based on a discrete group $G$ on a two-dimensional lattice. The Kitaev's QDMs (the toric code is one of them) have fundamental excitations like anyons whose fusion and braiding statistics are characterized by the modular tensor category.

A generalization of the topological quantum phases in three dimensions is believed to have more exotic phases beyond current theoretical developments. The ``fracton order", is one of the novel lattice models  discovered recently~\cite{Vijay2015AExcitations,PhysRevLett.94.040402,PhysRevA.83.042330,PhysRevB.88.125122,PhysRevB.92.235136,doi:10.1146/annurev-conmatphys-031218-013604,PhysRevX.8.031051,PhysRevB.97.165106,PhysRevX.9.021010,vijay2017generalization,PhysRevB.95.245126,BRAVYI2011839,PhysRevB.96.165105,PhysRevLett.119.257202,vijay2017isotropic,PhysRevB.96.165106,PhysRevB.96.224429,Slagle2018,bulmash2018generalized,10.21468/SciPostPhys.6.1.015,PhysRevB.97.041110,PhysRevB.97.134426,PhysRevLett.107.150504,PhysRevB.95.155133,PhysRevB.97.144106,PhysRevB.97.125101,PhysRevB.97.125102,PhysRevB.98.165140,PhysRevA.98.022332}.  These three-dimensional lattice models share common features with the two-dimensional counterparts such as ground state degeneracies on the torus and fractional excitations. However, some of the excitations, which are referred to as fractons, cannot be moved by local operators, and are intrinsically immobile. This unique feature has no counterparts in two dimensions. 
It is desired to develop new concepts and mathematical tools for understanding the properties of the fracton order.
Furthermore, the fracton order is related to tensor gauge theories~\cite{Pretko_2020} and is proposed to be applicable for quantum memory~\cite{PhysRevLett.111.200501,RevModPhys.88.045005}.

A natural extension of current existing theories of fractons is the
non-Abelian generalization. These non-Abelian fractons have potential applications to topological quantum computation and quantum memory.
A variety of the non-Abelian constructions,
including
gauging bilayer/permutation symmetries~\cite{Prem2019,BulmashGaugingDegeneracies}, cage-net models~\cite{PhysRevX.9.021010},
the layer constructions~\cite{vijay2017generalization,williamson2020designer, PhysRevB.99.155118} {and the topological defect networks~\cite{PhysRevResearch.2.043165,wang2020nonliquid}}
are proposed.
The former construction relies on gauging a global $\mathbb{Z}_2$/permutation symmetry of the ``Abelian" fracton models and the second one
requires flux-string condensation.  These models are restricted in a certain geometry (e.g., the cage-net model and the layer construction) or a specific layer construction (e.g, gauging bilayer fractons). 
Due to the restrictions from these proposals, a generalization to other non-Abelian fracton orders is not straightforward. To overcome this difficulty, we propose a general gauging procedure of constructing the non-Abelian fracton from gauging a pure matter theory on a lattice with a 
mixture of a subsystem symmetry and a global symmetry. This mixed symmetry can be a semidirect product of a subsystem symmetry and a global symmetry,
or a non-trivial extension of them.
The gauge principle on the lattice has tremendous success in constructing many models with topological order.
In particular, for a three-dimensional paramagnetic state with Abelian subsystem symmetries,
the gauged model in the exactly-solvable limit is the Abelian fracton lattice models~\cite{Shirley_2019}.
Following our gauge principle, one can obtain the (non-)Abelian fracton from gauging the (non-)Abelian versions of the mixed symmetry.
We demonstrate several examples of the gauged Hamiltonian in the exactly-solvable limit that host (non-)Abelian fractons and other excitations.
To the best of our knowledge, 
gauging the mixed symmetry on a lattice has not been discussed in the literature~\footnote{The continuous version of this mixed symmetry has been discussed in the field theory content~\cite{Wang2019Higher-RankEmbeddon,PhysRevResearch.2.043219,PhysRevResearch.3.013185}. Also, some fracton models can be constructed from gauging subsystem symmetries from the symmetry-protected topological order with a combination of global and subsystem symmtries~\cite{Stephen2020}.}.

Comparing with  gauging the bilayer/permutation fracton orders,
which the non-Abelian fractons are constructed from
(1) gauging subsystem symmetry of the matter field,
(2) taking the exactly-solvable limit, and (3) gauging the global symmetry, our gauging procedure only requires one-step gauging. 
This one-step gauging process is very general and allows us to construct various lattice models that host (non-)Abelian fracton orders.
We demonstrate four examples by gauging 
$G=\mathbb{Z}_3^\text{sub}\rtimes\mathbb{Z}_2^\text{glo}$,
$G=(\mathbb{Z}_2^\text{sub}\times\mathbb{Z}_2^\text{sub})\rtimes\mathbb{Z}_2^\text{glo}$,
$1\to \mathbb {Z}_2^\text {sub}\to G\to \mathbb {Z}_2^\text {glo}\to 1$, and
$1\to \mathbb {Z}_2^\text {sub}\to G\to K_4^\text {glo}\to 1$. The former two cases and the last one produce the non-Abelian fracton orders.
The third case is the non-trivial Abelian fracton order which has a hybrid structure~\cite{tantivasadakarn2021hybrid}. 

Besides the non-Abelian fracton orders can be systematically constructed from our one-step gauging process,
a direct analogy to the Kitaev's QDMs can be obtained.
The excitations in QDMs are characterized by the local operators around a vertex $v$ and a neighboring plaquette $p$. The algebra of these local operators is generated by $A_g,g\in G$ and $B_h,h\in G$, where $A_g$ is the gauge transformation at $v$, and $B_h$ is the projector onto the subspace that a product of group elements around a plaquette $p$ equals $h$.
    They satisfy
    \begin{align}
        &A_{g_1}A_{g_2}=A_{g_1 g_2},\quad
        B_{h_1}B_{h_2}=\delta_{h_1,h_2}B_{h_1}, \notag \\
        &A_gB_h =B_{ghg^{-1}}A_g, \quad
        \sum_{h} B_h =1.
        \label{eq:qdm}
    \end{align}
    The pure electric charges are classified by the irreducible representations (irreps) of $G$ in that $A_g$ acts on the Hilbert subspace with a specific charge according to that representations, while $B_h=\delta_{h,e}$ ($e$ is the identity) on this subspace.
    The pure magnetic fluxes are classified by the conjugacy classes of $G$ in that the Hilbert subspace with a specific flux is generated by the state with $B_h=\delta_{h,g}$, where $g$ runs over all elements of that conjugacy class.

In our one-step gauging procedure,
the local gauge transformations can generate a local symmetry $G^\text{local}$,
which allows us to identify the electric charges (including non-Abelian fractons) with the irreps of $G^\text{local}$.
In addition, the magnetic excitations together with the local gauge transformations form a subalgebra of the QDMs
in the constrained local Hilbert space. This algebraic structure allows us to identify the magnetic excitations with the conjugacy classes of $G^\text{local}$.
With the one-step gauging, we construct explicitly the underlying local $G^\text{local}=S_3$, $D_4$, $\mathbb{Z}_4$, and $Q_8$ symmetries,
by gauging 
$G=\mathbb{Z}_3^\text{sub}\rtimes\mathbb{Z}_2^\text{glo}$,
$G=(\mathbb{Z}_2^\text{sub}\times\mathbb{Z}_2^\text{sub})\rtimes\mathbb{Z}_2^\text{glo}$,
$1\to \mathbb {Z}_2^\text {sub}\to G\to \mathbb {Z}_2^\text {glo}\to 1$, and
$1\to \mathbb {Z}_2^\text {sub}\to G\to K_4^\text {glo}\to 1$,
respectively. 
We further identify the charges with the irreps of these symmetries,  
and the magnetic fluxes with the conjugacy classes of $G^\text{local}$ in these systems.
These excitations are summarized in Tables \ref{tab:z3}, \ref{tab:z2}, \ref{tab:z4}, and \ref{tab:q8}.



This paper is organized as follows:
In section \ref{sec:gauging}, we present the details of the general gauging procedure.
In section \ref{sec:QDMs}, we demonstrate the construction of (non-)Abelian fracton models by our gauging procedure.
We show that the electric and magnetic excitations satisfy the algebraic structure of the Kitaev's QDMs in the 
constrained Hilbert space. The algebraic structure of the electric charges and the magnetic fluxes allows us to classify
the (non-)Abelian fracton orders.
We construct two new non-Abelian fracton orders by
gauging $G=\mathbb{Z}_3^\text{sub}\rtimes\mathbb{Z}_2^\text{glo}$ and
$1\to \mathbb {Z}_2^\text {sub}\to G\to K_4^\text {glo}\to 1$.
 Our gauging procedure is very general. 
 It also reproduces the non-Abelian fracton in the gauged bilayer X-Cube code~\cite{Prem2019,BulmashGaugingDegeneracies}
 and also the non-trivial $\mathbb Z_4$ fracton order which has a hybrid structure discussed in Ref.~\cite{tantivasadakarn2021hybrid}.
Moreover, it can be easily generalized to more exotic mixed symmetries. One interesting example will be a
 generalization of the $\mathbb{Z}_N^\text{sub}\rtimes\mathbb{Z}_2^\text{glo}$, which is related to the continuous field-theoretic approach in  Refs.~[\onlinecite{Wang2019Higher-RankEmbeddon,PhysRevResearch.2.043219,PhysRevResearch.3.013185}]. 
Finally, we summarize our results in section \ref{sec:con}.

\section{General gauging procedure}\label{sec:gauging}

We start with an ``ungauged'' system, which can be described as consisting purely of the matter DOF on lattice sites.
The ungauged systems is described by $H=H_{o}+H_{n}$,
where $H_{o}$ is the onsite terms and $H_n=-\sum_c J_c c+\text{H.c.}$ is the non-onsite couplings with coupling strengths $J_c$.
The ungauged Hamiltonian has a symmetry $G$ such that the non-onsite couplings $c$ are the minimal symmetric couplings, e.g., the nearest neighbor interactions and the plaquette interactions.
Here, $G$ can be a global symmetry, a subsystem symmetry, or a mixture of them.
For a subsystem symmetry, $G$ is the group generated by the operation on each subsystem. 
For example, the subsystem $\mathbb{Z}_2$ spin-flip symmetry group of a cubic lattice with spin-$\frac{1}{2}$ on each site is denoted by $G=\mathbb{Z}_2^\text{sub}$.
 The group $\mathbb{Z}_2^\text{sub}$ is not actually isomorphic to the group $\mathbb{Z}_2$, but is a large group generated by the spin-flipping on every shifted coordinate planes of the cubic lattice. 

 

The gauging principle is to promote the symmetry $G$ to be local. The famous example is gauging the $U(1)$ symmetry.
The matter field $\psi(x)$ has a global $U(1)$ symmetry $\psi(x) \to \psi(x)e^{i \phi} $ such that the Hamiltonian $H= \int dx\psi^{\dagger}(x) (-i \partial_x) \psi(x)$ is invariant. While promoting the global $U(1)$ to be local, $\psi(x) \to \psi(x)e^{i \phi(x)}$, one need to introduce a gauge field $a(x)$ to ensure the Hamiltonian is invariant. The corresponding gauge transformation of the gauge field is $a(x) \to a(x)+\partial_x \phi(x)$. We also add the magnetic field and electric field terms in $H$ and promote the derivatives to the covariant derivatives.
With this example in mind,
we describe the general gauging procedure as follows:
\begin{enumerate}
    \item For each of the $G$-invariant minimal couplings $c$ in $H_n$, we define a gauge DOF $\tau_{c}$ with Hilbert space dimension equals the number of distinct eigenvalues of $c$.
    We label the computational basis of the $\tau_c$ with respect to the eigenvalues of the minimal couplings~\footnote{If the coupling $c$ has two distinct eigenvalues $+1$ and $-1$, we label the computational basis of $\tau_c$, $|0\rangle$ and $|1 \rangle$ respectively. If the coupling $c$ has three distinct eigenvalues $1$, $e^{i 2\pi/3}$, and $e^{-i 2\pi/3}$, we label the state
    of $\tau_c$, $|0\rangle$, $|1 \rangle$, and $|2\rangle$ respectively.}.
    These gauge fields $\tau_{c}$ are placed on the links or plaquettes for the corresponding minimal couplings $c$.
    We define $\supp c$ to be the set of vertices that $c$ acts on.
    When all the matter sites are in the same state such that the matter state is the simultaneous eigenstate of all $c$, the corresponding state of all the gauge DOF constitute a natural flat connection, and is denoted by $|0\rangle$.

    \item \label{step2}Construct the gauge transformation $A_{v,g}$:
    On the matter DOF, 
    $A_{v,g}$ acts on the matter DOF on the site $v$ by $g$ as the original symmetry transformation, but leave other matter DOF on other sites unchanged. 
    This action leads to changes of the eigenvalues of the non-onsite couplings $c$.
    To ensure the system is invariant under the gauge transformation, the gauge DOF must compensate the changes of the eigenvalues of $c$. 
    The gauge transformations need to satisfy $A_{v,g_1}A_{v,g_1}=A_{v,g_1g_2}$ and
    $[A_{v_1,g_1},A_{v_2,g_2}]=0$ when $v_1\neq v_2$. This means that $G$ is promoted to a local symmetry $G^\text{local}$ of the gauged system.
    We demonstrate a systematical way for construction the gauge transformation in {Appendix \ref{sec:a}}~\footnote{For the nontrivial mixture of the global and subsystem symmetries,  the branch cut from the twist defect charge leads to a complication of constructing the inverse transformation $g^{-1}$ acting on the gauge fields $\tau_c$. We give a systematical construction of $A_{v,g}$ of $G$ which can be a global symmetry, a subsystem symmetry, or a semidirect product of them. The former two symmetries do not suffer the complication and our construction reproduces the gauge transformation defined in Ref.\ [\onlinecite{Shirley_2019}].}.
    
    \item Construct the gauge flux:
    The existence of flux means that the gauge field is ``twisted'', which corresponds to the violation of some ``relations'' that should be satisfied when the connection is flat.
    Those ``relations'' come from the dependence between the couplings $c$ in the ungauged system. That is, some of $c$ can be combined (product or sum) to the identity.
    Similar to the ``minimal'' coupling, we can find the ``minimal'' relations  consists of local combination of the $c$ on a plaquette or a tube that produce the identity.
    We denote these relations by $R_r(\{c\}) =1$, where
    $r$ is the label of the local combination.
    We define $\supp r$ to be the union of $\supp c$ for all $c$ in the combination.
        Then the corresponding flux term is
        \begin{equation}            B_r=R_r(\{Z\})\prod_{\supp r'\subsetneq \supp r}P_{r'},
        \end{equation}
        where $R_r(\{Z\})$
        is the relation with $c$ replaced by the computational-basis operators (generalized $Z$) of the corresponding $\tau_c$,
        and $P_{r'}$ is the projector onto the $B_{r'}=1$ subspace~\footnote{The gauge flux $B_r$ may be attached by another gauge flux $B_{r'}$ in the submanifold of the manifold of $B_r$. The projector ensures no additional twisted of the gauge field from the gauge flux $B_r'$.  }.
        These projectors must be included when some smaller $r'$ is contained in the geometry of $r$ to ensure that the final Hamiltonian is gauge invariant~\footnote{
        In the approach of Ref.\ [\onlinecite{Prem2019}], the gauging map of the layer-swap symmetry automatically introduces such projectors on the flux term of the X-Cube code. Note that putting zero-flux projectors in some terms of the Hamiltonian is required for some systems even for global $\mathbb{Z}_2$ symmetry, such as the Levin-Gu model\cite{PhysRevB.86.115109}.}.

    \item
  
    Construct the gauged non-onsite terms $H_n(\tau)=-\sum_c J_c c(\tau)+\text{H.c.}$:
    The gauged coupling is defined to be a gauge-invariant (commutes with all $A_{v,g}$) operator acting on all the matter and gauge DOF within the geometry of $c$,
    \begin{equation}
        \supp c(\tau)\subseteq\supp c\cup \{\tau_{c'}\mid \supp c'\subseteq\supp c\}.
    \end{equation}
    When the gauge DOF of a state is trivial, $c(\tau)$ reduces to the ungauged coupling $c$:
    \begin{align}
    \label{eq:gauged_coup}
        &c(\tau)\left(|m\rangle_{\text{matter}}\otimes|0\rangle_{\text{gauge}}\right)   \notag\\
        =&\left(c|m\rangle_{\text{matter}}\right)\otimes|0\rangle_{\text{gauge}}.
    \end{align}

    For a general state $|\alpha\rangle$, 
    $c(\tau)|\alpha\rangle$ can be computed by expanding $|\alpha\rangle$ with the computational basis of the gauged DOF. For each term, we use a series of $A_{v,g}$ to rotate the gauge DOF of each term to be $|0\rangle$. We apply (\ref{eq:gauged_coup}) to the terms that succeed, and drop the terms that fail.

\item Construct the ``electric field'' terms $H_e$: The terms are the dynamical part of the gauge fields, which are gauge-invariant operators. 
\end{enumerate}
The final gauged Hamiltonian is
\begin{align}
H_{g}=H_o+ H_{n}(\tau)-\sum_r B_r+H_e,
\end{align}
and the physical Hilbert space has the constraint $A_{v,g}|\text{physical}\rangle=|\text{physical}\rangle$. The gauged Hamiltonian $H_g$ has a generalization of the confined-deconfined and the Higgsed-deconfined phase transition~\cite{
Fradkin1979}.

The exactly-solvable limit of the gauged Hamiltonian $H_g$ can be obtained by eliminating the non-onsite couplings and hopping terms of the fluxes. (taking both coupling and the hopping strengths to be vanishing).
\begin{equation}
    H_g=H_o-\sum_r B_r.
\end{equation}

If $G$ is a global or subsystem $\mathbb{Z}_2$ symmetry, then the above procedure reproduces the gauging process of Ref.~[\onlinecite{Shirley_2019}].
Also, if $G$ is a global symmetry with the regular representation on the matter on a 2D lattice and we eliminate the matter DOF by choosing the unitary gauge,
then the above procedure gives the Kitaev $G$-QDM introduced in Ref.~[\onlinecite{A.Yu.Kitaev2002}].
The idea of a semidirect product of global and subsystem symmetry in the continuous quantum field theories appeared in Refs.~[\onlinecite{Wang2019Higher-RankEmbeddon,PhysRevResearch.2.043219,PhysRevResearch.3.013185}].

    We argue that, if $G$ is a mixture of planar subsystem and global symmetries, then the resulting model contains both fractons and mobile particles (a hybrid order).
The argument is as follows: if $G$ contains a planar subsystem part, there must be some minimal plaquette-couplings; if $G$ also contains a global part not generated from the subsystem part, then there must also be minimal link-couplings.
Now the strings and membranes of the electric charges can be constructed from the minimal couplings:
If we pull the two ends of a link-coupling apart, or the four corners of a plaquette-coupling apart, the resulting operators are non-minimal couplings invariant under $G$.
The gauged version of these non-minimal couplings are the string (from link-couplings) and membrane (from plaquette-couplings) for the charges.
Having both types of couplings means that we have a hybrid order containing both mobile particles and fractons.
Moreover, if $G$ is not a direct product of a subsystem and a global symmetry, then the fusion rule of the charges will not be trivial.
It is because that the corresponding QDM will not be the direct product of the two types of charges corresponding to the global and subsystem ones. The correspondence with QDM will be discussed in the following examples.

In the following, we focus on the $H_g$ for  
$G=\mathbb{Z}_3^\text{sub}\rtimes\mathbb{Z}_2^\text{glo}$,
$G=(\mathbb{Z}_2^\text{sub}\times\mathbb{Z}_2^\text{sub})\rtimes\mathbb{Z}_2^\text{glo}$,
$1\to \mathbb {Z}_2^\text {sub}\to G\to \mathbb {Z}_2^\text {glo}\to 1$, and
$1\to \mathbb {Z}_2^\text {sub}\to G\to K_4^\text {glo}\to 1$.
In the exactly-solvable limit,
these gauged lattice models describe  (non-)Abelian fractons.

\section{classification of excitations with the quantum double models}
\label{sec:QDMs}
Before discussing the examples, we highlight our main results.
From the generic gauging principle, the (non-)Abelian fracton order can be obtained by gauging a mixed symmetry with a subsystem symmetry
and a global symmetry in the exactly-solvable limit.
Here we consider the case on a three-dimensional cubic lattice with a mixture of a global and a two-dimensional subsystem symmetry.
While promoting the mixed symmetry $G$ to be local,
the ``gauge fields'' are introduced on links and plaquettes. The local gauge transformations generate the local symmetry $G^\text{local}$. 
The magnetic flux operators defined from the ``gauge fields'' have two kinds of geometry,  one is plaquette-like and the other 
is tube-like. The operators are denoted as $B_p$ and $B_t$, where $p$ and $t$ are the label of the plaquette and the tube, respectively.
The gauged Hamiltonian in the exactly solvable-limit
hosts point-like and string-like excitations. 
The point-like excitations are created at the corners of the membrane operators or
the ends of the string operators. On the other hand, the string-like excitations are created at the boundaries of the membrane operators. 
The point-like excitations correspond to the electric charges are the excitations of the local gauge transformation $A^{\rm u}_{v, g}$.
Here the superscript ${\rm u}$ indicates the unitary gauge which eliminates the matter DOF.
 $v$ is the location of the vertex, and $g$ is the symmetry operation.
 These excitations are local with respect to the vertex $v$,
and can be specified from the local operators which form a representation of $G^\text{local}$ on the Hilbert space.
Hence we can identify electric charges of the $G$-fracton model
with the irreducible representation of $G^\text{local}$. 

For the magnetic fluxes, the excitations of $B_p$ are loop-like and the excitations of $B_t$ are point-like. Due to such difference in geometry,
a constraint of the local Hilbert space is needed for making possible comparison.
We consider the geometry of the plaquette $p_0$ and the tube $t_0$ adjacent to a chosen 
    vertex $v_0$:
\begin{equation}
    p_0=
    \renewcommand{\labelstyle}{\textstyle}
\begin{gathered}
\xymatrix@!0@=1.3em{
&\ar@{-}[ld]\ar@{-}[d]\ar@{-}[rrr]\ar@{}[rrd]&&&\ar@{-}[ld]\\
    {v_0}\ar@{-}[rrr]|{\bullet}\ar@{-}[ddd]|{\bullet}\ar@{}[ddr]&&&\ar@{-}[ddd]|{\bullet}\ar@{}[ddr]&\\
&&&&\\
&\ar@{-}[ld]\ar@{-}[rr]\ar@{-}[uu]\ar@{}[rrd]&&&\ar@{-}[uuu]\ar@{-}[l]\\
    \ar@{-}[rrr]|{\bullet}&&&\ar@{-}[ru]
}
\end{gathered},\quad
    t_0=
    \renewcommand{\labelstyle}{\textstyle}
\begin{gathered}
\xymatrix@!0@=1.3em{
&\ar@{-}[ld]\ar@{-}[d]\ar@{-}[rrr]\ar@{}[rrd]|{\bullet}&&&\ar@{-}[ld]\\
    {v_0}\ar@{-}[rrr]\ar@{-}[ddd]\ar@{}[ddr]|{\bullet}&&&\ar@{-}[ddd]\ar@{}[ddr]|{\bullet}&\\
&&&&\\
&\ar@{-}[ld]\ar@{-}[rr]\ar@{-}[uu]\ar@{}[rrd]|{\bullet}&&&\ar@{-}[uuu]\ar@{-}[l]\\
\ar@{-}[rrr]&&&\ar@{-}[ru]
}
\end{gathered},
\label{eq:geometry}
\end{equation} 
where the constituent of that geometry is indicated with dots.
The magnetic fluxes with the constraints of the local Hilbert space
 can be classified from the algebraic relation as the QDMs in Eq. (\ref{eq:qdm}). 
Define the projectors on this cube adjacent to $v_0$:
\begin{equation}
    P_\text{side}=
    \renewcommand{\labelstyle}{\textstyle}
\begin{gathered}
\xymatrix@!0@=1.3em{
&\ar@{-}[ld]\ar@{-}[d]\ar@{-}[rrr]\ar@{}[rrd]|{P_p}&&&\ar@{-}[ld]\\
    {v_0}\ar@{-}[rrr]\ar@{-}[ddd]\ar@{}[ddr]|{P_p}&&&\ar@{-}[ddd]\ar@{}[ddr]|{P_p}&\\
&&&&\\
&\ar@{-}[ld]\ar@{-}[rr]\ar@{-}[uu]\ar@{}[rrd]|{P_p}&&&\ar@{-}[uuu]\ar@{-}[l]\\
\ar@{-}[rrr]&&&\ar@{-}[ru]
}
\end{gathered},\quad
    P_\text{corner}=
    \renewcommand{\labelstyle}{\textstyle}
\begin{gathered}
\xymatrix@!0@=1.3em{
&\ar@{-}[ld]\ar@{-}[d]\ar@{-}[rrr]\ar@{}[rrd]|{P_p}\ar@{}[dddrrr]|{P_p}&&&\ar@{-}[ld]\\
    {v_0}\ar@{-}[rrr]\ar@{-}[ddd]\ar@{}[ddr]&&&\ar@{-}[ddd]\ar@{}[ddr]|{P_p}&\\
&&&&\\
&\ar@{-}[ld]\ar@{-}[rr]\ar@{-}[uu]\ar@{}[rrd]|{P_p}&&&\ar@{-}[uuu]\ar@{-}[l]\\
\ar@{-}[rrr]&&&\ar@{-}[ru]
}
\end{gathered} 
\label{eq:constraint}
\end{equation}
where $P_p$ denotes the projector onto the subspace without $B_p$ flux at the plaquette it is drawn on.
Imposing the constraint $P_\text{side}=1$ means there should be no $B_p$ flux
through the upper, lower, left, and right faces of the cube. This means that if there is $B_p$ flux through the front face ($P_{p_0}=0$), it must come out from the back face. Equivalently, if there is a loop-like excitation going through our region of discussion, the region must be at the side of the loop.
On the other hand, imposing $P_\text{corner}=1$ means there should be no $B_p$ flux through the upper, lower, right, and back faces of the cube.
This means that if there is $B_p$ flux through the front face ($P_{p_0}=0$), it must come out from the left face. Equivalently, if there is a loop-like excitation going through our region of discussion, the region must be at the corner of the loop.
Here, we emphasize again that the excitation of $B_{p_0}$ can be either at the side of the string excitation
which is specified by the protector $P_\text{side}=1$, or at the corner of the string excitation which is specified by the protector $P_\text{corner}=1$.

With certain constraint of the local Hilbert space, we find the flux operators can be mapped to a subalgebra of the QDM algebra in Eq.~(\ref{eq:qdm}) with gauge group $G^\text{local}$.
From all the examples we demonstrated, the subalgebra is enough to distinguish all of the conjugacy classes of $G^\text{local}$~\footnote{One may need to introduce additional fluxes to form
the subalgebra in some models, such as the the second and the third examples in Sec.~\ref{sec:z2} and Sec.~\ref{sec:z4}.}. This allows us to
identify the magnetic fluxes of the fracton model constructed by the mixed symmetry $G$ with the conjugacy classes of $G^\text{local}$.

    It may happen that certain constraint of the local Hilbert space does the above, but some other constraint does not, and corresponds (at least for the fluxes) to the QDM of another gauge group instead. In the example we study ($1\to\mathbb{Z}_2^\text{sub}\to G\to\mathbb{Z}_2^\text{glo}\to 1$), this situation leads to the fusion of two flexible loops into four lineons, a phenomenon first observed in Ref.~\cite{tantivasadakarn2021hybrid}.
    In general, we expect that such constraint-dependent QDMs in the Abelian cases\footnote{
The non-Abelian case is more tricky. Since some charges and point-like fluxes can be hidden along the string, two such strings can be fuse back into any such point line excitations along the string. However, the concept of fusion here may be problematic, since we need to apply an extended operator to fuse the string, and the same type of extended operators can also use to alter those hidden excitations, which makes the fusion result ambiguous.
Due to the complication, we do not consider the fusion of such objects in this article.
} lead to the fusion of excitations of one geometry into that of another geometry. In particular, the parts of the original geometry satisfying each constraint (e.g.\ sides, corners) fuse as the corresponding QDM.

Here we demonstrate the classifications of charges and fluxes from the general gauging procedure and the algebraic structure of QDMs with four examples in detail.

\subsection{Example: gauging \texorpdfstring{$G=\mathbb{Z}_3^\text{sub}\rtimes\mathbb{Z}_2^\text{glo}$}{Z\_3sub rx Z\_2glo}}\label{sec:z3}

The ungauged system is a three dimensional (3D) cubic lattice with a qutrit (labeled as $0$) and a qubit (labeled as $2$) on each site, corresponding to the quantum clock matter and the twist defect charge. The Hamiltonian is
    \begin{align}
        H_o&=-\sum_\text{sites}(XI+X^\dagger I+IX),\notag \\
        H_n&=-J_0\sum_\text{plaquettes}\fplaquetteop{c_0}{}{}{}{}{}{}{}{}
        -J_2\sum_\text{links}\flinkop{c_2}{}{}+\text{H.c.},
    \end{align}
    where the minimal couplings are
        \begin{align}
            \fplaquetteop{c_0}{}{}{}{}{}{}{}{}&:=\left(\fplaquetteop{}{}{}{}{}{Z[0]}{Z^\dagger I}{Z^\dagger I}{ZI}+\fplaquetteop{}{}{}{}{}{Z^\dagger[1]}{ZI}{ZI}{Z^\dagger I}\right),\notag \\
            \flinkop{c_2}{}{}&:=\flinkop{}{IZ}{IZ}.
        \end{align}
         Here $X$ and $Z$ are the (generalized) Pauli operators,  $[\psi]:=|\psi\rangle\langle\psi|$ is denoted as a projector, and the operators on the first and the second entries on the site act on the qutrit and the qubit, respectively.
The other two directions are represented as
        \begin{align}
\renewcommand{\labelstyle}{\textstyle}
\begin{gathered}
\xymatrix@=0.5cm{
    &\ar@{-}[dd]\\
    \ar@{-}[ur]\ar@{}[dr]|{c_0}&\\
    &\ar@{-}[dl]\\
    \ar@{-}[uu]&
}
\end{gathered}
           &:=\left(
           \begin{gathered}
\xymatrix@=0.2cm{
    &Z^\dagger I\ar@{-}[dd]\\
    Z[0]\ar@{-}[ur]\ar@{}[dr]&\\
    &Z I\ar@{-}[dl]\\
    Z^\dagger I\ar@{-}[uu]&
}
\end{gathered}
           +
           \begin{gathered}
\xymatrix@=0.2cm{
    &ZI\ar@{-}[dd]\\
    Z^\dagger[1]\ar@{-}[ur]\ar@{}[dr]&\\
    &Z^\dagger I\ar@{-}[dl]\\
    Z I\ar@{-}[uu]&
}
\end{gathered}
\right), \notag \\
\renewcommand{\labelstyle}{\textstyle}
\begin{gathered}
\xymatrix@=0.5cm{
    &\ar@{-}[rr]\ar@{}[dr]|{c_0}&&\ar@{-}[dl]\\
    \ar@{-}[ur]&&\ar@{-}[ll]&\\
}
\end{gathered}
           &:=\left(
           \begin{gathered}
\xymatrix@=0.2cm{
    &Z^\dagger I\ar@{-}[rr]\ar@{}[dr]&&ZI\ar@{-}[dl]\\
    Z[0]\ar@{-}[ur]&&Z^\dagger I\ar@{-}[ll]&\\
}
\end{gathered}\right.\notag\\
           &+\left.
           \begin{gathered}
\xymatrix@=0.2cm{
    &ZI\ar@{-}[rr]\ar@{}[dr]&&Z^\dagger I\ar@{-}[dl]\\
    Z^\dagger[1]\ar@{-}[ur]&&ZI\ar@{-}[ll]&\\
}
\end{gathered}
\right),
        \end{align}
These graphical notations of operators on the lattice define the minimal couplings and are included in the $\sum_\text{plaquettes}$ in $H_n$.
All graphical notations in this article follow these orientations.

    $G=\mathbb{Z}_3^\text{sub}\rtimes\mathbb{Z}_2^\text{glo}$ is generated by $g_P^{(0)}$ and $g^{(2)}$, where $g_P^{(0)}$ acts on the matter by mapping $|i\rangle$ to $|i+1\mod3\rangle$ on the qutrit on each site of a shifted coordinate plane $P$; $g^{(2)}$ acts on the matter by flipping the qubit and mapping $|i\rangle$ to $|-i\mod3\rangle$ on the qutrit on each site of the entire system. That is
    \begin{align}
    &g_P^{(0)}=\prod_{\substack{\text{sites}\in P}}XI,\quad g^{(2)}=\prod_{\substack{\text{sites}\in \text{all}}}SX,  \notag   \\
    &S=|0\rangle\langle 0|+|2\rangle\langle 1|+|1\rangle\langle 2|.
    \end{align}

The procedure of gauging works as follows:
\begin{enumerate}
    \item We put one qubit on each link (corresponds to $c_2$) and one qutrit on each plaquette (corresponds to $c_0$). 
             The generalized Pauli operators for the qutrits are denoted as  $X_0:=X$, $X_1:=X^\dagger$, $Z_0:=Z$,
             and $Z_1:=Z^\dagger$ for notational convenience. I.e., the generalized Pauli operators have an additional subscript for the gauge qutrits on each
             plaquettes.  
       \item 

Based on the general gauging procedure we described in Sec. \ref{sec:gauging}.\ref{step2}, the gauge transformations are
\begin{equation}
\vspace{-1em}
    A_{v,g_P^{(0)}}=\sum_{\substack{a,b,c,\\d,e,f\\=0,1}}
    \renewcommand{\labelstyle}{\textstyle}
\begin{smallgathered}
\xymatrix@!0{
    &&&&&\ar@{-}[d]&&&\\
    &\ar@{-}[rrr]|{\color{blue}[b]}&&&\ar@{-}[ru]\ar@{-}[ld]&\ar@{-}[dd]&&\ar@{-}[lll]&\\
    &&&\ar@{-}[ddd]|(.56){\color{orange}[f]}&&&&&\\
    &&\ar@{-}@[green!50!black][ld]&\ar@{-}@[green!50!black][l]&\ar@{-}@[green!50!black][l]&\ar@{-}@[green!50!black][l]\ar@{-}@[green!50!black][rr]&&\ar@{-}@[green!50!black][r]&\ar@{-}@[green!50!black][ld]\\
    &\ar@{-}[uuu]\ar@{-}[d]&&\ar@{-}[ll]|(.25){[c]}&X_{}^\dagger I\ar@{-}[uuu]|{[a]}\ar@{-}[rrr]\ar@{-}[d]\ar@{-}[l]\ar@{-}[ru]\ar@{-}[ld]|{[e]}\ar@{}[llluuu]|(.6){\color{blue}X_{a+b}^\dagger}\ar@{}[luuu]|(.45){\color{orange}X_{e+f}^\dagger}\ar@{}[ruuuu]|(.5){\color{orange}X_{a}}\ar@{}[rrruuu]|(.6){\color{blue}X_{a}}\ar@{}[llu]|(.7){\color{green!50!black}X_{c}}\ar@{}[rrrru]|(.5){\color{green!50!black}X_{0}^\dagger}\ar@{}[rrrddd]|(.6){\color{blue}X_{0}^\dagger}\ar@{}[rdd]|(.65){\color{orange}X_{0}^\dagger}\ar@{}[ldddd]|(.5){\color{orange}X_{e}}\ar@{}[lllddd]|(.6){\color{blue}X_{c}}\ar@{}[rrd]|(.7){\color{green!50!black}X_{e}}\ar@{}[lllld]|(.5){\color{green!50!black}X_{c+d}^\dagger}&\ar@{-}[u]&&\ar@{-}[uuu]\ar@{-}[ddd]&\\
    \ar@{-}@[green!50!black][ru]|{\color{green!50!black}[d]}&\ar@{-}[dd]&&\ar@{-}@[green!50!black][lll]\ar@{-}@[green!50!black][rrr]&\ar@{-}[dd]&\ar@{-}[u]&\ar@{-}@[green!50!black][ru]&&\\
                       &&&&&\ar@{-}[u]&&&\\
                       &\ar@{-}[rr]&&\ar@{-}[r]&\ar@{-}[ru]\ar@{-}[ld]&&&\ar@{-}[lll]&\\
                       &&&\ar@{-}[uuu]&&&&&
}
\end{smallgathered}\hspace{-0.5em}, \notag \label{eq:AvgP0z3}
\end{equation}
\vspace{-2em}
\begin{equation}
A_{v,g^{(2)}}=
    \renewcommand{\labelstyle}{\textstyle}
\begin{smallgathered}
\xymatrix@!0{
    &&&&&\ar@{-}[d]&&&\\
    &\ar@{-}[rrr]&&&\ar@{-}[ru]\ar@{-}[ld]&\ar@{-}[dd]&&\ar@{-}[lll]&\\
    &&&\ar@{-}[ddd]&&&&&\\
    &&\ar@{-}@[green!50!black][ld]&\ar@{-}@[green!50!black][l]&\ar@{-}@[green!50!black][l]&\ar@{-}@[green!50!black][l]\ar@{-}@[green!50!black][rr]&&\ar@{-}@[green!50!black][r]&\ar@{-}@[green!50!black][ld]\\
    &\ar@{-}[uuu]\ar@{-}[d]&&\ar@{-}[ll]|(.25){X}&S_{}X_{}\ar@{-}[uuu]|{X}\ar@{-}[rrr]|{X}\ar@{-}[d]\ar@{-}[l]\ar@{-}[ru]|{X}\ar@{-}[ld]|{X}\ar@{}[llluuu]\ar@{}[luu]\ar@{}[ruuuu]\ar@{}[rrruuu]\ar@{}[llu]\ar@{}[rrrru]|(.5){\color{green!50!black}S}\ar@{}[rrrddd]|(.6){\color{blue}S}\ar@{}[rdd]|(.65){\color{orange}S}\ar@{}[ldddd]\ar@{}[lllddd]\ar@{}[rrd]\ar@{}[lllld]&\ar@{-}[u]&&\ar@{-}[uuu]\ar@{-}[ddd]&\\
    \ar@{-}@[green!50!black][ru]&\ar@{-}[dd]&&\ar@{-}@[green!50!black][lll]\ar@{-}@[green!50!black][rrr]&\ar@{-}[dd]|(.25){X}&\ar@{-}[u]&\ar@{-}@[green!50!black][ru]&&\\
                       &&&&&\ar@{-}[u]&&&\\
                       &\ar@{-}[rr]&&\ar@{-}[r]&\ar@{-}[ru]\ar@{-}[ld]&&&\ar@{-}[lll]&\\
                       &&&\ar@{-}[uuu]&&&&&
}
\end{smallgathered}\hspace{-0.5em}.\label{eq:AvgP2z3}
\vspace{-1em}
\end{equation}
where  $v$ in $A_{v,g}$ indicates the site in the center.
The gauge transformations for other $g$ can be obtained by $A_{v,g_1 g_2}=A_{v,g_1} A_{v,g_2}$ from the above two cases. 
One should be notified that $i$ in $X^{(\dagger)}_{i}$ is integer modulo two. Note that these gauge transformations generate a local $\mathbb{Z}_3\rtimes\mathbb{Z}_2\cong S_3$ symmetry for the gauged system:
\begin{align}
    G^\text{local}=\Big\langle g^{(0)},g^{(2)}\mid (g^{(0)})^3=1,(g^{(2)})^2=1,\notag\\
    g^{(0)}g^{(2)}=g^{(2)}(g^{(0)})^2\Big\rangle\cong S_3.
\end{align}

\item There are two ways to combine the $c$'s to produce the identity.
    The first type is the product of four $c_2$ on the links:
    \begin{equation}
    \label{eq:Rp}
        R_p(\{c\})=\plaquetteop{}{c_2}{}=1,\quad p\in\text{plaquettes}
    \end{equation}
    The corresponding gauge field term is the plaquette operator of the link qubit.
    \begin{equation}
    \label{eq:Bp}
        B_p=R_p(\{Z\})=\plaquetteop{}{Z}{}.
    \end{equation}
    The second type is the tube-like product and sum of the $c_i$'s on the plaquettes (define $c_1=c_0^\dagger$ for notational convenience.)
\begin{align}
    \renewcommand{\labelstyle}{\textstyle}
    R_{t,i}(\{c\})=\sum_{a,b=0,1}
\begin{smallgathered}
\xymatrix@=0.45cm{
&\ar@{-}[ld]\ar@{-}[d]\ar@{-}[rrr]\ar@{}[rrd]|{c_i^\dagger}&&&\ar@{-}[ld]\\
    \ar@{-}[rrr]|{[a]_c}\ar@{-}[ddd]|{[b]_c}\ar@{}[ddr]|{c_{i}}&&&\ar@{-}[ddd]\ar@{}[ddr]|{c_{i+a}^\dagger}&\\
&&&&\\
&\ar@{-}[ld]\ar@{-}[rr]\ar@{-}[uu]\ar@{}[rrd]|{c_{i+b}}&&&\ar@{-}[uuu]\ar@{-}[l] \notag\\
\ar@{-}[rrr]&&&\ar@{-}[ru]
}
\end{smallgathered}
=1,\\ t\in\text{tubes},\, i=0,1
\end{align}
where $[a]_c:=\frac{1}{2}\left({c_2+(-1)^a}\right)$.
Here we denote $B_{t,1}=B_{t,0}^\dagger$. and the corresponding field terms are
\begin{align}
    B_{t,i}&=R_{t,i}(\{Z\})\prod_{\text{six faces}}P_p\notag\\
           &=
    \renewcommand{\labelstyle}{\textstyle}
    \sum_{a,b,i=0,1}
\begin{smallgathered}
\xymatrix@=0.45cm{
&\ar@{-}[ld]\ar@{-}[d]\ar@{-}[rrr]\ar@{}[rrd]|{Z_i^\dagger}&&&\ar@{-}[ld]\\
    \ar@{-}[rrr]|{[a]}\ar@{-}[ddd]|{[b]}\ar@{}[ddr]|{Z_{i}}&&&\ar@{-}[ddd]\ar@{}[ddr]|{Z_{i+a}^\dagger}&\\
&&&&\\
&\ar@{-}[ld]\ar@{-}[rr]\ar@{-}[uu]\ar@{}[rrd]|{Z_{i+b}}&&&\ar@{-}[uuu]\ar@{-}[l]\\
\ar@{-}[rrr]&&&\ar@{-}[ru]
}
\end{smallgathered}
\cdot \prod_{\text{six faces}}P_p.
\end{align}
where $P_p$ is the projector onto the subspace that $B_p=1$ and $p$ runs over the six faces of the smallest cube containing the tube $t$. 
One should be notified that the projectors are essential.
$B_{t,0}+B_{t,1}$ commutes with all $A_{v,g}$ (so that the Hamiltonian is gauge invariant), but would not do so if we exclude the projectors $P_p$.
\item The gauged couplings are
       \begin{align}
           \fplaquetteop{c_0(\tau)}{}{}{}{}{}{}{}{} &:=\sum_{\substack{a,b,c,d,e=0,1\\b+c+d+e=0}}\fplaquetteop{Z_a^\dagger}{[b]}{[c]}{[e]}{[d]}{Z_a[a]}{Z_{a+b} ^\dagger I}{Z_{a+e}^\dagger I}{Z_{a+b+c} I},& \notag\\
            \flinkop{c_2(\tau)}{}{} & :=\flinkop{Z}{IZ}{IZ} .&
        \end{align}

        Note that in $c_0(\tau)$, the gauge DOF from $c_2$ along the boundary of the plaquette is also included. This is in contrast to gauging pure global or pure subsystem symmetry, in which $c(\tau)$ is simply obtained by combining $c$ with the corresponding $\tau_c$. This reflects the fact that when global and subsystem symmetry is mixed together, the branch cut created by the global part may split the coupling of the subsystem part.

        The requirement $b+c+d+e=0$ in the summation reflects the fact that if there is a $\mathbb{Z}_2^\text{glo}$ flux in the plaquette ($B_p=-1$), then we cannot use $A_{v,g}$ to eliminate all gauge DOF along the plaquette boundary, giving the value $0$ for the gauged coupling.

\item The electric field terms are
\begin{equation}
    H_e=-g_0\sum_{p \text{ plaquettes}}\left(\fplaquetteop{X_0}{}{}{}{}{}{}{}{}+\fplaquetteop{X_{1}}{}{}{}{}{}{}{}{}\right)
{ -g_2\sum_\text{links}\flinkop{\tilde{X}}{}{},}
\end{equation}
{ where
\begin{equation}
    \renewcommand{\labelstyle}{\textstyle}
\begin{smallgathered}
\xymatrix@!0{
    &&\ar@{..}[dl]\ar@{..}[rrr]\ar@{}[drr]&&&\ar@{..}[dl]\\
    &\ar@{..}[dl]\ar@{..}[d]\ar@{-}[rrr]|{\tilde{X}}\ar@{}[drr]\ar@{}[dddrrr]&&&\ar@{..}[dl]\ar@{..}[ddd]&\\
    \ar@{..}[rrr]&\ar@{..}[dd]&&&&\\
    &&&&&\\
    &\ar@{..}[rrr]&&&&\\
}
\end{smallgathered}\hspace{-1.5em}
=
\sum_{\substack{a,b=0,1,2 \\ c, d,e=0,1}}\hspace{-0.8em}
    \renewcommand{\labelstyle}{\textstyle}
\begin{smallgathered}
\xymatrix@!0{
    &&\ar@{-}[dl]\ar@{-}[rrr]\ar@{}[drr]|{X_c^{a}}&&&\ar@{-}[dl]\\
    &\ar@{-}[dl]|{[d]}\ar@{-}[d]\ar@{-}[rrr]|{X[c]}\ar@{}[drr]|{X_{c+d}^{a}}\ar@{}[dddrrr]|{X_c^{a}X_{c+e}^b}&&&[a] I \ar@{-}[dl]\ar@{-}[ddd]|{[e]}&\\
    \ar@{-}[rrr]&\ar@{-}[dd]&&&&\\
    &&&&&\\
    &\ar@{-}[rrr]&&&[b] I&\\
}
\end{smallgathered}.
\end{equation}
}
\end{enumerate}

The decorated $\tilde{X}$ operator ensures the electric field is invariant under the gauge transformation $A_{v,g}$.
Similar to the toric code and the X-Cube code, in this example, the matter DOF can also be eliminate by choosing the unitary gauge. We regard the ``physical state'' satisfying the constraints as an equivalent class of computational basis state, and choose the representative such that all matter DOF are in the reference state $|00\rangle$.
In this case, the Hamiltonian becomes
\begin{align}
\label{eq:z2exactz3}
    H_\text{u}=
    &-\sum_{v\text{ sites}}\left(A_{v,g^{(0)}_P}^\text{u}+\left(A_{v,g^{(0)}_P}^\text{u}\right)^2+A_{v,g^{(2)}}^\text{u}\right)\notag\\
                    &-\sum_{t\text{ tubes}}\left(B_{t,0}+B_{t,1}\right)-\sum_{p\text{ plaquettes}}B_p\notag\\
&-\left(J_0\sum_{p \text{ plaquettes}}\fplaquetteop{Z_{0}}{}{}{}{}{}{}{}{}\cdot P_p
+J_2\sum_\text{links}\flinkop{Z}{}{}+\text{H.c.}\right)\notag\\
&-g_0\sum_{p \text{ plaquettes}}\left(\fplaquetteop{X_{0}}{}{}{}{}{}{}{}{}+\fplaquetteop{X_{1}}{}{}{}{}{}{}{}{}\right)
- g_2\sum_\text{links}\flinkop{X}{}{} ,
\end{align}
where $A_{v,g}^\text{u}$ is the gauge part of $A_{v,g}$ by removing the operation on the center vertex of (\ref{eq:AvgP0z3}).
The decorated $\tilde{X}$ operator becomes $X$ in the unitary gauge. 
The Hilbert space contains only the gauge DOF without constraints.
The above Hamiltonian $H_\text{u}$ describes the pure lattice gauge theory of $\mathbb{Z}_3^\text{sub}\rtimes\mathbb{Z}_2^\text{glo}$.

Note that this model can also be constructed in the way similar to Refs.~[\onlinecite{Prem2019,BulmashGaugingDegeneracies}] by gauging the charge conjugation symmetry of the $\mathbb{Z}_3$ X-Cube code~\cite{Slagle2018}. Also note that the combination of global and subsystem symmetry of this model is similar to the field-theoretic approach in Ref.~[\onlinecite{Wang2019Higher-RankEmbeddon}], where the $\mathbb{Z}_3^\text{sub}$ is the discrete version of $U(1)_{x_3}$ with constraints by the cubicity of the lattice.

In the exactly solvable limit ($J_0=J_2=g_0=g_2=0$), the system is deeply
in the deconfined phase and the fundamental excitations are fully mobile particles, fractons, and strings. 
These excitations can be categorized as electric charges or magnetic fluxes. 
The electric charges are the excitations from  $A_{v,g^{(0)}_P}^\text{u}$, $\left(A_{v,g^{(0)}_P}^\text{u}\right)^2$,
    and $A_{v,g^{(2)}}^\text{u}$. The magnetic fluxes are the excitations from $B_p$, $B_{t,0}$, and $B_{t,1}$. We discuss these excitations as follows.

\subsubsection{Electric charge excitations}

\begin{itemize}
\item $[f_0]$: The non-Abelian fracton, corresponding to the excitation of $A_{v,g^{(0)}_P}^\text{u}+\big(A_{v,g^{(0)}_P}^\text{u}\big)^2$ in (\ref{eq:z2exactz3}). It is created at the four corners of the membrane operator
    \begin{align}\label{eq:f0m}
            &M_{v,i}^{(m,n)} \notag\\
            =&
            \sum_{\substack{\{a_{kl}\},\\\{b_k\}=0,1}}
            \renewcommand{\labelstyle}{\textstyle}
            \begin{smallgathered}
            \xymatrix@=4.5em{
                \ar@{-}[r]|{[a_{11}]}\ar@{-}[d]|{[b_1]}\ar@{}[dr]|{Z_i} & \ar@{-}[r]|{\cdots}\ar@{}[dr]|{\cdots}\ar@{-}[d] & \ar@{-}[r]|{[a_{1n}]}\ar@{-}[d]\ar@{}[dr]|{Z_{\substack{i+a_{11}+\cdots\\+a_{1,n-1}}}} & \ar@{-}[d]\ar@{-}[r]\ar@{}[dr]|{Z_{\substack{i+a_{11}+\cdots\\+a_{1n}}}} & \ar@{-}[d] \\
                \ar@{-}[r]|{[a_{21}]}\ar@{}[dr]|{\vdots}\ar@{-}[d]|{\vdots} & \ar@{-}[r]|{\cdots}\ar@{}[dr]|{\ddots} & \ar@{-}[r]|{[a_{2n}]}\ar@{}[dr]|{\vdots} & \ar@{-}[r]\ar@{}[dr]|{\vdots} & \\
                \ar@{-}[r]|{[a_{m1}]}\ar@{-}[d]|{[b_m]}\ar@{}[dr]|{Z_{\substack{i+b_1+\cdots\\+b_{m-1}}}} & \ar@{-}[r]|{\cdots}\ar@{-}[d]\ar@{}[dr]|{\cdots} & \ar@{-}[r]|{[a_{mn}]}\ar@{-}[d]\ar@{}[dr]|{Z_{\substack{i+b_1+\cdots\\+b_{m-1}\\+a_{m1}+\cdots\\+a_{m,n-1}}}} & \ar@{-}[r]\ar@{-}[d]\ar@{}[dr]|{Z_{\substack{i+b_1+\cdots\\+b_{m-1}\\+a_{m1}+\cdots\\+a_{mn}}}} & \ar@{-}[d] \\
                \ar@{-}[r]|{[a_{m+1,1}]}\ar@{-}[d]\ar@{}[dr]|{Z_{\substack{i+b_1+\cdots\\+b_{m}}}} & \ar@{-}[r]|{\cdots}\ar@{-}[d]\ar@{}[dr]|{\cdots} & \ar@{-}[r]|{[a_{m+1,n}]}\ar@{-}[d]\ar@{}[dr]|{Z_{\substack{i+b_1+\cdots\\+b_{m}\\+a_{m+1,1}\\+\cdots+\\a_{m+1,n-1}}}} & \ar@{-}[r]\ar@{-}[d]\ar@{}[dr]|{Z_{\substack{i+b_1+\cdots\\+b_{m}\\+a_{m+1,1}\\+\cdots+\\a_{m+1,n}}}} & \ar@{-}[d]\\
                \ar@{-}[r] & & \ar@{-}[r] & \ar@{-}[r] &
            }
        \end{smallgathered},\hspace{-1em}
    \end{align}
        where $v$ is the upper left corner and $i=0,1$.
      Note that
        \begin{equation}
            A_{v,g^{(2)}}^\text{u}M_{v,0}^{(m,n)}|GS\rangle=M_{v,1}^{(m,n)}|GS\rangle,
        \end{equation}
        This means that the fracton created by $M_{v,0}^{(m,n)}$ and $M_{v,1}^{(m,n)}$ belongs to the same superselection sector, so is the same species $[f_0]$.
        Also note that this membrane operator only work when acting on a state with $B_p=+1$ for all plaquette $p$ on and near (at most one lattice spacing) the membrane.

        To show that it is non-Abelian, we consider the following two states
        \begin{align}
           |\psi_1\rangle&=M_{v,1}^{(L,L)}M_{v,0}^{(2L,2L)} M_{v,1}^{(3L,3L)}|GS\rangle, \notag \\
           |\psi_2\rangle&=M_{v,1}^{(L,L)}M_{v,0}^{(2L,2L)} M_{v,0}^{(3L,3L)}|GS\rangle,
        \end{align}
        where $L$ is large.
        The excitation patterns of the two states are both
        \begin{equation}\label{eq:f0}
           \renewcommand{\labelstyle}{\textstyle}
            \begin{smallgathered}
            \xymatrix@=2em{
                [f_0]\ar@{--}[r]\ar@{--}[d] & [f_0]\ar@{--}[d]\ar@{--}[r] & [f_0]\ar@{--}[dd]\ar@{--}[r] & [f_0]\ar@{--}[ddd]\\
                [f_0]\ar@{--}[r]\ar@{--}[d] & [f_0] & \\
                [f_0]\ar@{--}[rr]\ar@{--}[d] & & [f_0]\ar@{}[dr] \\
                [f_0]\ar@{--}[rrr] & & & [f_0]
            }
            \end{smallgathered}.
        \end{equation}
        Now, for the state  $|\psi_1\rangle$, the four $[f_0]$'s on the upper left square of Eq.~(\ref{eq:f0}) can be fused to the vacuum by applying an additional $M_{v,0}^{(L,L)}$.
        On the other hand, they cannot be fused to the vacuum for $|\psi_2\rangle$.
        This means that $[f_0]$ has non-trivial fusion rules, and therefore is non-Abelian.
        If we continue this pattern to apply $n$ $M_{v,i}$'s to the ground state, it can be shown that there are asymptotically $2^{n}$ fusion channels, which implies the quantum dimension of $[f_0]$ is $2$.

    \item $\phi$: The $\mathbb{Z}_2^\text{glo}$ charge, corresponding to the excitation of $A_{v,g^{(2)}}^\text{u}$ in (\ref{eq:z2exactz3}). It is an Abelian quasiparticle that can move freely in 3D, created at the end point of the string operator
        \begin{equation}\label{eq:phi}
            \renewcommand{\labelstyle}{\textstyle}
            \begin{gathered}
            \xymatrix@=4em{
                \ar@{-}[r]|Z & \ar@{-}[r]|Z & \ar@{}[r]|{\cdots} & \ar@{-}[r]|Z &
            }
            \end{gathered}.
        \end{equation}
\end{itemize}
We emphasize that these electric charge excitations 
are local with respect to the vertex $v$,
and can be specified from the local operators which form a representation of $S_3$ on the Hilbert space.
Hence we can identify electric charges of the $G=\mathbb{Z}^{\text{sub}}_3 \rtimes \mathbb{Z}^{\text{glo}}_2$ fracton model
with the irreducible representations of $G^\text{local} \cong S_3$ (left panel of Table \ref{tab:z3}). 
This identification is the same as the QDM with $S_3$ symmetry~\cite{A.Yu.Kitaev2002}.

\subsubsection{The magnetic flux excitations}
\begin{itemize}        
    \item $[e_d]$, where $d=x,y,z$: The non-Abelian lineons are constrained to move in the $d$ direction, corresponding to the excitation of $B_{t,0}+B_{t,1}$ in (\ref{eq:z2exactz3}). They are created at the endpoints at the string operator (the direction of the string is $d$)
        \begin{align}
        \label{eq:ed}
	    S_{v,i}^n=  
            \sum_{\substack{\{a_{k}\}\\=0,1}}
    \renewcommand{\labelstyle}{\textstyle}
\begin{smallgathered}
\xymatrix@=2em{
&\ar@{-}[ld]\ar@{-}[d]\ar@{}[rrd]&&&\ar@{-}[ld]\ar@{-}[d]\ar@{}[ddddrr]|{\cdots}\ar@{}[rrd]&&&\ar@{-}[ld]\ar@{-}[d]\ar@{}[rrd]&&&\ar@{-}[ld]\\
    \ar@{-}[rrr]|{[a_1]}\ar@{-}[ddd]\ar@{}[ddr]|{X_{i}}&&&\ar@{-}[rrr]|{\cdots}\ar@{-}[ddd]\ar@{}[ddr]|{X_{i+a_1}}&&&\ar@{-}[rrr]|{[a_n]}\ar@{-}[ddd]\ar@{}[ddr]|{X_{\substack{i+a_1\\+\cdots+\\a_{n-1}}}}&&&\ar@{-}[ddd]\ar@{}[ddr]|{X_{\substack{i+a_1\\+\cdots\\+a_{n}}}}&\\
&&&&&&&&&&\\
&\ar@{-}[ld]\ar@{-}[uu]\ar@{}[rrd]&&&\ar@{-}[ld]\ar@{-}[uu]\ar@{}[rrd]&&&\ar@{-}[ld]\ar@{-}[uu]\ar@{}[rrd]&&&\ar@{-}[uuu]\\
&&&&&&&&&\ar@{-}[ru]
}
\end{smallgathered},\hspace{-1em}
\end{align}
where $v$ is the upper left vertex and $i=0,1$.
Similar to $[f_0]$, we have
        \begin{equation}
            A_{v,g^{(2)}}^\text{u}S_{v,0}^{n}|GS\rangle=S_{v,1}^{n}|GS\rangle,
        \end{equation}
so $i=0,1$ results in the same species. Also, this string operator only works when acting on a state with $B_p=+1$ for all plaquette $p$ on and near (at most one lattice spacing) this string.

%

        To show that it is non-Abelian, we consider the following two states
        \begin{align}
           |\psi_1\rangle&=S_{v,1}^{L}S_{v,0}^{L} S_{v,1}^{L}|GS\rangle, \notag \\
           |\psi_2\rangle&=S_{v,1}^{L}S_{v,0}^{L} S_{v,0}^{L}|GS\rangle.
        \end{align}
        where $L$ is large.
        The excitation patterns of the two states are both
        \begin{equation}\label{eq:lineon_conf1z3}
           \renewcommand{\labelstyle}{\textstyle}
            \begin{gathered}
            \xymatrix@=3 em{
                [e_d]\ar@{--}[r] & [e_d]\ar@{--}[r] & [e_d]\ar@{--}[r] & [e_d].\\
            }
            \end{gathered}
        \end{equation}
        Now, for the state  $|\psi_1\rangle$, the two $[e_d]$'s on the left of (\ref{eq:lineon_conf1z3}) can be fused to the vacuum by applying an additional $S_{v,0}^L$.
        On the other hand, they cannot be fused to the vacuum for $|\psi_2\rangle$.
        This means that $[e_d]$ has non-trivial fusion rules, and therefore is non-Abelian.
        If we continue this pattern to apply $n$ $S_{v,i}$'s to the ground state, it can be shown that there are asymptotically $2^{n}$ fusion channels, which implies the quantum dimension of $[e_d]$ is $2$.

    \item $\sigma$: The $\mathbb{Z}_2^\text{glo}$ flux, which is a flexible string-like excitation (referred to the $\sigma$ string) corresponding to the excitation of $B_p$ in (\ref{eq:z2exactz3}). It is created along the boundary of the membrane operator
\begin{multline}
    \sum_{\substack{\{a_{ij}\}=0,1,2,\\\,\{b_{ij}\},\{c_{ij}\}\\=0,1}}
   \renewcommand{\labelstyle}{\textstyle}
\begin{smallgathered}
\xymatrix@=2em{
&\ar@{-}[ld]|{\tilde{X}^{(11)}}&&&\ar@{-}[ld]|{\tilde{X}^{(12)}}\ar@{}[ddddrr]|{}&&&\ar@{-}[ld]|{\tilde{X}^{(1n)}}\\
    \ar@{-}[ddd]|{[b_{11}]}\ar@{-}[rrr]|{[c_{11}]}\ar@{}[dddrrr]|{[a_{11}]}&&&\ar@{-}[ddd]|{[b_{12}]}\ar@{-}[rrr]|{\cdots}\ar@{}[dddrrr]|{\cdots}&&&\ar@{-}[ddd]|{[b_{1n}]}\ar@{-}[rrr]|{[c_{1n}]}\ar@{}[dddrrr]|{[a_{1n}]}&&&\ar@{-}[ddd]\\
&&&&&&&&&\\
&\ar@{-}[ld]|{\tilde{X}^{(21)}}\ar@{}[ddddrr]|{}&&&\ar@{-}[ld]|{\tilde{X}^{(22)}}\ar@{}[ddddrr]|{}&&&\ar@{-}[ld]|{\tilde{X}^{(2n)}}&&&\\
    \ar@{-}[ddd]|{\vdots}\ar@{-}[rrr]|{[c_{21}]}\ar@{}[dddrrr]|{\vdots}&&&\ar@{-}[ddd]|{\vdots}\ar@{-}[rrr]|{\cdots}\ar@{}[dddrrr]|{\ddots}&&&\ar@{-}[ddd]|{\vdots}\ar@{-}[rrr]|{[c_{2n}]}\ar@{}[dddrrr]|{\vdots}&&&\ar@{-}[ddd]|{\vdots}\\
&&&&&&&&&\\
&\ar@{-}[ld]|{\tilde{X}^{(m1)}}&&&\ar@{-}[ld]|{\tilde{X}^{(m2)}}&&&\ar@{-}[ld]|{\tilde{X}^{(mn)}}&&\\
    \ar@{-}[ddd]|{[b_{m1}]}\ar@{-}[rrr]|{[c_{m1}]}\ar@{}[dddrrr]|{[a_{m1}]}&&&\ar@{-}[ddd]|{[b_{m2}]}\ar@{-}[rrr]|{\cdots}\ar@{}[dddrrr]|{\cdots}&&&\ar@{-}[ddd]|{[b_{mn}]}\ar@{-}[rrr]|{[c_{mn}]}\ar@{}[dddrrr]|{[a_{mn}]}&&&\ar@{-}[ddd]\\
&&&&&&&&&\\
&&&&&&&&&\\
    \ar@{-}[rrr]&&&\ar@{-}[rrr]|{\cdots}&&&\ar@{-}[rrr]&&& \\
}
\end{smallgathered}\hspace{-1em},\hspace{-1em}
\end{multline}
where
\begin{align}
&\renewcommand{\labelstyle}{\textstyle}
\begin{smallgathered}
\xymatrix@C=3 em{
&\ar@{-}[ld]|{\tilde{X}^{(ij)}}\ar@{..}[rrr]\ar@{..}[d]\ar@{}[drr]&&&\ar@{..}[ld]\\
    \ar@{..}[rrr]\ar@{..}[ddd]\ar@{}[ddr]&\ar@{..}[dd]&&&\\
                &&&&\\
                &\ar@{..}[ld]&&&\\
                &&&&
}
\end{smallgathered}
\hspace{-1.5em}
=
   \renewcommand{\labelstyle}{\textstyle}
\begin{smallgathered}
\xymatrix@C=3 em{
&\ar@{-}[ld]|{X}\ar@{-}[rrr]\ar@{-}[d]\ar@{}[drr]|{X_0^{\tilde b^{(ij)}}S}&&&\ar@{-}[ld]\\
    \ar@{-}[rrr]\ar@{-}[ddd]\ar@{}[ddr]|{X_0^{\tilde c^{(ij)}}S}&\ar@{-}[dd]&&&\\
                &&&&\\
                &\ar@{-}[ld]&&&\\
                &&&&
}
\end{smallgathered}\notag \\
\tilde b^{(ij)}=&
a_{1j}(-1)^{b_{1j}+\cdots+ b_{i-1,j}}+a_{2j}(-1)^{b_{2j}+\cdots+ b_{i-1,j}} \notag \\
&+\cdots+a_{i-1,j}(-1)^{b_{i-1,j}}. \notag \\
\tilde c^{(ij)}=&
a_{i1}(-1)^{c_{i1}+\cdots+ c_{i,j-1}}+a_{i2}(-1)^{c_{i2}+\cdots+ c_{i,j-1}} \notag \\
&+\cdots+a_{i,j-1}(-1)^{c_{i,j-1}}.
\end{align}

  
\end{itemize}
The projectors on this decorated membrane operator ensures no additional excitations on the membrane.



\begin{table*}[t]
\centering
 \begin{tabular}{c c c  || c c c c c } 
 \hline
 Charge & Irrep of $G^\text{local}\cong S_3$ & Type  & Flux & $B_{p_0}$ & $B_{t_0,0}$ & Conj.\ class & Type \\ 
 \hline\hline
 Vacuum & Trivial & Vacuum  & Vacuum & $1$ & $1$ & \{1\} & Vacuum \\ 
 $\phi$ & Sign representation & Abelian fully mobile particle &  $\sigma$ & $-1$ & 0 & $\{ g^{(2)}, g^{(2)}g^{(0)}, g^{(2)}(g^{(0)})^2 \}$ & Flexible string \\ 
 $[f_0]$ & 2D representation & Non-Abelian fracton &  $[e_d]$ & $1$ & $e^{\frac{i2\pi}{3}}$ or $e^{-\frac{i2\pi}{3}}$  & $\{ g^{(0)},  (g^{(0)})^2  \}$ & Non-Abelian lineon \\
 \hline
 \end{tabular}
\caption{Left/right panel: Pure electric charges/magnetic fluxes  of the gauged $\mathbb{Z}_3^\text{sub}\rtimes\mathbb{Z}_2^\text{glo}$ model.}
\label{tab:z3}
\end{table*}
  
In the same spirit as the QDM, the magnetic fluxes would be the conjugacy classes of $G^\text{local}\cong S_3$.
    With the geometry defined in Eq. (\ref{eq:geometry}) and the constraints defined in Eq. (\ref{eq:constraint}), 
the flux operators can be mapped to a subalgebra of the QDM algebra for either $P_\text{side}=1$ or $P_\text{corner}=1$,
\begin{align}
    B_{p_0}\mapsto &B_{e}+B_{g^{(0)}}+B_{{g^{(0)}}^2}   \notag  \\
     &-B_{g^{(2)}}-B_{g^{(0)}g^{(2)}}-B_{{g^{(0)}}^2g^{(2)}} \notag \\
    B_{t_0,0}\mapsto & B_{e}+e^{\frac{i2\pi}{3}}B_{g^{(0)}}+e^{-\frac{i2\pi}{3}}B_{{g^{(0)}}^2} \notag \\
    B_{t_0,1}\mapsto& B_{e}+e^{-\frac{i 2\pi}{3}}B_{g^{(0)}}+e^{\frac{i 2\pi}{3}}B_{{g^{(0)}}^2},
\end{align}
 where $B_g$ is the QDM flux operator in Eq.~(\ref{eq:qdm}).
We can also identify the star operators $A_{v_0,g}^\text{u}$ with the QDM star operator $A_g$ Eq.~(\ref{eq:qdm}).
The map of $A$'s and $B$'s together forms an injective algebra homomorphism into the QDM algebra. In particular, they satisfy the relations (note that all the $B$'s commute)
\begin{align}
    &B_{p_0}^2=1,\quad B_{t_0,0}^2=B_{t_0,1},\quad B_{t_0,0}^3=B_{t_0,1}^3=\frac{1}{2}(1+B_{p_0}),  \notag\\
    &B_{p_0} B_{t_0,i}=B_{t_0,i},\quad i=0,1,   \notag \\
    &A_{g}B_{p_0}=B_{p_0}A_{g},\quad A_{g}B_{t_0,i}=B_{t_0,i}A_{g},\quad \forall g\in S_3,i=0,1.
\end{align}
In this way, the star and the flux operator of this fracton models is identified with a subalgebra of that of the corresponding QDM.

Note that this subalgebra is enough to distinguish all of the conjugacy classes of $S_3$ (more rigorously, if a state is the eigenstate of $B_{g_0}=1,B_{g}=0,g\neq g_0$ for some $g_0\in S_3$, then we can determine the conjugacy class of $g_0$ only by the eigenvalues of the image of $B_{p_0}$ and $B_{t_0,i}$.)
This allows us to identify the fluxes of this fracton model with the conjugacy classes of $S_3$.
The corresponding fluxes are listed in the right panel of Table \ref{tab:z3}.


\subsection{Example: gauging \texorpdfstring{$G=(\mathbb{Z}_2^\text{sub}\times\mathbb{Z}_2^\text{sub})\rtimes\mathbb{Z}_2^\text{glo}$}{(Z\_2subxZ\_2sub) rx Z\_2glo}}\label{sec:z2}

The ungauged system is a 3D cubic lattice with three qubits (labeled as $0,1,2$) on each site.
$0$ and $1$ labels are corresponding to the Ising spin of the first layer and the second layer. $2$ label is the twist defect charge in Refs.\ [\onlinecite{Prem2019,BulmashGaugingDegeneracies}]. 
For convenience, we combine the first and the second qubits together and  denote the operators by 
        $X_0 := XI$, $X_1:=IX$, $Z_0 := ZI$, and $Z_1:=IZ$. With these definitions, the Hamiltonian and the gauging procedure we describe in this example  will be similar as the first example, $G=\mathbb{Z}_3^\text{sub}\rtimes\mathbb{Z}_2^\text{glo}$.
The Hamiltonian is
    \begin{align}
        H_o=&-h\sum_\text{sites}(X_0I+X_1I+IX),\notag \\
        H_n=&-J_0\sum_\text{plaquettes}\fplaquetteop{c_0}{}{}{}{}{}{}{}{}-J_1\sum_\text{plaquettes}\fplaquetteop{c_1}{}{}{}{}{}{}{}{}    \notag\\
        &-J_2\sum_\text{links}\flinkop{c_2}{}{},
    \end{align}
    where the minimal couplings are
        \begin{align}
            \fplaquetteop{c_0}{}{}{}{}{}{}{}{}&  
            :=\left(\fplaquetteop{}{}{}{}{}{Z_0[0]}{Z_0I}{Z_0I}{Z_0I}+\fplaquetteop{}{}{}{}{}{Z_1[1]}{Z_1I}{Z_1I}{Z_1I}\right),\notag \\ 
            \fplaquetteop{c_1}{}{}{}{}{}{}{}{}& 
            :=\left(\fplaquetteop{}{}{}{}{}{Z_0[1]}{Z_0I}{Z_0I}{Z_0I}+\fplaquetteop{}{}{}{}{}{Z_1[0]}{Z_1I}{Z_1I}{Z_1I}\right),\notag \\
            \flinkop{c_2}{}{}&:=\flinkop{}{IZ}{IZ}.
        \end{align}
        Here $X$ and $Z$ are the Pauli operators,  $[\psi]:=|\psi\rangle\langle\psi|$ is denoted as a projector, and the operators on the first and second entries on the site act on the Ising spin on the first and second layers, and the twist defect charge, respectively.


    $G=(\mathbb{Z}_2^\text{sub}\times\mathbb{Z}_2^\text{sub})\rtimes\mathbb{Z}_2^\text{glo}$ is generated by $g_P^{(i)}$ and $g^{(2)}$, where $g_P^{(i)}$ acts on the matter by flipping (apply $X$) the $i$th qubit ($i=0,1$) on each site of a shifted coordinate plane $P$; $g^{(2)}$ acts on the matter by flipping the 3rd qubit and swapping the first two qubits on each site of the entire system.
    \begin{align}
	&g_P^{(0)}=\prod_{\substack{\text{sites}\in P}}X_0I,\quad g_P^{(1)}=\prod_{\substack{\text{sites}\in P}}X_1I,  \notag \\
	&g^{(2)}=\prod_{\substack{\text{sites}\in \text{all}}}\text{SWAP}X.
    \end{align}
    
The procedure of gauging works as follows:
\begin{enumerate}
    \item We put one qubit on each link (corresponds to $c_2$) and two qubits on each plaquette (corresponds to $c_0$ and $c_1$, and labeled by 0 and 1). These qubits are the gauge fields.
    \item 
Based on the general gauging procedure we described in Sec. \ref{sec:gauging}.\ref{step2}\ [details are demonstrated in Appendix \ref{sec:details}], the gauge transformations are
\begin{equation}
\vspace{-1em}
    A_{v,g_P^{(0)}}=\sum_{\substack{a,b,c,\\d,e,f\\=0,1}}
    \renewcommand{\labelstyle}{\textstyle}
\begin{smallgathered}
\xymatrix@!0{
    &&&&&\ar@{-}[d]&&&\\
    &\ar@{-}[rrr]|{\color{blue}[b]}&&&\ar@{-}[ru]\ar@{-}[ld]&\ar@{-}[dd]&&\ar@{-}[lll]&\\
    &&&\ar@{-}[ddd]|(.56){\color{orange}[f]}&&&&&\\
    &&\ar@{-}@[green!50!black][ld]&\ar@{-}@[green!50!black][l]&\ar@{-}@[green!50!black][l]&\ar@{-}@[green!50!black][l]\ar@{-}@[green!50!black][rr]&&\ar@{-}@[green!50!black][r]&\ar@{-}@[green!50!black][ld]\\
    &\ar@{-}[uuu]\ar@{-}[d]&&\ar@{-}[ll]|(.25){[c]}&X_{0}I\ar@{-}[uuu]|(.55){[a]}\ar@{-}[rrr]\ar@{-}[d]\ar@{-}[l]\ar@{-}[ru]\ar@{-}[ld]|{[e]}\ar@{}[llluuu]|(.6){\color{blue}X_{a+b}}\ar@{}[luuu]|(.46){\color{orange}X_{e+f}}\ar@{}[ruuuu]|(.5){\color{orange}X_{a}}\ar@{}[rrruuu]|(.6){\color{blue}X_{a}}\ar@{}[llu]|(.71){\color{green!50!black}X_{c}}\ar@{}[rrrru]|(.5){\color{green!50!black}X_{0}}\ar@{}[rrrddd]|(.6){\color{blue}X_{0}}\ar@{}[rdd]|(.65){\color{orange}X_{0}}\ar@{}[ldddd]|(.5){\color{orange}X_{e}}\ar@{}[lllddd]|(.6){\color{blue}X_{c}}\ar@{}[rrd]|(.7){\color{green!50!black}X_{e}}\ar@{}[lllld]|(.5){\color{green!50!black}X_{c+d}}&\ar@{-}[u]&&\ar@{-}[uuu]\ar@{-}[ddd]&\\
    \ar@{-}@[green!50!black][ru]|{\color{green!50!black}[d]}&\ar@{-}[dd]&&\ar@{-}@[green!50!black][lll]\ar@{-}@[green!50!black][rrr]&\ar@{-}[dd]&\ar@{-}[u]&\ar@{-}@[green!50!black][ru]&&\\
                       &&&&&\ar@{-}[u]&&&\\
                       &\ar@{-}[rr]&&\ar@{-}[r]&\ar@{-}[ru]\ar@{-}[ld]&&&\ar@{-}[lll]&\\
                       &&&\ar@{-}[uuu]&&&&&
}
\end{smallgathered}\hspace{-0.5em}, \notag \label{eq:AvgP0z2}
\end{equation}
\vspace{-2em}
\begin{equation}
A_{v,g^{(2)}}=
    \renewcommand{\labelstyle}{\textstyle}
\begin{smallgathered}
\xymatrix@!0{
    &&&&&\ar@{-}[d]&&&\\
    &\ar@{-}[rrr]&&&\ar@{-}[ru]\ar@{-}[ld]&\ar@{-}[dd]&&\ar@{-}[lll]&\\
    &&&\ar@{-}[ddd]&&&&&\\
    &&\ar@{-}@[green!50!black][ld]&\ar@{-}@[green!50!black][l]&\ar@{-}@[green!50!black][l]&\ar@{-}@[green!50!black][l]\ar@{-}@[green!50!black][rr]&&\ar@{-}@[green!50!black][r]&\ar@{-}@[green!50!black][ld]\\
    &\ar@{-}[uuu]\ar@{-}[d]&&\ar@{-}[ll]|(.25){X}&\text{SWAP}X\ar@{-}[uuu]|{X}\ar@{-}[rrr]|{X}\ar@{-}[d]\ar@{-}[l]\ar@{-}[ru]|{X}\ar@{-}[ld]|{X}\ar@{}[llluuu]\ar@{}[luu]\ar@{}[ruuuu]\ar@{}[rrruuu]\ar@{}[llu]\ar@{}[rrrru]|(.5){\color{green!50!black}\text{SWAP}}\ar@{}[rrrddd]|(.6){\color{blue}\text{SWAP}}\ar@{}[rdd]|(.65){\color{orange}\text{SWAP}}\ar@{}[ldddd]\ar@{}[lllddd]\ar@{}[rrd]\ar@{}[lllld]&\ar@{-}[u]&&\ar@{-}[uuu]\ar@{-}[ddd]&\\
    \ar@{-}@[green!50!black][ru]&\ar@{-}[dd]&&\ar@{-}@[green!50!black][lll]\ar@{-}@[green!50!black][rrr]&\ar@{-}[dd]|(.25){X}&\ar@{-}[u]&\ar@{-}@[green!50!black][ru]&&\\
                       &&&&&\ar@{-}[u]&&&\\
                       &\ar@{-}[rr]&&\ar@{-}[r]&\ar@{-}[ru]\ar@{-}[ld]&&&\ar@{-}[lll]&\\
                       &&&\ar@{-}[uuu]&&&&&
}
\end{smallgathered}\hspace{-0.5em}.\label{eq:AvgP2z2}
\vspace{-1em}
\end{equation}
where  $v$ in $A_{v,g}$ indicates the site in the center.
 One should be notified that $i$ in $X_{i}$ is integer modulo two.  On the other hand, the $X$ and $[a]$ on the link refer to the operators on the gauge field with respect to the twist defect charge.
The gauge transformations for other $g$ can be obtained by $A_{v,g_1 g_2}=A_{v,g_1} A_{v,g_2}$ from the above two cases. Note that they generate a local $(\mathbb{Z}_2\times\mathbb{Z}_2)\rtimes\mathbb{Z}_2\cong D_4$ symmetry for the gauged system:
\begin{align}
    G^\text{local}=\Big\langle g^{(0)},g^{(1)},g^{(2)}\mid (g^{(i)})^2=1,\notag\\
    g^{(0)}g^{(2)}=g^{(2)}g^{(1)}\Big\rangle\cong D_4.
\end{align}

\item There are two ways to combine the $c$'s to produce the identity.
    The first type is the product of four $c_2$ on the links, which has the identical structure as the first example in Eq.~(\ref{eq:Rp}).
    The corresponding gauge field term $B_p$ is the plaquette operator of the link qubit, 
    which also has the identical structure as the first example in Eq.~(\ref{eq:Bp}).
    The second type is the tube-like product and sum of the $c_i$'s on the plaquettes
\begin{align}
    \renewcommand{\labelstyle}{\textstyle}
    R_{t,i}(\{c\})&=\sum_{a,b=0,1}
\begin{gathered}
\xymatrix@=0.45cm{
&\ar@{-}[ld]\ar@{-}[d]\ar@{-}[rrr]\ar@{}[rrd]|{c_i}&&&\ar@{-}[ld]\\
    \ar@{-}[rrr]|{[a]_c}\ar@{-}[ddd]|{[b]_c}\ar@{}[ddr]|{c_{i}}&&&\ar@{-}[ddd]\ar@{}[ddr]|{c_{i+a}}&\\
&&&&\\
&\ar@{-}[ld]\ar@{-}[rr]\ar@{-}[uu]\ar@{}[rrd]|{c_{i+b}}&&&\ar@{-}[uuu]\ar@{-}[l]\\
\ar@{-}[rrr]&&&\ar@{-}[ru]
}
\end{gathered} \notag \\
&=1,\quad t\in\text{tubes},\quad i=0,1,
\end{align}
where $[a]_c:=\frac{1}{2}\left({c_2+(-1)^a}\right)$.
The corresponding field term is
\begin{align}
    B_{t,i}&=R_{t,i}(\{Z\})\prod_{\text{six faces}}P_p\notag\\
           &=
    \renewcommand{\labelstyle}{\textstyle}
    \sum_{a,b}
\begin{smallgathered}
\xymatrix@=0.45cm{
&\ar@{-}[ld]\ar@{-}[d]\ar@{-}[rrr]\ar@{}[rrd]|{Z_i}&&&\ar@{-}[ld]\\
    \ar@{-}[rrr]|{[a]}\ar@{-}[ddd]|{[b]}\ar@{}[ddr]|{Z_{i}}&&&\ar@{-}[ddd]\ar@{}[ddr]|{Z_{i+a}}&\\
&&&&\\
&\ar@{-}[ld]\ar@{-}[rr]\ar@{-}[uu]\ar@{}[rrd]|{Z_{i+b}}&&&\ar@{-}[uuu]\ar@{-}[l]\\
\ar@{-}[rrr]&&&\ar@{-}[ru]
}
\end{smallgathered}
\cdot \prod_{\text{six faces}}P_p,
\end{align}
where $P_p$ is the projector onto the subspace that $B_p=1$ and $p$ runs over the six faces of the smallest cube containing the tube $t$. 
One should be notified that the projectors are essential.
$B_{t,0}+B_{t,1}$ commutes with all $A_{v,g}$ (so that the Hamiltonian is gauge invariant), but would not do so if we exclude the projectors $P_p$.

\item The gauged coupling terms are
        \begin{align}
            \fplaquetteop{c_0(\tau)}{}{}{}{}{}{}{}{}&=\sum_{\substack{a,b,c,d,e=0,1\\b+c+d+e=0}}\fplaquetteop{Z_a}{[b]}{[c]}{[e]}{[d]}{Z_a[a]}{Z_{a+b} I}{Z_{a+e} I }{Z_{a+b+c} I}, \notag \\
                                                      \fplaquetteop{c_1(\tau)}{}{}{}{}{}{}{}{}&=\sum_{\substack{a,b,c,d,e=0,1\\b+c+d+e=0}}\fplaquetteop{Z_a}{[b]}{[c]}{[e]}{[d]}{Z_a[a+1]}{Z_{a+b} I }{Z_{a+e} I }{Z_{a+b+c} I}, \notag \\
                                                      \flinkop{c_2(\tau)}{}{}&=\flinkop{Z}{IZ}{IZ}.
        \end{align}
        Note that in $c_0(\tau)$ and $c_1(\tau)$, the gauge DOF from $c_2$ along the boundary of the plaquette is also included. This is in contrast to gauging pure global or pure subsystem symmetry, in which $c(\tau)$ is simply obtained by combining $c$ with the corresponding $\tau_c$. This reflects the fact that when global and subsystem symmetry is mixed together, the branch cut created by the global part may split the coupling of the subsystem part.
        In this case, we need to monitor the switching of the layer index created by the branch cut along the plaquette.        
        The requirement $b+c+d+e=0$ in the summation reflects the fact that if there is a $\mathbb{Z}_2^\text{glo}$ flux in the plaquette ($B_p=-1$), we will assign the value $0$ for the gauged coupling, which gives
        the correct gauge invariant $c_{i} (\tau)$.

\item The electric field terms are
\begin{align}
    H_e=&-g_0\sum_{p \text{ plaquettes}}\left(\fplaquetteop{X_0}{}{}{}{}{}{}{}{}+\fplaquetteop{X_1}{}{}{}{}{}{}{}{}\right)  \notag\\
    &-g_2\sum_\text{links}\flinkop{\tilde{X}}{}{},
\end{align}
where
\begin{equation}
    \renewcommand{\labelstyle}{\textstyle}
\begin{smallgathered}
\xymatrix@!0{
    &&\ar@{..}[dl]\ar@{..}[rrr]\ar@{}[drr]&&&\ar@{..}[dl]\\
    &\ar@{..}[dl]\ar@{..}[d]\ar@{-}[rrr]|{\tilde{X}}\ar@{}[drr]\ar@{}[dddrrr]&&&\ar@{..}[dl]\ar@{..}[ddd]&\\
    \ar@{..}[rrr]&\ar@{..}[dd]&&&&\\
    &&&&&\\
    &\ar@{..}[rrr]&&&&\\
}
\end{smallgathered}\hspace{-1.5em}
=
\sum_{\substack{a,b,c,d\\=0,1}}
    \renewcommand{\labelstyle}{\textstyle}
\begin{smallgathered}
\xymatrix@!0{
    &&\ar@{-}[dl]\ar@{-}[rrr]\ar@{}[drr]|{(X_0X_1)^{a+b}}&&&\ar@{-}[dl]\\
    &\ar@{-}[dl]\ar@{-}[d]\ar@{-}[rrr]|{X}\ar@{}[drr]|{(X_0X_1)^{a+b}}\ar@{}[dddrrr]|{(X_0X_1)^{a+b+c+d}}&&&{[ab] I} \ar@{-}[dl]\ar@{-}[ddd]&\\
    \ar@{-}[rrr]&\ar@{-}[dd]&&&&\\
    &&&&&\\
    &\ar@{-}[rrr]&&& {[cd] I}&\\
}
\end{smallgathered} .
\end{equation}
\end{enumerate}
Again, here $[ab] := |ab \rangle \langle  ab|$ is the projector on the first and second qubits.
The decorated $\tilde{X}$ operator ensures the electric field is invariant under the gauge transformation $A_{v,g}$.
Similar to the toric code and the X-Cube code, we can choose the unitary gauge to eliminate the matter DOF in this example. We regard the ``physical state'' satisfying the constraints as an equivalent class of computational basis state, and choose the representative such that all matter DOF are in the reference state $|000\rangle$.
In this case, the Hamiltonian becomes
\begin{align}
\label{eq:z2exact}
    H_\text{u}=&
    -\sum_{v\text{ sites}}\left(A_{v,g^{(0)}_P}^\text{u}+A_{v,g^{(1)}_P}^\text{u}+A_{v,g^{(2)}}^\text{u}\right)\notag\\
                    &-\sum_{t\text{ tubes}}\left(B_{t,0}+B_{t,1}\right)-\sum_{p\text{ plaquettes}}B_p\notag\\
                    &-\sum_{p \text{ plaquettes}}\left(J_0\fplaquetteop{Z_0}{}{}{}{}{}{}{}{}
                    +J_1\fplaquetteop{Z_1}{}{}{}{}{}{}{}{}\right)P_p
-J_2\sum_\text{links}\flinkop{Z}{}{}  \notag  \\
 &-g_0\sum_{p \text{ plaquettes}}\left(\fplaquetteop{X_0}{}{}{}{}{}{}{}{}+\fplaquetteop{X_1}{}{}{}{}{}{}{}{}\right)
-g_2\sum_\text{links}\flinkop{X}{}{},
\end{align}
and the Hilbert space contains only the gauge DOF without constraints.
The above Hamiltonian $H_\text{u}$ describes the pure lattice gauge theory of $(\mathbb{Z}_2^\text{sub}\times\mathbb{Z}_2^\text{sub})\rtimes\mathbb{Z}_2^\text{glo}$.

In the exactly solvable limit ($J_0=J_1=J_2=g_0=g_2=0$), the system is deeply
in the deconfined phase and the fundamental excitations are fully mobile particles, fractons, and strings. 
Similar to the previous case, the excitations can be categorized as electric charges (excitations of   $A_{v,g^{(0)}_P}^\text{u}$, $A_{v,g^{(1)}_P}^\text{u}$,   $A_{v,g^{(2)}}^\text{u}$) or magnetic fluxes (excitations of $B_{t,0}$, $B_{t,1}$, $B_{p}$).
We discuss these excitations as follows.

\subsubsection{Electric charge excitations}
\begin{itemize}
    \item $[f_0]$: The non-Abelian fracton, corresponding to the excitation of the first term in (\ref{eq:z2exact}). It is created at the four corners of the membrane operator, which has the identical structure as the first example in Eq. (\ref{eq:f0m}).
      Note that
        \begin{equation}
            A_{v,g^{(2)}}^\text{u}M_{v,0}^{(m,n)}|GS\rangle=M_{v,1}^{(m,n)}|GS\rangle,
        \end{equation}
        where $A_{v,g^{(2)}}^\text{u}$ is the summand of the third term of (\ref{eq:z2exact}) centered at $v$.
        This means that the fracton created by $M_{v,0}^{(m,n)}$ and $M_{v,1}^{(m,n)}$ belongs to the same superselection sector, so is the same species $[f_0]$.
        Also note that this membrane operator only work when acting on a state with $B_p=+1$ for all plaquette $p$ on and near (at most one lattice spacing) the membrane.

        To show that it is non-Abelian, we consider the following two states
        \begin{align}
           |\psi_1\rangle&=M_{v,0}^{(L,L)}M_{v,0}^{(2L,2L)} M_{v,0}^{(3L,3L)}|GS\rangle,  \notag \\
           |\psi_2\rangle&=M_{v,0}^{(L,L)}M_{v,0}^{(2L,2L)} M_{v,1}^{(3L,3L)}|GS\rangle,
        \end{align}
        where $L$ is large.
        The excitation patterns of the two states are the same as the first example in Eq.~(\ref{eq:f0}).
        Now, for the state  $|\psi_1\rangle$, the four $[f_0]$'s on the upper left square of Eq.~(\ref{eq:f0}) can be fused to the vacuum by applying an additional $M_{v,0}^{(L,L)}$.
        On the other hand, they cannot be fused to the vacuum for $|\psi_2\rangle$.
        This means that $[f_0]$ has non-trivial fusion rules, and therefore is non-Abelian.
        If we continue this pattern to apply $n$ $M_{v,i}$'s to the ground state, it can be shown that there are asymptotically $2^{n}$ fusion channels, which implies the quantum dimension of $[f_0]$ is $2$.
    \item $f_0^{(0)}f_0^{(1)}$: In addition to the non-Abelian fraction, this model also exhibits composite Abelian fractons, which 
are created at the four corners of $M_{v,0}^{(m,n)}M_{v,1}^{(m,n)}$.   
    \item $\phi$: The $\mathbb{Z}_2^\text{glo}$ charge, corresponding to the third term in Eq.~(\ref{eq:z2exact}). It is an Abelian quasiparticle that can move freely in 3D, created at the end point of the string operator, which has the identical structure as the first example in Eq. (\ref{eq:phi}).
\end{itemize}
 
These electric charge excitations are local with respect
to the vertex $v$, and can be specified from the local operators which form a representation of $D_4$ on the Hilbert space.
Hence we can identify electric charges of the $G=(\mathbb{Z}_2^\text{sub}\times\mathbb{Z}_2^\text{sub})\rtimes\mathbb{Z}_2^\text{glo}$ fracton model
with the irreducible representations of $G^\text{local} \cong D_4$ (left panel of Table~\ref{tab:z2}). 
This identification is the same as the QDM with $D_4$ symmetry~\cite{A.Yu.Kitaev2002}.   

\subsubsection{The magnetic flux excitations}
\begin{itemize} 
    \item $[e_d]$, where $d=x,y,z$: The non-Abelian lineons are constrained to move in the $d$ direction, corresponding to the forth and fifth terms in Eq.~(\ref{eq:z2exact}). They are created at the endpoints at the string operator (the direction of the string is $d$). The string operator has the identical structure as the first example in Eq.~(\ref{eq:ed}).
Similar to $[f_0]$, we have
        \begin{equation}
            A_{v,g^{(2)}}^\text{u}S_{v,0}^{n}|GS\rangle=S_{v,1}^{n}|GS\rangle,
        \end{equation}
so $i=0,1$ results in the same species. Also, this string operator only works when acting on a state with $B_p=+1$ for all plaquette $p$ on and near (at most one lattice spacing) this string.

%

       To show that it is non-Abelian,  we follow the same discussion in the first example. We consider these two states
        \begin{align}
           |\psi_1\rangle&=S_{v,0}^LS_{v,0}^{2L} S_{v,0}^{3L}|GS\rangle,  \notag \\
           |\psi_2\rangle&=S_{v,0}^LS_{v,0}^{2L} S_{v,1}^{3L}|GS\rangle,
        \end{align}
        where $L$ is large.
        The excitation patterns of the two states have the identical structure as the first example in Eq.~(\ref{eq:lineon_conf1z3}).
        Now, for the state  $|\psi_1\rangle$, the two $[e_d]$'s on the left of Eq.~(\ref{eq:lineon_conf1z3}) can be fused to the vacuum by applying an additional $S_{v,0}^L$.
        On the other hand, they cannot be fused to the vacuum for $|\psi_2\rangle$.
        This means that $[e_d]$ has non-trivial fusion rules, and therefore is non-Abelian.
        If we continue this pattern to apply $n$ $S_{v,i}$'s to the ground state, it can be shown that there are asymptotically $2^{n}$ fusion channels, which implies the quantum dimension of $[e_d]$ is $2$.

    \item $\sigma$: The $\mathbb{Z}_2^\text{glo}$ flux, which is a flexible string-like excitation (referred to the $\sigma$ string) corresponding to the fourth term in (\ref{eq:z2exact}). It is created along the boundary of the membrane operator
\begin{multline}
    \sum_{\substack{\{a_{ij}\},\\\{b_{ij}\}\\ =0,1}}
   \renewcommand{\labelstyle}{\textstyle}
\begin{smallgathered}
\xymatrix@=2.0 em{
&\ar@{-}[ld]|{\tilde{X}^{(11)}}&&&\ar@{-}[ld]|{\tilde{X}^{(12)}}\ar@{}[ddddrr]|{}&&&\ar@{-}[ld]|{\tilde{X}^{(1n)}}\\
    \ar@{-}[ddd]|{}\ar@{-}[rrr]|{}\ar@{}[dddrrr]|{[a_{11}b_{11}]}&&&\ar@{-}[ddd]|{}\ar@{-}[rrr]|{\cdots}\ar@{}[dddrrr]|{\cdots}&&&\ar@{-}[ddd]|{}\ar@{-}[rrr]|{}\ar@{}[dddrrr]|{[a_{1n}b_{1n}]}&&&\ar@{-}[ddd]\\
&&&&&&&&&\\
&\ar@{-}[ld]|{\tilde{X}^{(21)}}\ar@{}[ddddrr]|{}&&&\ar@{-}[ld]|{\tilde{X}^{(22)}}\ar@{}[ddddrr]|{}&&&\ar@{-}[ld]|{\tilde{X}^{(2n)}}&&&\\
    \ar@{-}[ddd]|{\vdots}\ar@{-}[rrr]|{}\ar@{}[dddrrr]|{\vdots}&&&\ar@{-}[ddd]|{\vdots}\ar@{-}[rrr]|{\cdots}\ar@{}[dddrrr]|{\ddots}&&&\ar@{-}[ddd]|{\vdots}\ar@{-}[rrr]|{}\ar@{}[dddrrr]|{\vdots}&&&\ar@{-}[ddd]|{\vdots}\\
&&&&&&&&&\\
&\ar@{-}[ld]|{\tilde{X}^{(m1)}}&&&\ar@{-}[ld]|{\tilde{X}^{(m2)}}&&&\ar@{-}[ld]|{\tilde{X}^{(mn)}}&&\\
    \ar@{-}[ddd]|{}\ar@{-}[rrr]|{}\ar@{}[dddrrr]|{[a_{m1}b_{m1}]}&&&\ar@{-}[ddd]|{}\ar@{-}[rrr]|{\cdots}\ar@{}[dddrrr]|{\cdots}&&&\ar@{-}[ddd]|{}\ar@{-}[rrr]|{}\ar@{}[dddrrr]|{[a_{mn}b_{mn}]}&&&\ar@{-}[ddd]\\
&&&&&&&&&\\
&&&&&&&&&\\
    \ar@{-}[rrr]&&&\ar@{-}[rrr]|{\cdots}&&&\ar@{-}[rrr]&&&\\
}
\end{smallgathered},
\end{multline}
where
\begin{align}
&   \renewcommand{\labelstyle}{\textstyle}
\begin{smallgathered}
\xymatrix@C=3. em{
&\ar@{-}[ld]|{\tilde{X}^{(ij)}}\ar@{..}[rrr]\ar@{..}[d]\ar@{}[drr]&&&\ar@{..}[ld]\\
    \ar@{..}[rrr]\ar@{..}[ddd]\ar@{}[ddr]&\ar@{..}[dd]&&&\\
                &&&&\\
                &\ar@{..}[ld]&&&\\
                &&&&
}
\end{smallgathered}
= 
   \renewcommand{\labelstyle}{\textstyle}
\begin{smallgathered}
\xymatrix@C=3. em{
&\ar@{-}[ld]|{X}\ar@{-}[rrr]\ar@{-}[d]\ar@{}[drr]|{\txt{$(X_0X_1)^{\tilde b^{(ij)}}\text{SWAP}$}}&&&\ar@{-}[ld]\\
    \ar@{-}[rrr]\ar@{-}[ddd]\ar@{}[ddr]|{\txt{$(X_0$$X_1)^{\tilde c^{(ij)}}$\\$\text{SWAP}$}}&\ar@{-}[dd]&&&\\
                &&&&\\
                &\ar@{-}[ld]&&& \\
                &&&&
}  
\end{smallgathered}, \notag \\
 &\tilde b^{(ij)}=
    a_{1j}+b_{1j}+\cdots+a_{i-1,j}+b_{i-1,j}, \notag\\
    &\tilde c^{(ij)}=
    a_{i1}+b_{i1}+\cdots+a_{i,j-1}+b_{i,j-1}.
\end{align}
The projectors on this decorated membrane operator ensures no additional excitations on the membrane.


\item $e_d^{(0)}e_d^{(1)}$: The composite Abelian lineon created by the two ends of $S_{v,0}^{n}S_{v,1}^{n}$

\end{itemize}

\begin{table*}[t]
\centering
 \begin{tabular}{c c c   || c c c c c } 
 \hline
Charge & Irrep of $G^\text{local}\cong D_4$ & Type  &  Flux & $B_{p_0}$ & $B_{t_0,0},B_{t_0,1},B_{t_0,2}$ & Conj.\ class  & Type    \\ 
 \hline\hline
 Vacuum & Trivial & Vacuum   &Vacuum & $1$ & $1,1,0$ & \{1\} & Vacuum  \\ 
 $\phi$ & $(1,1,-1)$ & Abelian fully mobile particle  & $\sigma$ & $-1$ & $0,0,1$  &  $\{g^{(2)},g^{(0)}g^{(1)}g^{(2)}\} $ & Flexible string  \\
 $f_0^{(0)}f_0^{(1)}$ & $(-1,-1,1)$ & Abelian fracton   & $e_d^{(0)}e_d^{(1)}$ & $1$ & $-1,-1,0$ & $\{g^{(0)}g^{(1)}\}$ & Abelian lineon   \\
 $\phi f_0^{(0)}f_0^{(1)}$ & $(-1,-1,-1)$ & Abelian fracton & $\sigma[e_d]$ & $-1$ & $0,0,-1$  & $\{g^{(1)}g^{(2)},g^{(0)}g^{(2)}\} $ & String and lineon   \\
 $[f_0]$ & 2D representation & Non-Abelian fracton & $[e_d]$ & $1$ & $1,-1,0$ or $-1,1,0$  & $\{g^{(0)},g^{(1)}\}$ & Non-Abelian lineon  \\ 
 \hline
 \end{tabular}
 \caption{Left/right panel: Pure electric charges/magnetic fluxes of the gauged $(\mathbb{Z}_2^\text{sub}\times\mathbb{Z}_2^\text{sub})\rtimes\mathbb{Z}_2^\text{glo}$ model. $(a,b,c)$ indicates the representation that $g^{(0)}\mapsto a,g^{(1)}\mapsto b,g^{(2)}\mapsto c$.} 
\label{tab:z2}
\end{table*}
In the same spirit as the QDM, the magnetic fluxes would be the conjugacy classes of $G^\text{local}\cong D_4$.
 With the geometry defined in Eq. (\ref{eq:geometry}) and the constraints defined in Eq. (\ref{eq:constraint}),
the flux operators can be mapped to a subalgebra of the QDM algebra for either $P_\text{side}=1$ or $P_\text{corner}=1$,
\begin{align}
    B_{p_0}&\mapsto B_{e}+B_{g^{(0)}}+B_{g^{(1)}}+B_{g^{(0)}g^{(1)}}\notag\\
    &-B_{g^{(2)}}-B_{g^{(0)}g^{(2)}}-B_{g^{(1)}g^{(2)}}-B_{g^{(0)}g^{(1)}g^{(2)}},\notag\\
   B_{t_0,0}&\mapsto B_{e}-B_{g^{(0)}}+B_{g^{(1)}}- B_{g^{(0)}g^{(1)}}, \notag\\
     B_{t_0,1}&\mapsto B_{e}+B_{g^{(0)}}-B_{g^{(1)}}- B_{g^{(0)}g^{(1)}},
\end{align}
where $B_g$ is the QDM flux operator in Eq.~(\ref{eq:qdm}).
We can also identify the star operators $A_{v_0,g}^\text{u}$ with the QDM star operator $A_g$ Eq.~(\ref{eq:qdm}).
The map of $A$'s and $B$'s together forms an injective algebra homomorphism into the QDM algebra. In particular, they satisfy the relations (note that all the $B$'s commute)
\begin{align}
    &B_{p_0}^2=1,\quad B_{t_0,0}^2=B_{t_0,1}^2=\frac{1}{2}(1+B_{p_0}), \notag\\
    &B_{p_0} B_{t_0,i}=B_{t_0,i},\quad i=0,1,   \notag\\
    &B_{p_0} B_{t_0,0}B_{t_0,1}=B_{t_0,0}B_{t_0,1},   \notag\\
    &A_{g}B_{p_0}=B_{p_0}A_{g}, \quad \forall g\in D_4,i=0,1,      \notag\\
    &A_{g}B_{t_0,i}=B_{t_0,i}A_{g},\quad g=e,g^{(0)},g^{(1)},g^{(0)}g^{(1)}, \quad i=0,1,      \notag\\
    &A_{g}B_{t_0,i}=B_{t_0,1-i}A_{g},  \notag  \\
    &g=g^{(2)},g^{(0)}g^{(2)},g^{(1)}g^{(2)},g^{(0)}g^{(1)}g^{(2)}, \quad i=0,1.   
\end{align}
In this way, the star and the flux operator of this fracton models is identified with a subalgebra of that of the corresponding QDM.

Unlike the $S_3$ case, this subalgebra is not enough to distinguish all of the conjugacy classes of $D_4$, since $\{g^{(2)},g^{(0)}g^{(1)}g^{(2)}\} $ and $\{g^{(1)}g^{(2)},g^{(0)}g^{(2)}\}$ both correspond to $B_{p_0}=-1$, $B_{t_0,0}=B_{t_0,1}=0$.
To avoid this situation, we define another flux operator which does not exist in the original Hamiltonian 
\begin{align}
    B_{t_0,2}=
    \renewcommand{\labelstyle}{\textstyle}
\begin{smallgathered}
\xymatrix@!0@=2.1em{
&\ar@{-}[ld]\ar@{-}[d]\ar@{-}[rrr]\ar@{}[rrd]|{Z_0Z_1}&&&\ar@{-}[ld]\\
    v_0\ar@{-}[rrr]\ar@{-}[ddd]\ar@{}[ddr]|{Z_0Z_1}&&&\ar@{-}[ddd]\ar@{}[ddr]|{Z_0Z_1}&\\
&&&&\\
&\ar@{-}[ld]\ar@{-}[rr]\ar@{-}[uu]\ar@{}[rrd]|{Z_0Z_1}&&&\ar@{-}[uuu]\ar@{-}[l]\\
    \ar@{-}[rrr]&&&\ar@{-}[ru]
}
\end{smallgathered}   
&\cdot  (1-P_{p_0}).
\end{align}
This flux operator is interpreted as a tube operator wrapping around $t$ twice as $B_{t,0}B_{t,1}$ when $B_{p_0}=-1$ ($p_0$ is the front plaquette).
Then we have the additional correspondence
\begin{align}
     B_{t_0,2}&\mapsto B_{g^{(2)}}-B_{g^{(0)}g^{(2)}}-B_{g^{(1)}g^{(2)}}+B_{g^{(0)}g^{(1)}g^{(2)}}
\end{align}
and additional relations
\begin{align}
    &B_{t_0,2}^2=\frac{1}{2}(1-B_{p_0}),\quad
    B_{p_0} B_{t_0,2}=-B_{t_0,2}, \notag  \\
    &B_{t_0,0}B_{t_0,1}B_{t_0,2}=0, \quad
    A_{g}B_{t_0,2}=B_{t_0,2}A_{g},\,\forall g\in D_4.
\end{align}
After including $B_{t_0,2}$, this subalgebra is enough to distinguish all of the conjugacy classes of $D_4$. 
To be more precise, if a state is the eigenstate of $B_{g_0}=1,B_{g}=0,g\neq g_0$ for some $g_0\in D_4$, then we can determine the conjugacy class of $g_0$ only by the eigenvalues of the image of $B_{p_0}$ and $B_{t_0,i}$.
This allows us to identify the fluxes of this fracton model with the conjugacy classes of $D_4$.
The corresponding fluxes is in the right panel of Table \ref{tab:z2}.


\subsection{ Example: gauging \texorpdfstring{$1\to\mathbb{Z}_2^\text{sub}\to G\to\mathbb{Z}_2^\text{glo}\to 1$}{1->Z\_2->G->Z\_2->1}}\label{sec:z4}
The ungauged system is a 3D cubic lattice with a $d=4$ qudit on each site. The Hamiltonian is
    \begin{align}
        H_o&=-\sum_\text{sites}(1 + X + X^2 + X^3), \notag \\
        H_n&=-J_0\sum_\text{plaquettes}\fplaquetteop{c_0}{}{}{}{}{}{}{}{}
        -J_1\sum_\text{links}\flinkop{c_1}{}{}+\text{H.c.},
    \end{align}
    where the minimal couplings are
\begin{align}
    \fplaquetteop{c_0}{}{}{}{}{}{}{}{}&:=f\Bigg(\fplaquetteop{\tilde c_0}{}{}{}{}{}{}{}{}\Bigg),\quad \fplaquetteop{\tilde c_0}{}{}{}{}{}{}{}{}=\fplaquetteop{}{}{}{}{}{Z}{Z^\dagger}{Z^\dagger}{Z}\label{eq:z4coupling}  \notag \\
            \flinkop{c_1}{}{}&:=\flinkop{}{Z^2}{Z^2}.
        \end{align}
        where $f(1)=f(i)=1$, $f(-1)=f(-i)=-1$.
     The construction of this function $f$, along with another choice of the coupling terms which are more similar to the previous cases, is discussed in Appendix \ref{sec:coupling}.
        Although $\tilde c_0$ is not a coupling in $H_n$, it can be obtained from $c_0$ and $c_1$:
        \begin{equation}
            \fplaquetteop{\tilde c_0}{}{}{}{}{}{}{}{}=\fplaquetteop{c_0}{}{}{}{}{}{}{}{}\sqrt{\fplaquetteop{}{c_1}{}{}{c_1}{}{}{}{}},
        \end{equation}
        where the square root takes the eigenvalue $-1$ of the operator inside it to $+i$.      

    $G$ is generated by $g_P^{(0)}$ and $g^{(1)}$, where
    \begin{equation}
    g_P^{(0)}=\prod_{\substack{\text{sites}\in P}}X^2,\quad g^{(1)}=\prod_{\substack{\text{sites}\in\text{all}}}X.
    \end{equation}
    Note that $g_P^{(0)}$'s generate a normal subgroup $\mathbb{Z}_2^{\text{sub}}$ of $G$, but $G$ is not a semidirect product of this $\mathbb{Z}_2^{\text{sub}}$ with $G/\mathbb{Z}_2^{\text{sub}}\cong \mathbb{Z}_2^{\text{glo}}$.
Nevertheless, $G$ can still be presented as the group extension
\begin{equation}
1\to\mathbb{Z}_2^\text{sub}\to G\to\mathbb{Z}_2^\text{glo}\to 1.
\end{equation}
In the notation of Ref.~\cite{tantivasadakarn2021hybrid}, this symmetry is called $(\mathbb{Z}_4,\mathbb{Z}_2)$.

The procedure of gauging works as follows:
\begin{enumerate}
    \item We put one qubit on each link (corresponds to $c_1$) and one qubit on each plaquette (corresponds to $c_0$).
Although we do not view $\tilde c_0$ as a minimal coupling term in $H_o$ and therefore do not put the corresponding gauge qudit, we can still define the ``dressed $Z$'' operator associated with $\tilde c_0$ on the plaquettes, which acts on the gauge qubits on both the plaquettes and the links:        
 \begin{equation}
            \fplaquetteop{\tilde Z}{}{}{}{}{}{}{}{}:=\fplaquetteop{Z}{}{}{}{}{}{}{}{}\sqrt{\fplaquetteop{}{Z}{}{}{Z}{}{}{}{}}
        \end{equation}
In addition, we define the ``dressed $X$'' operator associated to the link
\begin{equation}
    \renewcommand{\labelstyle}{\textstyle}
\begin{smallgathered}
\xymatrix@=1cm{
    \ar@{..}[r]\ar@{..}[d]\ar@{}[dr] & \ar@{..}[d] \\
    \ar@{-}[r]|{\tilde X}\ar@{..}[d]\ar@{}[dr] & \ar@{..}[d] \\
    \ar@{..}[r] &
}
\end{smallgathered}
:=
\sum_{a,b,c=0,1}
\begin{smallgathered}
\xymatrix@=1cm{
    \ar@{-}[r]|{[a]}\ar@{-}[d]\ar@{}[dr]|{X^{a+b+1}} & \ar@{-}[d] \\
    \ar@{-}[r]|{X[b]}\ar@{-}[d]\ar@{}[dr]|{X^{b+c}} & \ar@{-}[d] \\
    \ar@{-}[r]|{[c]} &
}
\end{smallgathered}.
\end{equation}
They satisfy the commutation relations
\begin{align}
    \linkop{Z}{}{}\linkop{\tilde X}{}{}&=-\linkop{\tilde X}{}{}\linkop{Z}{}{}   \notag \\
    \plaquetteop{\tilde Z}{}{}\plaquetteop{X}{}{}&=-\plaquetteop{X}{}{}\plaquetteop{\tilde Z}{}{}  \notag \\
    \renewcommand{\labelstyle}{\textstyle}
\fplaquetteop{\tilde Z}{}{}{}{}{}{}{}{}
\begin{gathered}
\xymatrix@=0.7 cm{
    \ar@{-}[r]|{\tilde X}\ar@{..}[d]\ar@{}[dr] & \ar@{..}[d] \\
    \ar@{..}[r] &
}
\end{gathered}  
&=
+i
    \renewcommand{\labelstyle}{\textstyle}
\begin{gathered}
\xymatrix@=0.7 cm{
    \ar@{-}[r]|{\tilde X}\ar@{..}[d]\ar@{}[dr] & \ar@{..}[d]   \notag \\
    \ar@{..}[r] &
}
\end{gathered}
\fplaquetteop{\tilde Z}{}{}{}{}{}{}{}{}\\
    \renewcommand{\labelstyle}{\textstyle}
\fplaquetteop{\tilde Z}{}{}{}{}{}{}{}{}
\begin{gathered}
\xymatrix@=0.7 cm{
    \ar@{..}[r]\ar@{..}[d]\ar@{}[dr] & \ar@{..}[d] \\
    \ar@{-}[r]|{\tilde X} &
}
\end{gathered}
&=
-i
    \renewcommand{\labelstyle}{\textstyle}
\begin{gathered}
\xymatrix@=0.7 cm{
    \ar@{..}[r]\ar@{..}[d]\ar@{}[dr] & \ar@{..}[d] \\
    \ar@{-}[r]|{\tilde X} &
}
\end{gathered}
\fplaquetteop{\tilde Z}{}{}{}{}{}{}{}{}
,
\end{align}
and all other combinations commute.

 \item
Based on the general gauging procedure we described in Sec. \ref{sec:gauging}.\ref{step2}, the gauge transformations are
\begin{align} 
\vspace{-2em}
    &A_{v,g^{(1)}}=
    \renewcommand{\labelstyle}{\textstyle}
\begin{smallgathered}
\xymatrix@=0.5cm{
    &&&&\\
    &&&&\\
     &&X\ar@{-}[dl]|{\tilde X}\ar@{-}[ll]|{\tilde X}\ar@{-}[rr]|{\tilde X^\dagger}\ar@{-}[uu]|{\tilde X}\ar@{-}[dd]|{\tilde X^\dagger}\ar@{-}[ur]|{\tilde X^\dagger}&&\\
     &&&&\\
     &&&& 
}  \notag 
\end{smallgathered},  \\
  &  A_{v,g_P^{(0)}}
    =A_{v,g^{(1)}}^2=
    \renewcommand{\labelstyle}{\textstyle}
\begin{smallgathered}
\xymatrix@=1.5em@!0{
    &&&&&\ar@{-}[d]&&&\\
    &\ar@{-}[rrr]|{\color{blue}}&&&\ar@{-}[ru]\ar@{-}[ld]&\ar@{-}[dd]|(.25){\color{orange}}&&\ar@{-}[lll]|{\color{blue}}&\\
    &&&\ar@{-}[ddd]|{\color{orange}}&&&&&\\
    &&\ar@{-}@[green!50!black][ld]|{\color{green!50!black}}&\ar@{-}@[green!50!black][l]&\ar@{-}@[green!50!black][l]&\ar@{-}@[green!50!black][l]\ar@{-}@[green!50!black][rr]&&\ar@{-}@[green!50!black][r]&\ar@{-}@[green!50!black][ld]|{\color{green!50!black}}\\
    &\ar@{-}[uuu]\ar@{-}[d]&&\ar@{-}[ll]|(.25){}&X^2\ar@{-}[uuu]|{}\ar@{-}[rrr]|{}\ar@{-}[d]\ar@{-}[l]\ar@{-}[ru]|{}\ar@{-}[ld]|{}\ar@{}[llluuu]|(.6){\color{blue}X}\ar@{}[luu]|(.68){\color{orange}X}\ar@{}[ruuuu]|(.5){\color{orange}X}\ar@{}[rrruuu]|(.6){\color{blue}X}\ar@{}[llu]|(.7){\color{green!50!black}X}\ar@{}[rrrru]|(.5){\color{green!50!black}X}\ar@{}[rrrddd]|(.6){\color{blue}X}\ar@{}[rdd]|(.65){\color{orange}X}\ar@{}[ldddd]|(.5){\color{orange}X}\ar@{}[lllddd]|(.6){\color{blue}X}\ar@{}[rrd]|(.7){\color{green!50!black}X}\ar@{}[lllld]|(.5){\color{green!50!black}X}&\ar@{-}[u]&&\ar@{-}[uuu]\ar@{-}[ddd]&\\
    \ar@{-}@[green!50!black][ru]|{\color{green!50!black}}&\ar@{-}[dd]&&\ar@{-}@[green!50!black][lll]\ar@{-}@[green!50!black][rrr]&\ar@{-}[dd]|(.25){}&\ar@{-}[u]|{\color{orange}}&\ar@{-}@[green!50!black][ru]|{\color{green!50!black}}&&\\
                       &&&&&\ar@{-}[u]&&&\\
                       &\ar@{-}[rr]|(.75){\color{blue}}&&\ar@{-}[r]&\ar@{-}[ru]\ar@{-}[ld]&&&\ar@{-}[lll]|{\color{blue}}&\\
                       &&&\ar@{-}[uuu]|{\color{orange}}&&&&&
}
\end{smallgathered}.\hspace{-1em}
\label{eq:AvgP2z4}
\vspace{-1em}
\end{align}

Note that these gauge transformations generate a local $\mathbb{Z}_4$ symmetry for the gauged system:
\begin{equation}
    G^\text{local}=\Big\langle g^{(0)}\mid \left(g^{(0)}\right)^4=1\Big\rangle\cong \mathbb{Z}_4.
\end{equation}
We will denote $\left(g^{(0)}\right)^n$ simply by $n$ for $n=0,1,2,3$, as the usual notation for $\mathbb{Z}_4$.

\item There are two ways to combine the $c$'s to produce the identity.
    The first type is the product of four $c_1$ on the links, which has the identical structure
    as the first example in Eq. (\ref{eq:Rp}) [replacing $c_2$ by $c_1$].
    The corresponding gauge field term $B_p$ is the plaquette operator of the link qubit,
    which has the identical structure as the first example in Eq. (\ref{eq:Bp}).
    The second type is the tube-like product of $\tilde c_0$
\begin{equation}
    \renewcommand{\labelstyle}{\textstyle}
    R_{t}(\{c\})=
\begin{smallgathered}
\xymatrix@=0.2cm{
&\ar@{-}[ld]\ar@{-}[d]\ar@{-}[rrr]\ar@{}[rrd]|{\tilde c_0^\dagger}&&&\ar@{-}[ld]\\
    \ar@{-}[rrr]\ar@{-}[ddd]\ar@{}[ddr]|{\tilde c_0}&&&\ar@{-}[ddd]\ar@{}[ddr]|{\tilde c_0^\dagger}&\\
&&&&\\
&\ar@{-}[ld]\ar@{-}[rr]\ar@{-}[uu]\ar@{}[rrd]|{\tilde c_0}&&&\ar@{-}[uuu]\ar@{-}[l]\\
\ar@{-}[rrr]&&&\ar@{-}[ru]
}
\end{smallgathered}
=1,\quad t\in\text{tubes},
\end{equation}

The corresponding field term is
\begin{align}
    B_{t}&=R_{t}(\{Z\})\prod_{\text{six faces}}P_p
    =
    \renewcommand{\labelstyle}{\textstyle}
\begin{smallgathered}
\xymatrix@=0.2cm{
&\ar@{-}[ld]\ar@{-}[d]\ar@{-}[rrr]\ar@{}[rrd]|{\tilde Z^\dagger}&&&\ar@{-}[ld]\\
    \ar@{-}[rrr]\ar@{-}[ddd]\ar@{}[ddr]|{\tilde Z}&&&\ar@{-}[ddd]\ar@{}[ddr]|{\tilde Z^\dagger}&\\
&&&&\\
&\ar@{-}[ld]\ar@{-}[rr]\ar@{-}[uu]\ar@{}[rrd]|{\tilde Z}&&&\ar@{-}[uuu]\ar@{-}[l]\\
\ar@{-}[rrr]&&&\ar@{-}[ru]
}
\end{smallgathered}\prod_{\text{six faces}}P_p
\end{align}

Note that, unlike the non-Abelian cases, the projectors $P_p$ are not necessary. 
This is because that $R_{t}(\{Z\})$ commutes with all $A_{v,g}$.

\item The gauged couplings are
       \begin{align}
           \fplaquetteop{c_0(\tau)}{}{}{}{}{}{}{}{}:=f\left(\fplaquetteop{\tilde Z}{}{}{}{}{Z}{Z^\dagger}{Z^\dagger}{Z}\right)\cdot P_p
        \end{align}
        Again, in this case the projector $P_p$ is not necessary.

\item The electric field terms are
\begin{align}
    H_e&=-g_0\sum_{p \text{ plaquettes}}\fplaquetteop{X}{}{}{}{}{}{}{}{}
-g_1\sum_\text{links}\left(\flinkop{\tilde X}{}{}+\flinkop{\tilde X^\dagger}{}{}\right).
\end{align}

\end{enumerate}

Similar to the toric code and the X-Cube code, we can choose the unitary gauge to eliminate the matter DOF in this example. We regard the ``physical state'' satisfying the constraints as an equivalent class of computational basis state, and choose the representative such that all matter DOF are in the reference state $|0\rangle$.
In this case, the Hamiltonian becomes
\begin{align}\label{eq:z4exact}
    H_\text{u}=&
    -\sum_{v\text{ sites}}\left(1+A_{v,g^{(1)}}^\text{u}+A_{v,g^{(0)}_P}^\text{u} +\left(A_{v,g^{(1)}}^\text{u}\right)^\dagger \right)\notag\\
    &-\sum_{t\text{ tubes}}\left(B_{t}+B_{t}^\dagger\right)-\sum_{p\text{ plaquettes}}B_p\notag\\
&-\left(J_0\sum_{p \text{ plaquettes}}f\left(\fplaquetteop{\tilde Z}{}{}{}{}{}{}{}{}\right)
+J_1\sum_\text{links}\flinkop{Z}{}{}+\text{H.c.}\right)\notag\\
&-g_0\sum_{p \text{ plaquettes}}\fplaquetteop{X}{}{}{}{}{}{}{}{}
-g_1\sum_\text{links}\left(\flinkop{\tilde X}{}{}+\flinkop{\tilde X^\dagger}{}{}\right),
\end{align}
and the Hilbert space contains only the gauge DOF without constraints.
The above Hamiltonian $H_\text{u}$ describes the pure lattice gauge theory of $G$.
Note that if we replace $\tilde X^{(\dagger)}$ and $\tilde Z^{(\dagger)}$ by the usual Pauli operators $X$ and $Z$ on the corresponding links and plaquettes, the resulting model will be just the tensor product of an X-cube code and a 3D toric code.

In the exactly solvable limit ($J_0=J_1=g_0=g_1=0$), the system is deeply
in the deconfined phase and the fundamental excitations are fully mobile particles, fractons, and strings. 
These excitations are electric charges which correspond to excitations of $A^\text{u}_{v, g^{(0)}_P}$, $A^\text{u}_{v, g^{(1)}}$, $\left(A^\text{u}_{v, g^{(1)}} \right)^\dagger$,
and the magnetic fluxes which are the excitations of $B_p$, $B_t$, $B_t^\dagger$. We discuss these excitations as follows.

\subsubsection{Electric charge excitations}
\begin{itemize}
    \item $e$: The Abelian fracton, corresponding to the excitation of the first line of (\ref{eq:z4exact}), with the eigenvalues of the four terms being $1$, $i$, $-1$ and $-i$, respectively.
        It is created at the upper left and lower right corners of the membrane operator
        \begin{equation}
            \renewcommand{\labelstyle}{\textstyle}
            \begin{smallgathered}
            \xymatrix@=3.5 em{
                \ar@{-}[r]\ar@{-}[d]\ar@{}[dr]|{\tilde Z} & \ar@{}[dr]|{\cdots}\ar@{-}[d] & \ar@{-}[r]\ar@{-}[d]\ar@{}[dr]|{\tilde Z} & \ar@{-}[d]\ar@{-}[r]\ar@{}[dr]|{\tilde Z} & \ar@{-}[d] \\
                \ar@{-}[r]\ar@{}[dr]|{\vdots} & \ar@{}[dr]|{\ddots} & \ar@{-}[r]\ar@{}[dr]|{\vdots} & \ar@{-}[r]\ar@{}[dr]|{\vdots} & \\
                \ar@{-}[r]\ar@{-}[d]\ar@{}[dr]|{\tilde Z} & \ar@{-}[d]\ar@{}[dr]|{\cdots} & \ar@{-}[r]\ar@{-}[d]\ar@{}[dr]|{\tilde Z} & \ar@{-}[r]\ar@{-}[d]\ar@{}[dr]|{\tilde Z} & \ar@{-}[d] \\
                \ar@{-}[r]\ar@{-}[d]\ar@{}[dr]|{\tilde Z} & \ar@{-}[d]\ar@{}[dr]|{\cdots} & \ar@{-}[r]\ar@{-}[d]\ar@{}[dr]|{\tilde Z} & \ar@{-}[r]\ar@{-}[d]\ar@{}[dr]|{\tilde Z} & \ar@{-}[d]\\
                \ar@{-}[r] & & \ar@{-}[r] & \ar@{-}[r] &
            }
            \end{smallgathered},
        \end{equation}
       with its antiparticle $e^3$ created at the upper right and the lower left corner.
    \item $e^2$: The combination of two $e$'s, which is an Abelian fully mobile particle.
It corresponds to the excitation of the first line of Eq.~\ref{eq:z4exact},  with the eigenvalues of the four terms being $1$, $-1$, $1$ and $-1$, respectively.
Besides the $e^2$ excitation can be created at the edge of the square of the membrane for $e$, it can also be created at the endpoints of the string operator, which has the identical structure as 
the first example Eq. (\ref{eq:phi}).
 Indeed, the square of the membrane for $e$ is just two copies of this string, one at the top and one at the bottom of the membrane.
\end{itemize}
 The excitations of $A_{v,g^{(1)}}^\text{u}$, $A_{v,g^{(0)}_P}^\text{u}$, and $\left(A_{v,g^{(1)}}^\text{u}\right)^\dagger$ are referred to the electric charge excitations. These excitations are local with respect
to the vertex $v$,
and can be specified from the local operators which form a representation of $\mathbb{Z}_4$ on the Hilbert space.
Hence we can identify electric charges of the $G$ fracton model
with the irreducible representations of $G^\text{local} \cong \mathbb{Z}_4$ (left panel of Table~\ref{tab:z4}). 
This identification is the same as the QDM with $\mathbb{Z}_4$ symmetry~\cite{A.Yu.Kitaev2002}.

\subsubsection{The magnetic flux excitations}
\begin{itemize} 
    \item $m$: The flexible string-like excitation corresponding to the excitation $B_p=-1$. It is created at the boundary of the membrane operator
 \begin{equation}
   \renewcommand{\labelstyle}{\textstyle}
\begin{smallgathered}
\xymatrix@=1.7 em{
&\ar@{-}[ld]|{\tilde X}&&&\ar@{-}[ld]|{\tilde X}\ar@{}[ddddrr]|{\cdots}&&&\ar@{-}[ld]|{\tilde X}\\
&&&&&&&\\
&&&&&&&\\
&\ar@{-}[ld]|{\tilde X}\ar@{}[ddddrr]|{\vdots}&&&\ar@{-}[ld]|{\tilde X}\ar@{}[ddddrr]|{\ddots}&&&\ar@{-}[ld]|{\tilde X}\\
&&&&&&&\\
&&&&&&&\\
&\ar@{-}[ld]|{\tilde X}&&&\ar@{-}[ld]|{\tilde X}&&&\ar@{-}[ld]|{\tilde X}\\
&&&&&&&\\
}
\end{smallgathered}
\end{equation}
Note that this membrane also excites the $B_t$ operators. If we look at the $R_t(\{  Z \})$ (without the projectors), the excitation has eigenvalues $\pm i$ at the corners of the membrane, but still have $1$ otherwise.
The $m^3$ string can be created by the similar membrane operator with replacing $\tilde{X} \to \tilde{X}^\dagger$.
\item $m^2_d$, where $d=x,y,z$: The Abelian lineon constrained to move in the $d$ direction, corresponding to the excitation having $B_t=-1$. It is created at the endpoints at the string operator (the direction of the string is $d$)
        \begin{equation}
    \renewcommand{\labelstyle}{\textstyle}
\begin{smallgathered}
\xymatrix@=0.7 em{
&\ar@{-}[ld]\ar@{}[rrd]&&&\ar@{-}[ld]\ar@{}[ddddrr]|{\cdots}\ar@{}[rrd]&&&\ar@{-}[ld]\ar@{}[rrd]&&&\ar@{-}[ld]\\
    \ar@{-}[ddd]\ar@{}[ddr]|{X}&&&\ar@{}[rrr]\ar@{-}[ddd]\ar@{}[ddr]|{X}&&&\ar@{-}[ddd]\ar@{}[ddr]|{X}&&&\ar@{-}[ddd]\ar@{}[ddr]|{X}&\\
&&&&&&&&&&\\
&\ar@{-}[ld]\ar@{-}[uuu]\ar@{}[rrd]&&&\ar@{-}[ld]\ar@{-}[uuu]\ar@{}[rrd]&&&\ar@{-}[ld]\ar@{-}[uuu]\ar@{}[rrd]&&&\ar@{-}[uuu]\\
&&&&&&&&&\ar@{-}[ru]
}
\end{smallgathered},
\end{equation}
It is also created at the corners of the square of the membrane for $m$.

\end{itemize}


\begin{table*}[t]
\centering
 \begin{tabular}{c c c   || c c c c c c } 
 \hline
 Charge & Irrep of $G^\text{local}\cong Z_4$ & Type & Flux & $B_{p_0}$ & $B_{t_0}$& $B_{t_0}^{'}$ & Conj. class  & Type  \\ 
 \hline\hline
 Vacuum & Trivial & Vacuum & Vacuum & $1$ & $1$ & $1$ & $\{0\}$ & Vacuum  \\ 
 $e$ & $1\mapsto i$ & Abelian fracton & $m$ & $-1$ & $0$ & $i$ & $\{1\}$ & Flexible string \\
 $e^2$ & $1\mapsto -1$ & Abelian fully mobile particle  & $m^2_d$ & $1$ & $-1$ & $-1$  & $\{2\}$ & Abelian lineon \\
 $e^3$ & $1\mapsto -i$ & Abelian fracton & $m^3$ & $-1$ & $0$ & $-i$ & $\{3\}$ & Flexible string \\
 \hline
 \end{tabular}
 \caption{Left/right panel: Pure electric charges/magnetic fluxes of the gauged $1\to\mathbb{Z}_2^\text{sub}\to G\to\mathbb{Z}_2^\text{glo}\to 1$ model. 
 This correspondence of fluxes with the conjugacy classes is meaningful only if we choose $P_\text{corner}=1$, that is, the point of interest must be at the corner if there is a flexible string. }
\label{tab:z4}
\end{table*}

        In the same spirit as the QDM, the magnetic fluxes would be the conjugacy classes of $G^\text{local}\cong \mathbb{Z}_4$, that is, elements of $\mathbb{Z}_4$.
 With the geometry defined in Eq. (\ref{eq:geometry}) and the constraints defined in Eq. (\ref{eq:constraint}), 
the flux operators can be mapped to a subalgebra of the QDM algebra for only $P_\text{corner}=1$,
\begin{align}
    B_{p_0}&\mapsto B_{0}-B_{1}+B_{2}-B_{3}  \notag \\
    B_{t_0}&\mapsto B_{0}-B_{2}
\end{align}
where $B_g$ is the QDM flux operator in Eq.~(\ref{eq:qdm}).
We can also identify the star operators $A_{v_0,g}^\text{u}$ with the QDM star operator $A_g$ Eq.~(\ref{eq:qdm}).
The map of $A$'s and $B$'s together forms an injective algebra homomorphism into the QDM algebra. In particular, they satisfy the relations (note that all the $B$'s commute)
\begin{align}
    &B_{p_0}^2=1, \quad B_{t_0}^2=\frac{1}{2}(1+B_{p_0}), \quad B_{p_0} B_{t_0}=B_{t_0}  \notag\\
    &A_{g}B_{p_0}=B_{p_0}A_{g},\quad A_{g}B_{t_0}=B_{t_0}A_{g}, \quad \forall g\in \mathbb{Z}_4  
\end{align}
In this way, the star and the flux operator of this fracton models is identified with a subalgebra of that of the corresponding QDM.

Unlike the $S_3$ case, this subalgebra is not enough to distinguish all of the conjugacy classes of $\mathbb{Z}_4$, since $\{1\}$ and $\{3\}$ both correspond to $B_{p_0}=-1$, $B_{t_0}=0$.
To avoid this situation, we note that in this Abelian case, the zero-flux projectors in the definition of $B_{t_0}$ is not necessary. Hence we can define the flux operator without the projectors. (this does not mean that we need to modified the Hamiltonian)
\begin{equation}
    B_{t_0}^{'}=
    \renewcommand{\labelstyle}{\textstyle}
\begin{smallgathered}
\xymatrix@!0@=1.3em{
&\ar@{-}[ld]\ar@{-}[d]\ar@{-}[rrr]\ar@{}[rrd]|{\tilde Z^\dagger}&&&\ar@{-}[ld]\\
    v_0\ar@{-}[rrr]\ar@{-}[ddd]\ar@{}[ddr]|{\tilde Z}&&&\ar@{-}[ddd]\ar@{}[ddr]|{\tilde Z^\dagger}&\\
&&&&\\
&\ar@{-}[ld]\ar@{-}[rr]\ar@{-}[uu]\ar@{}[rrd]|{\tilde Z}&&&\ar@{-}[uuu]\ar@{-}[l]\\
\ar@{-}[rrr]&&&\ar@{-}[ru]
}
\end{smallgathered}.
\end{equation}
Then we have the correspondence
\begin{align}
    B_{t_0}^{'}&\mapsto B_0+iB_1-B_2-iB_3 
\end{align}
and additional relations
\begin{align}
    &B_{t_0}^{'2}=B_{p_0},\,
    B_{p_0} B_{t_0}^{'}=B_{t_0}^{'\dagger},\,
    B_{t_0}B_{t_0}^{'}=\frac{1}{2}(1+B_{p_0}), \notag \\
    &A_{g}B_{t_0}^{'}=B_{t_0}^{'}A_{g},\,\forall g\in \mathbb{Z}_4
\end{align}

After including $B_{t_0}^{'}$, the corresponding subalgebra is enough to distinguish all of the elements of $\mathbb{Z}_4$ (more rigorously, if a state is the eigenstate of $B_{g_0}=1,B_{g}=0,g\neq g_0$ for some $g_0\in \mathbb{Z}_4$, then we can determine $g_0$ only by the eigenvalues of the image of $B_{p_0}$, $B_{t_0}$ and $B_{t_0}^{'}$. In this case, $B_{t_0}^{'}$ alone is enough to distinguish them.)
This allows us to identify the fluxes of this fracton model with the conjugacy classes of $\mathbb{Z}_4$ in the situation that $P_\text{corner}=1$.
The corresponding fluxes is in the right panel of Table \ref{tab:z4}.

    However, if we choose $P_\text{side}=1$ instead, then we have $B_{t_0}^{'2}=1$ on the fracton side instead of $B_{t_0}^{'2}=B_{p_0}$,
which implies that the previous mapping into $\mathbb{Z}_4$ QDM is no longer a homomorphism.
Curiously, it is isomorphic to the $\mathbb{Z}_2\times\mathbb{Z}_2$ QDM algebra by the mapping
\begin{align}
    B_{p_0}&\mapsto B_{(0,0)}-B_{(0,1)}+B_{(1,0)}-B_{(1,1)},  \notag \\
    B_{t_0}&\mapsto B_{(0,0)}-B_{(1,0)},  \notag \\
     B_{t_0}^{'}&\mapsto B_{(0,0)}+B_{(0,1)}-B_{(1,0)} - B_{(1,1)}.
\end{align}
Here the first $\mathbb{Z}_2$ is associated to $m^2_d$, and the second $\mathbb{Z}_2$ is associated to $m$.
This explains why when we apply the square of the membrane for the string excitation $m$, we create four lineons at the corners but nothing at the edge of the membrane. The fusion rules of the fluxes behave like the $\mathbb{Z}_4$ QDM ($m\times m=m^2$) only at the corners, while at the edge of a string excitation, they behave like a $\mathbb{Z}_2\times\mathbb{Z}_2$ QDM ($m\times m=0$) instead.
This phenomenon was first observed in Ref.~\cite{tantivasadakarn2021hybrid}, and we expect that it is general for gauged Abelian models with constraint-dependent QDM correspondence (see discussion before the first example).

\subsection{Example: gauging \texorpdfstring{$1\to\mathbb{Z}_2^\text{sub}\to G\to K_4^\text{glo}\to 1$}{1->Z\_2->G->K\_4->1}}\label{sec:q8}

The ungauged system is a 3D cubic lattice with three qubits on each site.
We use the following correspondence of the qubits with the elements of $Q_8$:
\begin{align}
    &|000\rangle\leftrightarrow 1,\quad
    |001\rangle\leftrightarrow i,\quad
    |010\rangle\leftrightarrow j,\quad
    |011\rangle\leftrightarrow k, \notag \\
    &|100\rangle\leftrightarrow -1,\quad
    |101\rangle\leftrightarrow -i,\quad
    |110\rangle\leftrightarrow -j,\quad
    |111\rangle\leftrightarrow -k.
\end{align}

For notational convenience, we combine the last two qubits together and define $X_1= XI$,  $X_2= IX$,  $Z_1= ZI$,  $Z_2= IZ$, and $II = I$.
The Hamiltonian is
    \begin{align}
        H_o&=-\sum_\text{sites}(IX_1 + IX_2 + XI), \notag \\
        H_n&=-J_0\sum_\text{plaquettes}\fplaquetteop{c_0}{}{}{}{}{}{}{}{}-J_1\sum_\text{links}\flinkop{c_1}{}{}-J_2\sum_\text{links}\flinkop{c_2}{}{},
    \end{align}
    where the minimal couplings are
        \begin{align}
            \fplaquetteop{c_0}{}{}{}{}{}{}{}{}    
          :=&\sum_{\substack{\mathbf a=00,01, \\ 10,11}}
         (-1)^{ a^{(1)}+a^{(2)}+a^{(1)}a^{(2)}}\fplaquetteop{}{}{}{}{}{Z[\mathbf a]}{Z\mathbf Z^{\mathbf a}}{Z\mathbf Z^{\mathbf a}}{Z\mathbf Z^{\mathbf a}}, \notag \\
          \flinkop{c_1}{}{}:=& \flinkop{}{IZ_1}{IZ_1},\quad\flinkop{c_2}{}{}:=\flinkop{}{IZ_2}{IZ_2},
        \end{align}
             where $\mathbf{a}:=a^{(1)}a^{(2)}$ with $a^{(1)}, a^{(2)}=0,1$, and  $\mathbf Z^{\mathbf a}:=Z_1^{a^{(1)}}Z_2^{a^{(1)}+a^{(2)}}$ on the second entry acts on the second and the third qubits.
             The projector $[\mathbf a]:=[a^{(1)}a^{(2)}]= | a^{(1)}a^{(2)} \rangle \langle a^{(1)}a^{(2)}|$ is defined on the second entry acting on the second and the third qubits.
    $G$ is generated by $g_P^{(-)}$, $g^{(i)}$, $g^{(j)}$, $g^{(k)}$ where
    \begin{align}
        &g_P^{(-)}=\prod_{\substack{\text{sites}\\\in P}}XI, \quad 
          g^{(i)}=\prod_{\substack{\text{sites}\\\text{(all)}}}\sum_{a, b=0,1}X^b\left( X_2[a b] \right),\notag \\
        &g^{(j)}=\prod_{\substack{\text{sites}\\\text{(all)}}}\sum_{a,b=0,1}X^{a+b}\left(  X_1[a b] \right), \notag \\
        &g^{(k)}=\prod_{\substack{\text{sites}\\\text{(all)}}}\sum_{a, b=0,1}X^a \left(  X_1 X_2 [ab]  \right),
    \end{align}
    where $\left( \cdots \right)$ acts on the second and third qubits. 
   These generators are the subsystem multiplication of $-1\in Q_8$ and the global multiplication of $i,j,k\in Q_8$, respectively.
    Note that $g_P^{(-)}$'s generate a normal subgroup $\mathbb{Z}_2^{\text{sub}}$ of $G$, but $G$ is not a semidirect product of this $\mathbb{Z}_2^{\text{sub}}$ with $G/\mathbb{Z}_2^{\text{sub}}\cong K_4^{\text{glo}}$.
 Nevertheless, $G$ can still be presented as the group extension
\begin{equation}
1\to\mathbb{Z}_2^\text{sub}\to G\to K_4^\text{glo}\to 1.
\end{equation}
In the notation of Ref.~\cite{tantivasadakarn2021hybrid}, this symmetry is called $(Q_8,\mathbb{Z}_2)$.

The procedure of gauging works as follows:
\begin{enumerate}
    \item We put two qubits on each link (corresponds to $c_1$ and $c_2$) and one qubit on each plaquette (corresponds to $c_0$).
    \item
Based on the general gauging procedure we described in Sec. \ref{sec:gauging}.\ref{step2}, the gauge transformations are
\begin{widetext}
\vspace{-4em}
\begin{equation}
    A_{v,g^{(i)}}=\sum_{a,\ldots,z=0,1}
    \renewcommand{\labelstyle}{\textstyle}
\begin{smallgathered}
\xymatrix@C=5.4em@!0{
    &&&&&\ar@{-}[d]&&&\\
    &\ar@{-}[rrr]|{\color{blue}[mn]}&&&\ar@{-}[ru]\ar@{-}[ld]&\ar@{-}[dd]|(.25){\color{orange}}&&\ar@{-}[lll]|{\color{blue}}&\\
    &&&\ar@{-}[ddd]|{\color{orange}[uv]}&&&&&\\
    &&\ar@{-}@[green!50!black][ld]|{\color{green!50!black}}&\ar@{-}@[green!50!black][l]&\ar@{-}@[green!50!black][l]&\ar@{-}@[green!50!black][l]\ar@{-}@[green!50!black][rr]&&\ar@{-}@[green!50!black][r]&\ar@{-}@[green!50!black][ld]|{\color{green!50!black}[st]}\\
    &\ar@{-}[uuu]\ar@{-}[d]&&\ar@{-}[ll]|(.25){X_2[ab]}&X^z \left( X_2[yz] \right) \ar@{-}[uuu]|(.7){X_2[cd]}\ar@{-}[rrr]|{X_2[ef]}\ar@{-}[d]\ar@{-}[l]\ar@{-}[ru]|(.6){X_2[gh]}\ar@{-}[ld]|(.6){X_2[ij]}\ar@{}[llluuu]|(.6){\color{blue}X^{m+n+c+d+1}}\ar@{}[luuu]|(.5){\color{orange}X^{u+v+i+j+1}}\ar@{}[ruuuu]|(.45){\color{orange}X^{c+d+1}}\ar@{}[rrruuu]|(.6){\color{blue}X^{c+d+1}}\ar@{}[llu]|(.7){\color{green!50!black}X^{a+b+1}}\ar@{}[rrrru]|(.5){\color{green!50!black}X^{h+t}}\ar@{}[rrrddd]|(.6){\color{blue}X^{f+p}}\ar@{}[rdd]|(.65){\color{orange}X^{l+x}}\ar@{}[ldddd]|(.5){\color{orange}X^{i+j+1}}\ar@{}[lllddd]|(.6){\color{blue}X^{a+b+1}}\ar@{}[rrd]|(.72){\color{green!50!black}X^{i+j+1}}\ar@{}[lllld]|(.5){\color{green!50!black}X^{q+r+a+b+1}}&\ar@{-}[u]&&\ar@{-}[uuu]\ar@{-}[ddd]&\\
    \ar@{-}@[green!50!black][ru]|{\color{green!50!black}[qr]}&\ar@{-}[dd]&&\ar@{-}@[green!50!black][lll]\ar@{-}@[green!50!black][rrr]&\ar@{-}[dd]|(.25){X_2[kl]}&\ar@{-}[u]|{\color{orange}[wx]}&\ar@{-}@[green!50!black][ru]|{\color{green!50!black}}&&\\
                       &&&&&\ar@{-}[u]&&&\\
                       &\ar@{-}[rr]|(.75){\color{blue}}&&\ar@{-}[r]&\ar@{-}[ru]\ar@{-}[ld]&&&\ar@{-}[lll]|{\color{blue}[op]}&\\
                       &&&\ar@{-}[uuu]|{\color{orange}}&&&&&
}
\vspace{-2em}
\end{smallgathered}
\notag,\label{eq:AvgPiq8}
\end{equation}
\begin{equation}
    A_{v,g^{(j)}}=\sum_{a,\ldots,z=0,1}
    \renewcommand{\labelstyle}{\textstyle}
\begin{smallgathered}
\xymatrix@C=5.4em@!0{
    &&&&&\ar@{-}[d]&&&\\
    &\ar@{-}[rrr]|{\color{blue}[mn]}&&&\ar@{-}[ru]\ar@{-}[ld]&\ar@{-}[dd]|(.25){\color{orange}}&&\ar@{-}[lll]|{\color{blue}}&\\
    &&&\ar@{-}[ddd]|{\color{orange}[uv]}&&&&&\\
    &&\ar@{-}@[green!50!black][ld]|{\color{green!50!black}}&\ar@{-}@[green!50!black][l]&\ar@{-}@[green!50!black][l]&\ar@{-}@[green!50!black][l]\ar@{-}@[green!50!black][rr]&&\ar@{-}@[green!50!black][r]&\ar@{-}@[green!50!black][ld]|{\color{green!50!black}[st]}\\
    &\ar@{-}[uuu]\ar@{-}[d]&&\ar@{-}[ll]|(.25){X_1[ab]}&X^{y+z} \left( X_1[yz] \right) \ar@{-}[uuu]|{X_1[cd]}\ar@{-}[rrr]|{X_1[ef]}\ar@{-}[d]\ar@{-}[l]\ar@{-}[ru]|(.65){X_1[gh]}\ar@{-}[ld]|(.6){X_1[ij]}\ar@{}[llluuu]|(.6){\color{blue}X^{m+c+1}}\ar@{}[luu]|(.65){\color{orange}X^{u+i+1}}\ar@{}[ruuuu]|(.5){\color{orange}X^{c+1}}\ar@{}[rrruuu]|(.6){\color{blue}X^{c+1}}\ar@{}[llu]|(.7){\color{green!50!black}X^{a+1}}\ar@{}[rrrru]|(.5){\color{green!50!black}X^{g+h+s+t}}\ar@{}[rrrddd]|(.6){\color{blue}X^{e+f+o+p}}\ar@{}[rddd]|(.45){\color{orange}X^{k+l+w+x}}\ar@{}[ldddd]|(.5){\color{orange}X^{i+1}}\ar@{}[lllddd]|(.6){\color{blue}X^{a+1}}\ar@{}[rrd]|(.7){\color{green!50!black}X^{i+1}}\ar@{}[lllld]|(.5){\color{green!50!black}X^{q+a+1}}&\ar@{-}[u]&&\ar@{-}[uuu]\ar@{-}[ddd]&\\
    \ar@{-}@[green!50!black][ru]|{\color{green!50!black}[qr]}&\ar@{-}[dd]&&\ar@{-}@[green!50!black][lll]\ar@{-}@[green!50!black][rrr]&\ar@{-}[dd]|(.4){X_1[kl]}&\ar@{-}[u]|{\color{orange}[wx]}&\ar@{-}@[green!50!black][ru]|{\color{green!50!black}}&&\\
                       &&&&&\ar@{-}[u]&&&\\
                       &\ar@{-}[rr]|(.75){\color{blue}}&&\ar@{-}[r]&\ar@{-}[ru]\ar@{-}[ld]&&&\ar@{-}[lll]|{\color{blue}[op]}&\\
                       &&&\ar@{-}[uuu]|{\color{orange}}&&&&&
}
\vspace{-2em}
\end{smallgathered}
,\label{eq:AvgPjq8}
\end{equation}
\end{widetext}
\begin{equation}
    A_{v,g_p^{(-)}}=A_{v,g^{(i)}}^2=A_{v,g^{(j)}}^2=
    \renewcommand{\labelstyle}{\textstyle}
\begin{smallgathered}
\xymatrix@=1.5em@!0{
    &&&&&\ar@{-}[d]&&&\\
    &\ar@{-}[rrr]|{\color{blue}}&&&\ar@{-}[ru]\ar@{-}[ld]&\ar@{-}[dd]|(.25){\color{orange}}&&\ar@{-}[lll]|{\color{blue}}&\\
    &&&\ar@{-}[ddd]|{\color{orange}}&&&&&\\
    &&\ar@{-}@[green!50!black][ld]|{\color{green!50!black}}&\ar@{-}@[green!50!black][l]&\ar@{-}@[green!50!black][l]&\ar@{-}@[green!50!black][l]\ar@{-}@[green!50!black][rr]&&\ar@{-}@[green!50!black][r]&\ar@{-}@[green!50!black][ld]|{\color{green!50!black}}\\
    &\ar@{-}[uuu]\ar@{-}[d]&&\ar@{-}[ll]|(.25){}&X I \ar@{-}[uuu]|{}\ar@{-}[rrr]|{}\ar@{-}[d]\ar@{-}[l]\ar@{-}[ru]|{}\ar@{-}[ld]|{}\ar@{}[llluuu]|(.6){\color{blue}X}\ar@{}[luu]|(.68){\color{orange}X}\ar@{}[ruuuu]|(.5){\color{orange}X}\ar@{}[rrruuu]|(.6){\color{blue}X}\ar@{}[llu]|(.7){\color{green!50!black}X}\ar@{}[rrrru]|(.5){\color{green!50!black}X}\ar@{}[rrrddd]|(.6){\color{blue}X}\ar@{}[rdd]|(.65){\color{orange}X}\ar@{}[ldddd]|(.5){\color{orange}X}\ar@{}[lllddd]|(.6){\color{blue}X}\ar@{}[rrd]|(.7){\color{green!50!black}X}\ar@{}[lllld]|(.5){\color{green!50!black}X}&\ar@{-}[u]&&\ar@{-}[uuu]\ar@{-}[ddd]&\\
    \ar@{-}@[green!50!black][ru]|{\color{green!50!black}}&\ar@{-}[dd]&&\ar@{-}@[green!50!black][lll]\ar@{-}@[green!50!black][rrr]&\ar@{-}[dd]|(.25){}&\ar@{-}[u]|{\color{orange}}&\ar@{-}@[green!50!black][ru]|{\color{green!50!black}}&&\\
                       &&&&&\ar@{-}[u]&&&\\
                       &\ar@{-}[rr]|(.75){\color{blue}}&&\ar@{-}[r]&\ar@{-}[ru]\ar@{-}[ld]&&&\ar@{-}[lll]|{\color{blue}}&\\
                       &&&\ar@{-}[uuu]|{\color{orange}}&&&&&
}
\end{smallgathered}\hspace{-1em},\hspace{-1em}\label{eq:AvgP2q8}
\end{equation}
One should be notified that $X^\alpha \left( X_{1(2)} [\beta\gamma] \right)$ acts on the center vertex with the first entry $X^\alpha$ acting on the first qubit and  the  $X_{1(2)} [\beta\gamma]$ acting on the second and third qubits.
These gauge transformations generate a local $Q_8$ symmetry for the gauged system:
\begin{align}
    G^\text{local}=\Big\langle g^{(i)},g^{(j)},g^{(k)}\mid \left(g^{(i)}\right)^2=\left(g^{(j)}\right)^2 \notag \\
    =\left(g^{(k)}\right)^2=g^{(i)}g^{(j)}g^{(k)}\Big\rangle\cong Q_8.
\end{align}
We will use the usual notation for $Q_8$, denoting $g^{(i)}$ by $i$, $\left(g^{(i)}\right)^2$ by $-1$, etc.

\item There are two ways to combine the $c$'s to produce the identity.
    The first type is the product of four $c_1$ or $c_2$ on the links:
    \begin{align}
        R_{p,i}(\{c\})=&\plaquetteop{}{c_i}{}=1,\notag\\ 
       & p\in\text{plaquettes}, \quad i=1,2.
    \end{align}
    The corresponding gauge field term is the plaquette operator of the link qubit.
    \begin{align}
        B_{p,i}=R_{p,i}(\{Z\})=\plaquetteop{}{Z_i}{}.
    \end{align}
    The second type is the tube-like product
\begin{align}
    \renewcommand{\labelstyle}{\textstyle}
    R_{t}(\{c\})=&
    \sum_{a,b,c,d=0,1}
\begin{smallgathered}
\xymatrix@=0.55cm{
&\ar@{-}[ld]\ar@{-}[d]\ar@{-}[rrr]\ar@{}[rrd]|{c_0}&&&\ar@{-}[ld]|{c_1^ac_2^{a+b}\hspace{0.5em}}\\
    \ar@{-}[rrr]|{[ab]_c}\ar@{-}[ddd]|{[cd]_c}\ar@{}[ddr]|{c_0}&&&\ar@{-}[ddd]\ar@{}[ddr]|{c_0}&\\
&&&&\\
&\ar@{-}[ld]|{\hspace{0.5em}c_1^cc_2^{c+d}}\ar@{-}[rr]\ar@{-}[uu]\ar@{}[rrd]|{c_0}&&&\ar@{-}[uuu]\ar@{-}[l]\\
    \ar@{-}[rrr]&&&\ar@{-}[ru]|{\hspace{3.5em}c_1^{a+c}c_2^{a+b+c+d}}
}
\end{smallgathered}\hspace{-2em}
=1, \notag \\ 
& t \in\text{tubes}.
\end{align}
The corresponding field term is
\begin{align}
    B_{t} &=R_{t}(\{Z\})\prod_{\text{six faces}}P_{p,1}P_{p,2} \notag \\
   & =
    \renewcommand{\labelstyle}{\textstyle}
    \sum_{\mathbf a,\mathbf b=00,01,10,11}
\begin{smallgathered}
\xymatrix@=0.55cm{
&\ar@{-}[ld]\ar@{-}[d]\ar@{-}[rrr]\ar@{}[rrd]|{Z}&&&\ar@{-}[ld]|{\mathbf Z^{\mathbf a}}\\
    \ar@{-}[rrr]|{[\mathbf a]}\ar@{-}[ddd]|{[\mathbf b]}\ar@{}[ddr]|{Z}&&&\ar@{-}[ddd]\ar@{}[ddr]|{Z}&\\
&&&&\\
&\ar@{-}[ld]|{\mathbf Z^{\mathbf b}}\ar@{-}[rr]\ar@{-}[uu]\ar@{}[rrd]|{Z}&&&\ar@{-}[uuu]\ar@{-}[l]\\
    \ar@{-}[rrr]&&&\ar@{-}[ru]|{\mathbf Z^{\mathbf a+\mathbf b}}
}
\end{smallgathered}
\cdot \prod_{\text{six faces}}P_{p,1}P_{p,2},
\end{align}

\item The gauged coupling terms are
        \begin{align}
             \fplaquetteop{c_0(\tau)}{}{}{}{}{}{}{}{}  
            =& \sum_{\substack{\mathbf a,\mathbf b,\mathbf c,\mathbf d,\mathbf e\\=00,01,10,11\\ \mathbf b+\mathbf c+\mathbf d+\mathbf e=00}}  
             (-1)^{a^{(1)} (a^{(2)} + b^{(1)} + b^{(2)} + d^{(1)} + d^{(2)}+1)}    \notag\\
             & \times  (-1)^{a^{(2)} (b^{(2)} + d^{(2)}+1)}
          \fplaquetteop{Z}{[\mathbf b]}{[\mathbf c]}{[\mathbf e]}{[\mathbf d]}{Z[\mathbf a]}{Z\mathbf Z^{\mathbf a + \mathbf b}}{Z\mathbf Z^{\mathbf a+\mathbf e}}{Z\mathbf Z^{\mathbf a+\mathbf b+\mathbf c}}, \notag\\
            \flinkop{c_1(\tau)}{}{} = &\flinkop{Z_1}{I{Z_1}}{I{Z_1}},\quad
            \flinkop{c_2(\tau)}{}{}=\flinkop{Z_2}{I{Z_2}}{I{Z_2}}.
        \end{align}

\item the electric field terms are
\begin{equation}
    H_e=-g_0\sum_{p \text{ plaquettes}}\fplaquetteop{X}{}{}{}{}{}{}{}{}
    -g_1\sum_\text{links}\flinkop{\tilde{X}^{(i)}}{}{}
    -g_2\sum_\text{links}\flinkop{\tilde{X}^{(j)}}{}{},
\end{equation}
where
\vspace{-2em}
\begin{equation}
   \renewcommand{\labelstyle}{\textstyle}
\begin{smallgathered}
\xymatrix@!0{
    &\ar@{-}[ld]|{\tilde{X}^{(i)}}\\
    &&
}
\hspace{-3em}
\end{smallgathered}
:=\sum_{\substack{a,\ldots,g\\=0,1 \\ \alpha,\beta, \gamma = 0,1 }}
   \renewcommand{\labelstyle}{\textstyle}
\begin{smallgathered}
\xymatrix@!0{
&&&&\ar@{-}[ld]\ar@{}[ldddd]|{X^c}&&&\\
&&&\ar@{-}[ddd]&&&&\\
&&&&&&&\\
    &\ar@{-}[dl]\ar@{}[drr]|{X^{d+e}}&&\ar@{-}[ll]& {I [\alpha c]}\ar@{-}[l]\ar@{-}[uuu]\ar@{-}[ld]|{X_2 }\ar@{-}[rrr]|{[\beta a]}\ar@{-}[d]\ar@{}[drr]|{X^{a+f+g}}&&&\ar@{-}[ld]\\
{I [de] }\ar@{-}[rrr]&&& {I [fg] }\ar@{-}[rrr]\ar@{-}[ddd]\ar@{}[ddr]|{X^b}&\ar@{-}[dd]|(.25){[\gamma b]}&&&\\
&&&&&&&\\
&&&&\ar@{-}[ld]&&&\\
&&&&&&&
}
\end{smallgathered}, \notag
\end{equation}
\vspace{-2em}
\begin{equation}
   \renewcommand{\labelstyle}{\textstyle}
\begin{smallgathered}
\xymatrix@!0{
    &\ar@{-}[ld]|{\tilde{X}^{(j)}}\\
    &&
}
\hspace{-3em}
\end{smallgathered}
:=\sum_{\substack{a,\ldots,h\\=0,1 \\ \alpha, \beta = 0, 1}}
   \renewcommand{\labelstyle}{\textstyle}
\begin{smallgathered}
\xymatrix@!0{
&&&&\ar@{-}[ld]\ar@{}[ldddd]|{X^{e+f}}&&&\\
&&&\ar@{-}[ddd]&&&&\\
&&&&&&&\\
&\ar@{-}[dl]\ar@{}[drr]|{X^{g}}&&\ar@{-}[ll]&{I [ef] }\ar@{-}[l]\ar@{-}[uuu]\ar@{-}[ld]|{X_1}\ar@{-}[rrr]|{[ab]}\ar@{-}[d]\ar@{}[drr]|{X^{a+b+h}}&&&\ar@{-}[ld]\\
    {I [g \alpha] }\ar@{-}[rrr]&&&{I [h \beta] }\ar@{-}[rrr]\ar@{-}[ddd]\ar@{}[ddr]|{X^{c+d}}&\ar@{-}[dd]|(.25){[cd]}&&&\\
&&&&&&&\\
&&&&\ar@{-}[ld]&&&\\
&&&&&&&
}
\end{smallgathered}
\end{equation}
\vspace{-2em}
\end{enumerate}

The decorated $\tilde{X}^{(i)}$ and $\tilde{X}^{(j)}$ operators ensure the electric field is invariant under the gauge transformation $A_{v,g}$.
Similar to the toric code and the X-Cube code, we can choose the unitary gauge to eliminate the matter DOF in this example. We regard the ``physical state'' satisfying the constraints as an equivalent class of computational basis state, and choose the representative such that all matter DOF are in the reference state $|000\rangle$.

In this case, the Hamiltonian becomes
\begin{align}
\label{eq:Q8exact}
H_\text{u} 
=  & -\sum_{v\text{ sites}}\left(A_{v,g^{(i)}}^\text{u}+A_{v,g^{(j)}}^\text{u} +A_{v,g^{(-)}_P}^\text{u}\right)\notag\\
    &-\sum_{t\text{ tubes}}B_{t}-\sum_{p\text{ plaquettes}}\left(B_{p,1}+B_{p,2}\right)\notag\\
    &-J_0\sum_{p \text{ plaquettes}}\fplaquetteop{Z}{}{}{}{}{}{}{}{}\cdot P_{p,1}P_{p,2}
                    -\sum_\text{links}\left(J_1\flinkop{ Z_1}{}{}+J_2\flinkop{ Z_2}{}{}\right)\notag\\
 &-g_0\sum_{p \text{ plaquettes}}\fplaquetteop{X}{}{}{}{}{}{}{}{}
   -g_1\sum_\text{links}\flinkop{\tilde{X}_\text{u}^{(i)}}{}{}
 -g_2\sum_\text{links}\flinkop{\tilde{X}_\text{u}^{(j)}}{}{},
\end{align}
where the dressed operators in the unitary gauge are~\footnote{We can also choose a different set of dressed operators such that they simply become a single $IX$ and a $XI$ on the link, as in the cases of $S_3$ and $D_4$. Here our choice of the dressed operators is to be in consistent with the membrane operators for the string excitations. }
\begin{align}
&\renewcommand{\labelstyle}{\textstyle}
\begin{smallgathered}
\xymatrix@=1.5 em{
&\ar@{-}[ld]|{\tilde X_\text{u}^{(i)}}\ar@{..}[rrr]\ar@{..}[d]&&&\ar@{..}[ld]\\
    \ar@{..}[rrr]\ar@{..}[ddd]&\ar@{..}[dd]&&&\\
                &&&&\\
                &\ar@{..}[ld]&&&\\
                &&&&
}
\end{smallgathered}
    =\sum_{a,b,\alpha,\beta=0,1}
   \renewcommand{\labelstyle}{\textstyle}
\begin{smallgathered}
\xymatrix@=1.5 em{
&\ar@{-}[ld]|{X_2}\ar@{-}[rrr]|{[\alpha a]}\ar@{-}[d]\ar@{}[drr]|{X^a}&&&\ar@{-}[ld]\\
    \ar@{-}[rrr]\ar@{-}[ddd]\ar@{}[ddr]|{X^b}&\ar@{-}[dd]|(.25){[\beta b]}&&&\\
                &&&&\\
                &\ar@{-}[ld]&&&\\
                &&&&
}
\end{smallgathered}, \notag \\ 
&\renewcommand{\labelstyle}{\textstyle}
\begin{smallgathered}
\xymatrix@=1.5 em{
&\ar@{-}[ld]|{\tilde X_\text{u}^{(j)}}\ar@{..}[rrr]\ar@{..}[d]&&&\ar@{..}[ld]\\
    \ar@{..}[rrr]\ar@{..}[ddd]&\ar@{..}[dd]&&&\\
                &&&&\\
                &\ar@{..}[ld]&&&\\
                &&&&
}
\end{smallgathered} 
=\sum_{\substack{a,b,c,d=0,1}}
   \renewcommand{\labelstyle}{\textstyle}
\begin{smallgathered}
\xymatrix@=1.5 em{
&\ar@{-}[ld]|{X_1}\ar@{-}[rrr]|{[ab]}\ar@{-}[d]\ar@{}[drr]|{X^{a+b}}&&&\ar@{-}[ld]\\
    \ar@{-}[rrr]\ar@{-}[ddd]\ar@{}[ddr]|{X^{c+d}}&\ar@{-}[dd]|(.25){[cd]}&&&\\
                &&&&\\
                &\ar@{-}[ld]&&&\\
                &&&&
}
\end{smallgathered} ,
\end{align}
and $\tilde{X}_\text{u}^{(k)}= \tilde{X}_\text{u}^{(i)} \tilde{X}_\text{u}^{(j)}$.
The Hilbert space contains only the gauge DOF without constraints.
The above Hamiltonian $H_\text{u}$ describes the pure lattice gauge theory of $G$.
In the exactly solvable limit ($J_0=J_1=J_2=g_0=g_1=g_2$), the system is deeply
in the deconfined phase and the fundamental excitations are fully mobile particles, fractons, and strings. 
These excitations are electric charges which are the excitations of $A^\text{u}_{v, g^{(i)}}$, $A^\text{u}_{v, g^{(j)}}$, $A^\text{u}_{v, g_p^{(-)}}$
and magnetic fluxes which are the excitations of $B_t$, $B_{p,1}$, $B_{p,2}$. We discuss these excitations in the following. 

\subsubsection{Electric change excitations}
\begin{itemize}

    \item $\phi^{(q)}, q=i,j,k$: An Abelian quasiparticle that can move freely in 3D, corresponding to $A_{v,g^{(q')}}=-1,q'\neq q$. For $q=i,j,k$, it is created at the end point of the string operator
        \begin{equation}
            \renewcommand{\labelstyle}{\textstyle}
            \begin{gathered}
            \xymatrix@=4em{
                \ar@{-}[r]|{Z_1} & \ar@{-}[r]|{Z_1} & \ar@{}[r]|{\cdots} & \ar@{-}[r]|{Z_1} &
            }
            \end{gathered}
            , \notag
        \end{equation}
        \begin{equation}
            \renewcommand{\labelstyle}{\textstyle}
            \begin{gathered}
            \xymatrix@=4em{
                \ar@{-}[r]|{Z_2} & \ar@{-}[r]|{Z_2} & \ar@{}[r]|{\cdots} & \ar@{-}[r]|{Z_2} &
            }
            \end{gathered}
            ,   \notag
        \end{equation}
        \begin{equation}
            \renewcommand{\labelstyle}{\textstyle}
            \begin{gathered}
            \xymatrix@=4em{
                \ar@{-}[r]|{Z_1Z_2} & \ar@{-}[r]|{Z_1Z_2 } & \ar@{}[r]|{\cdots} & \ar@{-}[r]|{Z_1Z_2} &
            }
            \end{gathered}
            ,
        \end{equation}
        respectively.
    \item $[f_0]$: The non-Abelian fracton, created at the four corners of the membrane operator
        \begin{align}
            &M_{v,\mathbf b_0}^{(m,n)}  \notag\\
            =&
                \sum_{\substack{\{\mathbf b_k\},\\ \{\mathbf c_k\}\\=00,01,\\10,11}}   
            \renewcommand{\labelstyle}{\textstyle}
            \begin{smallgathered}
            \xymatrix@=4em@!0{
                \ar@{-}[r]|{}\ar@{-}[d]|{\txt{$[\mathbf b_1]$\\$\mathbf Z^{\mathbf b_0}$}}\ar@{}[dr]|{\tilde Z} & \ar@{-}[r]|{\cdots}\ar@{}[dr]|{\cdots}\ar@{-}[d] & \ar@{-}[r]|{}\ar@{-}[d]\ar@{}[dr]|{\tilde Z} & \ar@{-}[d]\ar@{-}[r]|{}\ar@{}[dr]|{\tilde Z} & \ar@{-}[d]|{\txt{$[\mathbf c_1]$}} \\
                \ar@{-}[r]|{}\ar@{}[dr]|{\vdots}\ar@{-}[d]|{\vdots} & \ar@{-}[r]|{\cdots}\ar@{}[dr]|{\ddots}\ar@{-}[d]|{\vdots} & \ar@{-}[r]|{}\ar@{}[dr]|{\vdots}\ar@{-}[d]|{\vdots} & \ar@{-}[r]|{}\ar@{}[dr]|{\vdots}\ar@{-}[d]|{\vdots} & \ar@{-}[d]|{\vdots} \\
                \ar@{-}[r]|{}\ar@{-}[d]|{\txt{$[\mathbf b_{m-1}]$\\$\mathbf Z^{\mathbf b_0+\cdots+\mathbf b_{m-2}}$}\hspace{3em}}\ar@{}[dr]|{\tilde Z} & \ar@{-}[r]|{\cdots}\ar@{-}[d]\ar@{}[dr]|{\cdots} & \ar@{-}[r]|{}\ar@{-}[d]\ar@{}[dr]|{\tilde Z} & \ar@{-}[r]|{}\ar@{-}[d]\ar@{}[dr]|{\tilde Z} & \ar@{-}[d]|{\hspace{3em}\txt{$[\mathbf c_{m-1}]$\\$\mathbf Z^{\mathbf c_1+\cdots+\mathbf c_{m-2}}$}} \\
                \ar@{-}[r]|{}\ar@{-}[d]|{\mathbf Z^{\mathbf b_0+\cdots+\mathbf b_{m-1}}\hspace{3em}}\ar@{}[dr]|{\tilde Z} & \ar@{-}[r]|{\cdots}\ar@{-}[d]\ar@{}[dr]|{\cdots} & \ar@{-}[r]|{}\ar@{-}[d]\ar@{}[dr]|{\tilde Z} & \ar@{-}[r]|{}\ar@{-}[d]\ar@{}[dr]|{\tilde Z} & \ar@{-}[d]|{\hspace{3em}\txt{$\mathbf Z^{\mathbf c_1+\cdots+\mathbf c_{m-1}}$}}\\
                \ar@{-}[r] & & \ar@{-}[r] & \ar@{-}[r] &
            }
        \end{smallgathered}\hspace{-2em},
        \end{align}
        where $v$ is the upper left corner
        \begin{equation}
            \fplaquetteop{\tilde Z}{}{}{}{}{}{}{}{}=\sum_{\mathbf{a}=00,01,10,11}\fplaquetteop{Z}{[\mathbf{a}]}{\mathbf Z^{\mathbf a}}{}{}{}{}{}{}
        \end{equation}
        and addition of boldface letters are componentwise and modulo 2. 
        The membranes for different choices of $\mathbf b_0$ differ by an additional string for some $\phi^{(q)}$ along the left edge of the membrane.
      Note that
      \begin{align}
          A_{v,g^{(i)}}^\text{u}M_{v,00}^{(m,n)}|GS\rangle&=-M_{v,01}^{(m,n)}|GS\rangle, \notag \\
          A_{v,g^{(j)}}^\text{u}M_{v,00}^{(m,n)}|GS\rangle&=-M_{v,10}^{(m,n)}|GS\rangle,\notag \\
          A_{v,g^{(k)}}^\text{u}M_{v,00}^{(m,n)}|GS\rangle&=M_{v,11}^{(m,n)}|GS\rangle.
      \end{align}
        This means that the fracton created by $M_{v,\mathbf b_0}^{(m,n)}, \mathbf b_0=00,01,10,11$ belongs to the same superselection sector, so is the same species $[f_0]$.
        Also note that this membrane operator only work when acting on a state with $B_{p,i}=+1$ for all plaquette $p$ on and near (at most one lattice spacing) the membrane.
        To show that it is non-Abelian, we consider the following two states
        \begin{align}
           |\psi_1\rangle&=M_{v,00}^{(L,L)}M_{v,00}^{(2L,2L)} M_{v,00}^{(3L,3L)}|GS\rangle  \notag \\
           |\psi_2\rangle&=M_{v,00}^{(L,L)}M_{v,00}^{(2L,2L)} M_{v,01}^{(3L,3L)}|GS\rangle,
        \end{align}
        where $L$ is large.
        The excitation patterns of the two states are the same as the first example in Eq.~(\ref{eq:f0}).  
        Now, for the state  $|\psi_1\rangle$, the four $[f_0]$'s on the upper left square of Eq.~(\ref{eq:f0}) can be fused to the vacuum by applying an additional $M_{v,00}^{(L,L)}$.
        On the other hand, they cannot be fused to the vacuum for $|\psi_2\rangle$.
        This means that $[f_0]$ has non-trivial fusion rules, and therefore is non-Abelian.
        If we continue this pattern to apply $n$ $M_{v,i}$'s to the ground state, it can be shown that there are asymptotically $2^{n}$ fusion channels, which implies the quantum dimension of $[f_0]$ is $2$.
\end{itemize}
 The excitations of $A_{v,g^{(i)}}^\text{u}$, $A_{v,g^{(j)}}^\text{u}$, and $A_{v,g^{(-)}_P}^\text{u}$
 are referred to the electric charge excitations. These excitations are local with respect
to the vertex $v$,
and can be specified from the local operators which form a representation of $Q_8$ on the Hilbert space.
Hence we can identify electric charges of the $G$ fracton model
with the irreducible representations of $G^\text{local} \cong Q_8$ (left panel of Table~\ref{tab:q8}). 
This identification is the same as the quantum double model (QDM) with $Q_8$ symmetry~\cite{A.Yu.Kitaev2002}.

\subsubsection{The magnetic flux excitations}
\begin{itemize}
    \item $e_d$, the Abelian lineon, corresponding to $B_t=-1$. It is created at the endpoints of the string
\begin{equation}
    \renewcommand{\labelstyle}{\textstyle}
\begin{smallgathered}
\xymatrix@=0.5 em{
&\ar@{-}[ld]\ar@{}[rrd]&&&\ar@{-}[ld]\ar@{}[ddddrr]|{\cdots}\ar@{}[rrd]&&&\ar@{-}[ld]\ar@{}[rrd]&&&\ar@{-}[ld]\\
    \ar@{-}[ddd]\ar@{}[ddr]|{X}&&&\ar@{}[rrr]\ar@{-}[ddd]\ar@{}[ddr]|{X}&&&\ar@{-}[ddd]\ar@{}[ddr]|{X}&&&\ar@{-}[ddd]\ar@{}[ddr]|{X}&\\
&&&&&&&&&&\\
&\ar@{-}[ld]\ar@{-}[uuu]\ar@{}[rrd]&&&\ar@{-}[ld]\ar@{-}[uuu]\ar@{}[rrd]&&&\ar@{-}[ld]\ar@{-}[uuu]\ar@{}[rrd]&&&\ar@{-}[uuu]\\
&&&&&&&&&\ar@{-}[ru]
}
\end{smallgathered},
\end{equation}

    \item $\sigma^{(q)},q=i,j,k$: The flexible string-like excitation corresponding to the excitation $(B_{p,1},B_{p,2})=(-1,1),(1,-1),(-1,-1)$, respectively. 
 \begin{equation}
   \renewcommand{\labelstyle}{\textstyle}
\begin{smallgathered}
\xymatrix@=1.7em{
&\ar@{-}[ld]|{\tilde X_\text{u}^{(q)}}&&&\ar@{-}[ld]|{\tilde X_\text{u}^{(q)}}\ar@{}[ddddrr]|{\cdots}&&&\ar@{-}[ld]|{\tilde X_\text{u}^{(q)}}\\
&&&&&&&\\
&&&&&&&\\
&\ar@{-}[ld]|{\tilde X_\text{u}^{(q)}}\ar@{}[ddddrr]|{\vdots}&&&\ar@{-}[ld]|{\tilde X_\text{u}^{(q)}}\ar@{}[ddddrr]|{\ddots}&&&\ar@{-}[ld]|{\tilde X_\text{u}^{(q)}}\\
&&&&&&&\\
&&&&&&&\\
&\ar@{-}[ld]|{\tilde X_\text{u}^{(q)}}&&&\ar@{-}[ld]|{\tilde X_\text{u}^{(q)}}&&&\ar@{-}[ld]|{\tilde X_\text{u}^{(q)}}\\
&&&&&&&\\
}
\end{smallgathered}
\end{equation}
where $\tilde X_\text{u}^{(i)}$ and $\tilde X_\text{u}^{(j)}$ are defined before and
\begin{equation}
    \tilde X_\text{u}^{(k)}=\tilde X_\text{u}^{(i)}\tilde X_\text{u}^{(j)}
\end{equation}

\end{itemize}

\begin{table*}[t]
\centering
 \begin{tabular}{c c c || c c c c c c} 
 \hline
 Charge & Irrep of $G^\text{local}\cong Q_8$ & Type & Flux & $B_{p_0,1}$ & $B_{p_0,2}$ & $B_{t_0}$ & Conj.\ class  & Type \\ 
 \hline\hline
 Vacuum & Trivial & Vacuum &  Vacuum & $1$ & $1$ & $1$ & \{1\} & Vacuum  \\ 
 $\phi^{(i)}$ & $(i,j)\mapsto(1,-1)$ & Abelian fully mobile particle &  $\sigma^{(i)}$ & $-1$ & $1$ & 0 & $\left\{i,-i\right\}$ & Flexible string \\
 $\phi^{(j)}$ & $(i,j)\mapsto(-1,1)$ & Abelian fully mobile particle &  $\sigma^{(j)}$ & $1$ & $-1$ & 0 & $\left\{j,-j\right\}$ & Flexible string \\
 $\phi^{(k)}$ & $(i,j)\mapsto(-1,-1)$ & Abelian fully mobile particle & $\sigma^{(k)}$ & $-1$ & $-1$ & 0 & $\left\{k,-k\right\}$ & Flexible string \\
 $[f_0]$ & 2D representation & Non-Abelian fracton &  $e_d$ & $1$ & $1$ & $-1$ & $\left\{-1\right\}$ & Abelian lineon \\
 \hline
 \end{tabular}
 \caption{Pure electric charges of the gauged $1\to\mathbb{Z}_2^\text{sub}\to G\to K_4^\text{glo}\to 1$ model.}
\label{tab:q8}
\end{table*}

In the same spirit as the QDM, the magnetic fluxes would be the conjugacy classes of $G^\text{local}\cong Q_8$.
 With the geometry defined in Eq. (\ref{eq:geometry}) and the constraints defined in Eq. (\ref{eq:constraint}), 
the flux operators can be mapped to a subalgebra of the QDM algebra for either $P_\text{side}=1$ or $P_\text{corner}=1$,
\begin{align}
    B_{p_0,1}&\mapsto B_{1}+B_{i}-B_{j}-B_{k}+B_{-1}+B_{-i}-B_{-j}-B_{-k},\notag \\
    B_{p_0,2}&\mapsto B_{1}-B_{i}+B_{j}-B_{k}+B_{-1}-B_{-i}+B_{-j}-B_{-k}, \notag \\
    B_{t_0}& \mapsto B_{1}-B_{-1},
\end{align}
where $B_g$ is the QDM flux operator in Eq.~(\ref{eq:qdm}).
We can also identify the star operators $A_{v_0,g}^\text{u}$ with the QDM star operator $A_g$ Eq.~(\ref{eq:qdm}).
The map of $A$'s and $B$'s together forms an injective algebra homomorphism into the QDM algebra. In particular, they satisfy the relations (note that all the $B$'s commute)
\begin{align}
    &B_{p_0,1}^2=B_{p_0,2}^2=1, \notag\\ 
    &B_{p_0,1}B_{t_0}=B_{p_0,2}B_{t_0}=B_{p_0,1}B_{p_0,2}B_{t_0}=B_{t_0},  \notag\\ 
    &B_{t_0}^2=\frac{1}{4}\left(1+B_{p_0,1}+B_{p_0,2}+B_{p_0,1}B_{p_0,2}\right),  \notag\\ 
    &A_{g}B_{p_0,i}=B_{p_0,i}A_{g},\,A_{g}B_{t_0}=B_{t_0}A_{g},\,\forall g\in Q_8,i=1,2
\end{align}
In this way, the star and the flux operator of this fracton models is identified with a subalgebra of that of the corresponding QDM.

Note that this subalgebra is enough to distinguish all of the conjugacy classes of $Q_8$ (more rigorously, if a state is the eigenstate of $B_{g_0}=1,B_{g}=0,g\neq g_0$ for some $g_0\in Q_8$, then we can determine the conjugacy class of $g_0$ only by the eigenvalues of the image of $B_{p_0}$ and $B_{t_0,i}$.)
This allows us to identify the fluxes of this fracton model with the conjugacy classes of $Q_8$.%
The corresponding fluxes are listed in the right panel of Table \ref{tab:q8}.

\section{Conclusion}
\label{sec:con}
In this paper, we demonstrate a systematical gauging procedure on a lattice with pure matter fields. 
This general procedure works not only for a semidirect product of the global symmetry and the subsystem symmetry,
but also for a non-trivial extension of them. 
We give four examples of gauging 
$G=\mathbb{Z}_3^{\text{sub}}\rtimes \mathbb{Z}_2^{\text{glo}}$,
$G=(\mathbb{Z}_2^{\text{sub}}\times \mathbb{Z}_2^{\text{sub}})\rtimes \mathbb{Z}_2^{\text{glo}}$,
$1\to \mathbb {Z}_2^\text {sub}\to G\to \mathbb {Z}_2^\text {glo}\to 1$, and
$1\to \mathbb {Z}_2^\text {sub}\to G\to K_4^\text {glo}\to 1$. The former two cases and the last one produce the non-Abelian fracton orders. 
By using a one-step gauging, we give a transparent identification of electric charges with the irreducible representations of $G^\text{local}$, which include the non-Abelian fracton orders. Furthermore, to compare the magnetic excitations with different geometry (tubes and plaquettes), we set a specific constraint on the
local Hilbert space. We observe that the magnetic excitations satisfy the subalgebra of the QDMs, which allows us to identify the 
magnetic excitations as the conjugacy classes of  $G^\text{local}$. Our gauging procedure is very general, and can be easily extended to more exotic symmetries, and can produce different types of non-Abelian version of fractons.

Before we close the discussion, we would like to point out some future directions.
\begin{enumerate}
    \item If one applies the symmetry of ``(1D subsystem) $\rtimes$ (2D subsystem) $\rtimes$ global'' or ``fractal $\rtimes$ global'' form, we expect the gauged Hamiltonian in the exactly-solvable limit can produce new types of non-Abelian fractons.

\item The fully gauged Hamiltonian contains matter and gauge fields, which may contain partial confined-deconfined or Higgsed-deconfined transition for some of the excitations. One can also consider other types of matter fields such as the Majorana fermion.

\item The quotient superselection sector (QSS)~\cite{SHIRLEY2019167922} is a method to identify fracton species. We would like to understand the deep relation between a non-Abelian generalization of QSS and the corresponding electric charges with the irreps of $G^{\text{local}}$.


\item Although we associate the magnetic fluxes with the conjugacy class of $G^\text{local}$ in the constrained local Hilbert space from
 the subalgebra of the QDMs, some magnetic fluxes (e.g., $B_{t_0,2}$ in the second example) cannot be obtained from our
 gauging procedure. A direct
identification of the fluxes with structures of $G$ and $G^\text{local}$ from the general gauging process is still desired.

\item The ungauging map~\cite{aleks2018ungauging} is the map from the zero-flux subspace of a pure lattice gauge system back to the matter Hilbert space with the symmetry. We expect that it can be generalized to non-Abelian symmetries.

\end{enumerate}

{\it Note added.} 
While preparing the update of this paper, we recently learnt of a related work by
Tantivasadakarn, Ji, and Vijay~\cite{Tantivasadakarn2021}, which they construct a 
solvable lattice model labelled as $(Q_8, \mathbb{Z}_2)$ in their terminology, which is our fourth example.

\section*{Acknowledgements}
We thank Yu-An Chen and Sheng-Jie Huang for insightful discussions.
The work was supported by the Young Scholar Fellowship Program under MOST Grant for the Einstein Program MOST 110-2636-M-007 -007.
We also acknowledge support from the NCTS.

\begin{appendix}
\section{Systematical construction of the gauge transformation}\label{sec:a}

We assume that each site has the same local Hilbert space with a chosen computational basis. We define a reference state $|0\rangle$ for this computational basis. A tensor product of computational basis states will also be referred to as a computational basis state. 
Furthermore, the computational basis states are chosen to be eigenstates of the local operators in the non-local coupling $c$. For example,
in the transverse-field Ising model, the local operator in
the non-local coupling $Z_{i+1}Z_i$ is the Pauli matrix $Z$. The computational basis $|0\rangle$ and $|1 \rangle$ are the eigenstates of $Z$ with corresponding eigenvalues $+1$ and $-1$, respectively.
The gauge transformation $A_{v,g}$ acts on the site $v$ with the matter and gauge fields associated with $v$. $A_{v,g}$ acts on the matter field on the site $v$ as the original symmetry transformation $g\in G$. Matter fields on other sites remain the same.  
$A_{v,g}$ acting on the gauge field needs to compensate the change of the $A_{v,g}$ acts on the matter field. 
This compensation then defines the gauge transformation $A_{v,g}$ acting on the gauge fields,
$A_{v,g}\bigotimes_{c}|\tau_c \rangle = \bigotimes_{c}|\tau_c'\rangle$.  Together with matter field, the gauge transformation is 
           \begin{multline}
A_{v,g}\left(\bigotimes_{s \neq v}|m_{s}\rangle\otimes|m_v\rangle\otimes\bigotimes_{c}|\tau_c\rangle\right)\\=\bigotimes_{s\neq v}|m_{s}\rangle\otimes g|m_{v}\rangle\otimes\bigotimes_{c\in C_{v}}|\tau'_c\rangle\otimes\bigotimes_{c\notin C_{v}}|\tau_c\rangle.
\end{multline}
Here $|m_{s(v)}\rangle$ are the state of matter fields, and $C_v$ being a subset of the couplings associated with the site $v$ (we will state the requirements for $C_v$ later). 
We now demonstrate a systematical way to find $|\tau_c'\rangle$ after the gauge transformation, which defines $A_{v,g}$ acting on $|\tau_c\rangle$.
 \begin{enumerate}
            \item\label{step:a} 
            We construct a virtual state of matter fields $|\tilde{m}\rangle =\bigotimes_{s \neq v}|\tilde{m}_{s}\rangle\otimes|\tilde{m}_v\rangle$ associated with $|\tau_c \rangle$ by the coupling $c$. This virtual state of matter fields $| \tilde{m}\rangle$
            is chosen to satisfy
            \begin{equation}\label{Eq.36}
    |\tilde m_{v}\rangle=|0\rangle,
\quad
    c |\tilde m\rangle=\tau_c |\tilde m\rangle,\,\forall c\in C_{v}.
\end{equation}
            That means for a state of gauge field $|\tau_c\rangle$, we can find a corresponding virtual matter state $|\tilde{m}_v\rangle$ that satisfies Eq. (\ref{Eq.36}).
            \item 
            $|\tau_c' \rangle$ associates with another virtual state $| \tilde{m}'\rangle$. Since $|\tau_c' \rangle$ is the compensation of the changes of $A_{v,g}$ acting on the matter field, the virtual state $|\tilde{m}'\rangle$ is obtained by
            applying $g^{-1}$ to the state $|\tilde{m}\rangle$ on $v$,
            $|\tilde{m}'\rangle = \bigotimes_{s \neq v} |m_s\rangle \otimes g^{-1} |\tilde{m}_v\rangle$.
            \item\label{step:c} 
            The eigenvalue of $c$ for the virtual state $|\tilde{m}'_c\rangle$, $c |\tilde{m}'_c\rangle = \tau_c' |\tilde{m}'_c\rangle$
            then specifies the state $|\tau_c' \rangle$. This step determines the gauge transformation
            $A_{v,g}$ on $|\tau_c\rangle$.
            It is required that $|\tilde m\rangle$ and $|\tilde m'\rangle $ are eigenstates of $c$  with the same eigenvalue, $\forall c\notin C_v$ (otherwise one should choose a larger $C_v$).
        \end{enumerate}

The above definition automatically implies $A_{v,g_1}A_{v,g_2}=A_{v,g_1g_2}$, and $C_v$ is chosen to be the minimal set such that $A_{v,g}$ is well-defined for all $g$ and that $[A_{v_1,g_1},A_{v_2,g_2}]=0$ when $v_1\neq v_2$. This means that $G$ is promoted to a local symmetry of the gauged system.
In the cases of gauging pure global or pure subsystem symmetries, we simply have $C_v=\{c\mid v\in \supp c\}$. However, when global and subsystem symmetries are mixed together, the choice of $C_v$ becomes nontrivial. See the example below.

\subsection{Specific example for obtaining \texorpdfstring{$A_{v,g}$}{A\_vg} for \texorpdfstring{$(\mathbb{Z}_2^\text{sub}\times\mathbb{Z}_2^\text{sub})\rtimes\mathbb{Z}_2^\text{glo}$}{(Z\_2subxZ\_2sub) rx Z\_2glo}}\label{sec:details}
 The reference state is $|000\rangle$.
 At each vertex $v$ [the center vertex in Eq.~(\ref{eq:support_nonabelian})], the support $C_v$ of the gauge transformation is chosen to contain twelve $c_0$ and $c_1$ couplings on the plaquettes,
         and nine $c_2$ couplings on the links,
        
\begin{equation}\label{eq:support_nonabelian}
    C_v=\left\{
    \renewcommand{\labelstyle}{\textstyle}
\begin{smallgathered}
\xymatrix@=0.7cm{
    &&&&&\ar@{-}[d]&&&\\
    &\ar@{-}[rrr]|{\color{blue}c_2,}&&&\ar@{-}[ru]\ar@{-}[ld]&\ar@{-}[dd]&&\ar@{-}[lll]&\\
    &&&\ar@{-}[ddd]|(.55){\color{orange}c_2,}&&&&&\\
    &&\ar@{-}@[green!50!black][ld]&\ar@{-}@[green!50!black][l]&\ar@{-}@[green!50!black][l]&\ar@{-}@[green!50!black][l]\ar@{-}@[green!50!black][rr]&&\ar@{-}@[green!50!black][r]&\ar@{-}@[green!50!black][ld]\\
    &\ar@{-}[uuu]\ar@{-}[d]&&\ar@{-}[ll]|(.25){c_2,}&\ar@{-}[uuu]|{c_2,}\ar@{-}[rrr]|{c_2,}\ar@{-}[d]\ar@{-}[l]\ar@{-}[ru]|{c_2,}\ar@{-}[ld]|{c_2,}\ar@{}[llluuu]|(.6){\color{blue}c_0,c_1,}\ar@{}[luu]|(.68){\color{orange}c_0,c_1,}\ar@{}[ruuuu]|(.5){\color{orange}c_0,c_1,}\ar@{}[rrruuu]|(.6){\color{blue}c_0,c_1,}\ar@{}[llu]|(.7){\color{green!50!black}c_0,c_1,}\ar@{}[rrrru]|(.5){\color{green!50!black}c_0,c_1,}\ar@{}[rrrddd]|(.6){\color{blue}c_0,c_1,}\ar@{}[rdd]|(.65){\color{orange}c_0,c_1,}\ar@{}[ldddd]|(.5){\color{orange}c_0,c_1,}\ar@{}[lllddd]|(.6){\color{blue}c_0,c_1,}\ar@{}[rrd]|(.7){\color{green!50!black}c_0,c_1,}\ar@{}[lllld]|(.5){\color{green!50!black}c_0,c_1,}&\ar@{-}[u]&&\ar@{-}[uuu]\ar@{-}[ddd]&\\
    \ar@{-}@[green!50!black][ru]|{\color{green!50!black}c_2,}&\ar@{-}[dd]&&\ar@{-}@[green!50!black][lll]\ar@{-}@[green!50!black][rrr]&\ar@{-}[dd]|(.25){c_2,}&\ar@{-}[u]&\ar@{-}@[green!50!black][ru]&&\\
                       &&&&&\ar@{-}[u]&&&\\
                       &\ar@{-}[rr]&&\ar@{-}[r]&\ar@{-}[ru]\ar@{-}[ld]&&&\ar@{-}[lll]&\\
                       &&&\ar@{-}[uuu]&&&&&
}
\end{smallgathered}
\right\}.
\end{equation}

Here we demonstrate the calculation of $A_{v,g_P^{(0)}}$, where $P$ is any of the three shifted coordinate planes containing $v$. Since $g_P^{(0)}$ acts on $v$ by the operator $X_0$, $A_{v,g_P^{(0)}}$ also acts on $v$ by $X_0$. Also, by definition, $A_{v,g_P^{(0)}}$ acts on other sites trivially.
Now we demonstrate how to calculate its action on the gauge DOF.
Suppose we want to calculate $A_{v,g_P^{(0)}}$ acting on a computation basis state in which the part of the state around $v$ is
\begin{equation}\label{eq:demo_state}
\bigotimes_{c\in C_{v}}|\tau_c\rangle=
    \renewcommand{\labelstyle}{\textstyle}
\begin{smallgathered}
\xymatrix@=0.5cm{
    &&&&&\color{orange}\ar@{-}[d]&&&\\
    &\color{blue}\ar@{-}[rrr]|{\color{blue}|0\rangle}&&&\ar@{-}[ru]\ar@{-}[ld]&\ar@{-}[dd]&&\color{blue}\ar@{-}[lll]&\\
    &&&\color{orange}\ar@{-}[ddd]|(.55){\color{orange}|1\rangle}&&&&&\\
    &&\color{green!50!black}\ar@{-}@[green!50!black][ld]&\ar@{-}@[green!50!black][l]&\ar@{-}@[green!50!black][l]&\ar@{-}@[green!50!black][l]\ar@{-}@[green!50!black][rr]&&\ar@{-}@[green!50!black][r]&\color{green!50!black}\ar@{-}@[green!50!black][ld]\\
    &\ar@{-}[uuu]\ar@{-}[d]&&\ar@{-}[ll]|(.25){|0\rangle}&\ar@{-}[uuu]|{|1\rangle}\ar@{-}[rrr]|{|0\rangle}\ar@{-}[d]\ar@{-}[l]\ar@{-}[ru]|{|1\rangle}\ar@{-}[ld]|{|1\rangle}\ar@{}[llluuu]|(.6){\color{blue}|01\rangle}\ar@{}[luu]|(.68){\color{orange}|01\rangle}\ar@{}[ruuuu]|(.5){\color{orange}|10\rangle}\ar@{}[rrruuu]|(.6){\color{blue}|11\rangle}\ar@{}[llu]|(.7){\color{green!50!black}|00\rangle}\ar@{}[rrrru]|(.5){\color{green!50!black}|01\rangle}\ar@{}[rrrddd]|(.6){\color{blue}|10\rangle}\ar@{}[rdd]|(.65){\color{orange}|11\rangle}\ar@{}[ldddd]|(.5){\color{orange}|00\rangle}\ar@{}[lllddd]|(.6){\color{blue}|01\rangle}\ar@{}[rrd]|(.7){\color{green!50!black}|10\rangle}\ar@{}[lllld]|(.5){\color{green!50!black}|11\rangle}&\ar@{-}[u]&&\ar@{-}[uuu]\ar@{-}[ddd]&\\
    \color{green!50!black}\ar@{-}@[green!50!black][ru]|{\color{green!50!black}|1\rangle}&\ar@{-}[dd]&&\ar@{-}@[green!50!black][lll]\ar@{-}@[green!50!black][rrr]&\ar@{-}[dd]|(.25){|0\rangle}&\ar@{-}[u]&\color{green!50!black}\ar@{-}@[green!50!black][ru]&&\\
                       &&&&&\color{orange}\ar@{-}[u]&&&\\
                       &\color{blue}\ar@{-}[rr]&&\ar@{-}[r]&\ar@{-}[ru]\ar@{-}[ld]&&&\color{blue}\ar@{-}[lll]&\\
                       &&&\color{orange}\ar@{-}[uuu]&&&&&
}
\end{smallgathered},
\end{equation}
where $v$ is the center vertex. 
Note that we only include the states of the $\tau_c$'s corresponds to couplings $c\in C_v$ in Eq.~(\ref{eq:support_nonabelian}). We will discuss this choice of $C_v$ below.
Now the steps works as follows:

\begin{enumerate}
    \item Consider the virtual situation that the matter is in the state (only the state relevant to $C_v$ is shown)

\begin{equation}\label{eq:demo_matter}
|\tilde{m}\rangle=
    \renewcommand{\labelstyle}{\textstyle}
\begin{smallgathered}
\xymatrix@=0.0cm{
    &&&&&\color{orange}|010\rangle\ar@{-}[d]&&&\\
    &\color{blue}|101\rangle\ar@{-}[rrr]|{\color{blue}}&&&|001\rangle\ar@{-}[ru]\ar@{-}[ld]&\ar@{-}[dd]&&\color{blue}|110\rangle\ar@{-}[lll]&\\
    &&&\color{orange}|010\rangle\ar@{-}[ddd]|(.55){\color{orange}}&&&&&\\
    &&\color{green!50!black}|000\rangle\ar@{-}@[green!50!black][ld]&\ar@{-}@[green!50!black][l]&\ar@{-}@[green!50!black][l]&|001\rangle\ar@{-}@[green!50!black][l]\ar@{-}@[green!50!black][rr]&&\ar@{-}@[green!50!black][r]&\color{green!50!black}|010\rangle\ar@{-}@[green!50!black][ld]\\
    &|000\rangle\ar@{-}[uuu]\ar@{-}[d]&&\ar@{-}[ll]|(.25){}&|000\rangle\ar@{-}[uuu]|{}\ar@{-}[rrr]|{}\ar@{-}[d]\ar@{-}[l]\ar@{-}[ru]|{}\ar@{-}[ld]|{}\ar@{}[llluuu]|(.6){\color{blue}}\ar@{}[luu]|(.68){\color{orange}}\ar@{}[ruuuu]|(.5){\color{orange}}\ar@{}[rrruuu]|(.6){\color{blue}}\ar@{}[llu]|(.7){\color{green!50!black}}\ar@{}[rrrru]|(.5){\color{green!50!black}}\ar@{}[rrrddd]|(.6){\color{blue}}\ar@{}[rdd]|(.65){\color{orange}}\ar@{}[ldddd]|(.5){\color{orange}}\ar@{}[lllddd]|(.6){\color{blue}}\ar@{}[rrd]|(.7){\color{green!50!black}}\ar@{}[lllld]|(.5){\color{green!50!black}}&\ar@{-}[u]&&|000\rangle\ar@{-}[uuu]\ar@{-}[ddd]&\\
    \color{green!50!black}|111\rangle\ar@{-}@[green!50!black][ru]|{\color{green!50!black}}&\ar@{-}[dd]&&|001\rangle\ar@{-}@[green!50!black][lll]\ar@{-}@[green!50!black][rrr]&\ar@{-}[dd]|(.25){}&\ar@{-}[u]&\color{green!50!black}|011\rangle\ar@{-}@[green!50!black][ru]&&\\
                                                               &&&&&\color{orange}|110\rangle\ar@{-}[u]&&&\\
                                                               &\color{blue}|010\rangle\ar@{-}[rr]&&\ar@{-}[r]&|000\rangle\ar@{-}[ru]\ar@{-}[ld]&&&\color{blue}|100\rangle\ar@{-}[lll]&\\
                                                               &&&\color{orange}|001\rangle\ar@{-}[uuu]&&&&&
}
\end{smallgathered}.
\end{equation}
This satisfies the requirements, since the state of $v$ is the reference state $|000\rangle$ and one can check that the eigenvalues of the minimal couplings are in correspondence with $|\tau_c\rangle$. For example, $|\tilde m\rangle$ is the $(+1)$-eigenstate of the $c_0$ coupling on the upper left blue plaquette and is the $(-1)$-eigenstate of the $c_1$ on that plaquette, which corresponds to the state $|01\rangle$ of the corresponding $\tau_{c_0}$ and $\tau_{c_1}$ shown in Eq.~(\ref{eq:demo_state}) (Recall that we label the computational basis of $\tau_c$ as $|0\rangle$ and $|1\rangle$ corresponding to the $+1$ and $-1$ eigenvalues of $c$).
Of course, this is not the only virtual matter state that satisfies the requirement, but any of such states should give the same result. See the discussion on $C_v$ below.

\item The action of $\left(g_P^{(0)}\right)^{-1}$ on $v$ is $X_0$, which maps $v$ on the virtual matter to $|100\rangle$. Also, other sites of the virtual matter are unchanged. So the virtual matter state is changed to
\begin{equation}\label{eq:demo_matter2}
|\tilde{m}'\rangle=
    \renewcommand{\labelstyle}{\textstyle}
\begin{smallgathered}
\xymatrix@=0.0cm{
    &&&&&\color{orange}|010\rangle\ar@{-}[d]&&&\\
    &\color{blue}|101\rangle\ar@{-}[rrr]|{\color{blue}}&&&|001\rangle\ar@{-}[ru]\ar@{-}[ld]&\ar@{-}[dd]&&\color{blue}|110\rangle\ar@{-}[lll]&\\
    &&&\color{orange}|010\rangle\ar@{-}[ddd]|(.55){\color{orange}}&&&&&\\
    &&\color{green!50!black}|000\rangle\ar@{-}@[green!50!black][ld]&\ar@{-}@[green!50!black][l]&\ar@{-}@[green!50!black][l]&|001\rangle\ar@{-}@[green!50!black][l]\ar@{-}@[green!50!black][rr]&&\ar@{-}@[green!50!black][r]&\color{green!50!black}|010\rangle\ar@{-}@[green!50!black][ld]\\
    &|000\rangle\ar@{-}[uuu]\ar@{-}[d]&&\ar@{-}[ll]|(.25){}&|100\rangle\ar@{-}[uuu]|{}\ar@{-}[rrr]|{}\ar@{-}[d]\ar@{-}[l]\ar@{-}[ru]|{}\ar@{-}[ld]|{}\ar@{}[llluuu]|(.6){\color{blue}}\ar@{}[luu]|(.68){\color{orange}}\ar@{}[ruuuu]|(.5){\color{orange}}\ar@{}[rrruuu]|(.6){\color{blue}}\ar@{}[llu]|(.7){\color{green!50!black}}\ar@{}[rrrru]|(.5){\color{green!50!black}}\ar@{}[rrrddd]|(.6){\color{blue}}\ar@{}[rdd]|(.65){\color{orange}}\ar@{}[ldddd]|(.5){\color{orange}}\ar@{}[lllddd]|(.6){\color{blue}}\ar@{}[rrd]|(.7){\color{green!50!black}}\ar@{}[lllld]|(.5){\color{green!50!black}}&\ar@{-}[u]&&|000\rangle\ar@{-}[uuu]\ar@{-}[ddd]&\\
    \color{green!50!black}|111\rangle\ar@{-}@[green!50!black][ru]|{\color{green!50!black}}&\ar@{-}[dd]&&|001\rangle\ar@{-}@[green!50!black][lll]\ar@{-}@[green!50!black][rrr]&\ar@{-}[dd]|(.25){}&\ar@{-}[u]&\color{green!50!black}|011\rangle\ar@{-}@[green!50!black][ru]&&\\
                                                               &&&&&\color{orange}|110\rangle\ar@{-}[u]&&&\\
                                                               &\color{blue}|010\rangle\ar@{-}[rr]&&\ar@{-}[r]&|000\rangle\ar@{-}[ru]\ar@{-}[ld]&&&\color{blue}|100\rangle\ar@{-}[lll]&\\
                                                               &&&\color{orange}|001\rangle\ar@{-}[uuu]&&&&&
}
\end{smallgathered}.
\end{equation}

\item Now we calculate the eigenvalues of the minimal couplings for $|\tilde{m}'\rangle$, and correspond them back to the state of $\tau_c$. The result is
\begin{equation}\label{eq:demo_state2}
\bigotimes_{c\in C_{v}}|\tau'_c\rangle=
    \renewcommand{\labelstyle}{\textstyle}
\begin{smallgathered}
\xymatrix@=0.5cm{
    &&&&&\color{orange}\ar@{-}[d]&&&\\
    &\color{blue}\ar@{-}[rrr]|{\color{blue}|0\rangle}&&&\ar@{-}[ru]\ar@{-}[ld]&\ar@{-}[dd]&&\color{blue}\ar@{-}[lll]&\\
    &&&\color{orange}\ar@{-}[ddd]|(.55){\color{orange}|1\rangle}&&&&&\\
    &&\color{green!50!black}\ar@{-}@[green!50!black][ld]&\ar@{-}@[green!50!black][l]&\ar@{-}@[green!50!black][l]&\ar@{-}@[green!50!black][l]\ar@{-}@[green!50!black][rr]&&\ar@{-}@[green!50!black][r]&\color{green!50!black}\ar@{-}@[green!50!black][ld]\\
    &\ar@{-}[uuu]\ar@{-}[d]&&\ar@{-}[ll]|(.25){|0\rangle}&\ar@{-}[uuu]|{|1\rangle}\ar@{-}[rrr]|{|0\rangle}\ar@{-}[d]\ar@{-}[l]\ar@{-}[ru]|{|1\rangle}\ar@{-}[ld]|{|1\rangle}\ar@{}[llluuu]|(.6){\color{blue}|00\rangle}\ar@{}[luu]|(.68){\color{orange}|11\rangle}\ar@{}[ruuuu]|(.5){\color{orange}|11\rangle}\ar@{}[rrruuu]|(.6){\color{blue}|10\rangle}\ar@{}[llu]|(.7){\color{green!50!black}|10\rangle}\ar@{}[rrrru]|(.5){\color{green!50!black}|11\rangle}\ar@{}[rrrddd]|(.6){\color{blue}|00\rangle}\ar@{}[rdd]|(.65){\color{orange}|01\rangle}\ar@{}[ldddd]|(.5){\color{orange}|01\rangle}\ar@{}[lllddd]|(.6){\color{blue}|11\rangle}\ar@{}[rrd]|(.7){\color{green!50!black}|11\rangle}\ar@{}[lllld]|(.5){\color{green!50!black}|10\rangle}&\ar@{-}[u]&&\ar@{-}[uuu]\ar@{-}[ddd]&\\
    \color{green!50!black}\ar@{-}@[green!50!black][ru]|{\color{green!50!black}|1\rangle}&\ar@{-}[dd]&&\ar@{-}@[green!50!black][lll]\ar@{-}@[green!50!black][rrr]&\ar@{-}[dd]|(.25){|0\rangle}&\ar@{-}[u]&\color{green!50!black}\ar@{-}@[green!50!black][ru]&&\\
                       &&&&&\color{orange}\ar@{-}[u]&&&\\
                       &\color{blue}\ar@{-}[rr]&&\ar@{-}[r]&\ar@{-}[ru]\ar@{-}[ld]&&&\color{blue}\ar@{-}[lll]&\\
                       &&&\color{orange}\ar@{-}[uuu]&&&&&
}
\end{smallgathered},
\end{equation}
which is defined as the result of $A_{v,g_P^{(0)}}$ acting on the state in Eq.~(\ref{eq:demo_state}).

    Note that for a minimal coupling $c\notin C_v$ [corresponding to a state not shown in Eq.~(\ref{eq:demo_state})], the eigenvalues for $|\tilde m\rangle$ and $|\tilde m'\rangle$ are the same. This is important when we discuss the choice of $C_v$ below.

\end{enumerate}

The above steps give the result of $A_{v,g_P^{(0)}}$ acting on a specific computation basis state. To obtain the full expression of $A_{v,g_P^{(0)}}$, one needs to consider its action on all possible states for gauge DOF (that is, change the state of Eq.~(\ref{eq:demo_state}) to any possible combinations of $0$ and $1$. Then one should obtain the first line of Eq.~(\ref{eq:AvgP0z2}).

Now we discuss our choice of $C_v$.
Firstly, one can show that if we choose a different $|\tilde m\rangle$ that also corresponds to (\ref{eq:demo_state}), then we still obtain the same Eq.~(\ref{eq:demo_state2}). Secondly, in the last step above, the eigenvalues of $c\notin C_v$ are not changed. This is true in general for this choice of $C_v$, and therefore $A_{v,g}$ is well-defined. Also one can check [see Eq.(\ref{eq:AvgP0z2})] that $A_{v,g}$ on different sites commute.

Now we discuss why this choice is minimal. That is, if we delete some couplings from $C_v$, then the above procedure will fail.
Suppose we delete the $c_2$ in the upper left blue horizontal link in Eq.~(\ref{eq:support_nonabelian}). That is, we exclude the state on the horizontal blue link in Eq.~(\ref{eq:demo_state}).
Then Eq.~(\ref{eq:demo_matter}) still satisfies the requirements. In addition to that, if we modify the upper left corner from $|101\rangle$ to $|100\rangle$ in Eq.~(\ref{eq:demo_matter}), it also satisfies the requirements. However, if we use the latter choice of $|\tilde m\rangle$, than the upper left plaquette of the resulting Eq.~(\ref{eq:demo_state2}) will be $|11\rangle$ instead of $|00\rangle$. That is, different choices of virtual matter lead to different result, so this $C_v$ does not work.

Next, suppose we delete the $c_0$ in the upper left blue horizontal link in Eq.~(\ref{eq:support_nonabelian}). Note that in the last step, the eigenvalue of this $c_0$ still changes in respond to the change of the virtual state of $v$. Then since this $c_0$ is not in $C_v$, it violates the requirement that all eigenvalues of $c\notin C_v$ is not changed.

Other deletion of elements in $C_v$ violates the requirements in similar ways. Note that some deletion still works when calculating $A_{v,g_P^{(0)}}$ but not when calculating $A_{v,g^{(2)}}$ (we require that a single $C_v$ works for any $g$ around a vertex.) Therefore, the chosen $C_v$ is minimal.

Note that this is not the only choice of $C_v$, but other choice are essentially the same. If we require that $A_{v,g}$ is symmetric under cyclic permutations of $x,y,z$ axis, then there are only two possible $C_v$ related by reflection.

\section{Discussion on the minimal coupling terms for \texorpdfstring{$1\to\mathbb{Z}_2^\text{sub}\to G\to\mathbb{Z}_2^\text{glo}\to 1$}{1->Z\_2->G->Z\_2->1}\label{sec:coupling}
}

Here we discuss the choice of the $c_0$ coupling in Eq. (\ref{eq:z4coupling}).
first we notice that 
\begin{align}
    \fplaquetteop{\tilde c_0}{}{}{}{}{}{}{}{}=\fplaquetteop{}{}{}{}{}{Z}{Z^\dagger}{Z^\dagger}{Z},\quad
            \flinkop{c_1}{}{}&:=\flinkop{}{Z^2}{Z^2}
\end{align}
        are valid coupling terms invariant under $G$, which are just the usual coupling terms in the pure global and pure subsystem case, respectively. 
        However, in this mixture of global and system symmetry, $\tilde c_0$ is not a good choice for the following reason.
        Suppose we use $\tilde c_0$ and $c_1$ in $H_n$ instead. Then the couplings satisfy the relation
        \begin{equation}
            \fplaquetteop{\tilde c_0^2}{c_1}{}{}{c_1}{}{}{}{}=1.
        \end{equation}
        In our gauging process, this corresponds to a flux operator.
        However, unlike the flux on a square plaquette and tube, this ``flux'' does not have a non-contractible geometry. So the above relation should be interpreted as unnecessary redundancy of the couplings rather than something that allows curvature to be inserted in the resulting gauge connection.
        To avoid this redundancy, we defined the coupling $c_0$ that drops out the part of $\tilde c_1$ that can be obtained from $c_0$:
\begin{align}
    \fplaquetteop{\tilde c_0}{}{}{}{}{}{}{}{}=+1&\iff\fplaquetteop{}{c_1}{}{}{c_1}{}{}{}{}=+1,\,\fplaquetteop{c_0}{}{}{}{}{}{}{}{}=+1\\
    \fplaquetteop{\tilde c_0}{}{}{}{}{}{}{}{}=+i&\iff\fplaquetteop{}{c_1}{}{}{c_1}{}{}{}{}=-1,\,\fplaquetteop{c_0}{}{}{}{}{}{}{}{}=+1\\
    \fplaquetteop{\tilde c_0}{}{}{}{}{}{}{}{}=-1&\iff\fplaquetteop{}{c_1}{}{}{c_1}{}{}{}{}=+1,\,\fplaquetteop{c_0}{}{}{}{}{}{}{}{}=-1\\
    \fplaquetteop{\tilde c_0}{}{}{}{}{}{}{}{}=-i&\iff\fplaquetteop{}{c_1}{}{}{c_1}{}{}{}{}=-1,\,\fplaquetteop{c_0}{}{}{}{}{}{}{}{}=-1
\end{align}
which is equivalent to what we defined in the text.

\subsection{Another choice of the couplings}

Here we discuss another choice of the minimal couplings for the $1\to\mathbb{Z}_2^\text{sub}\to G\to\mathbb{Z}_2^\text{glo}\to 1$ model which reproduces exactly  the Hamiltonian of the fractonic hybrid X-Cube code in \cite{tantivasadakarn2021hybrid}
upon gauging (except the zero-flux projectors, which only change some excitation energies and can be dropped in this Abelian case).

The spirit behind this choice is the same as in Sec.~\ref{sec:z2} and Sec.~\ref{sec:z3}. First we split the local Hilbert space into a tensor product of ``subsystem'' and ``global'' part, then put a projector on the ``global'' part at a special ``reference corner'' of the plaquette (here we choose the upper left corner) to decide the operation on the rest of the operator.
Nevertheless, we can do this by splitting the matter qudit in to two qubits
\begin{align}
    &|0\rangle\mapsto|0\rangle\otimes|0\rangle, \quad
    |1\rangle\mapsto|0\rangle\otimes|1\rangle, \quad\notag\\
    &|2\rangle\mapsto|1\rangle\otimes|0\rangle, \quad
    |3\rangle\mapsto|1\rangle\otimes|1\rangle.
\end{align}
Then $g_P^{(0)}$ acts as $XI$ on each site of $P$, which allows us to regard the first qubit as the ``subsystem'' charge and the second as the ``global'' charge.
The minimal couplings can then be constructed similar to the previous two cases as
    \begin{align}
        H_n&=-J_0\sum_\text{plaquettes}\fplaquetteop{c_0}{}{}{}{}{}{}{}{}
        -J_1\sum_\text{links}\flinkop{c_1}{}{}+\text{H.c.},\\
            \fplaquetteop{c_0}{}{}{}{}{}{}{}{}&:=\left(\fplaquetteop{}{}{}{}{}{Z[0]}{Z I}{Z I}{ZI}-\fplaquetteop{}{}{}{}{}{Z[1]}{ZZ}{ZZ}{ZZ}\right), \notag \\
\\
            \flinkop{c_1}{}{}&:=\flinkop{}{IZ}{IZ}.
    \end{align}

    We can proceed our gauging procedure on the original cubic lattice as in Sec.~\ref{sec:z2} and Sec.~\ref{sec:z3}, with a nontrivial choice of the support $C_v$ (that is, not simply $C_v=\{c\mid v\in \supp c\}$) when we do the systematical construction of $A_{v,g}$.

    In the approach of Ref.~\cite{tantivasadakarn2021hybrid}, they avoid the nontrivial choice of $C_v$ by introducing diagonal couplings on the lattice:
\begin{equation}
    \renewcommand{\labelstyle}{\textstyle}
\begin{gathered}
\xymatrix@=1cm{
    \ar@{..}[r]\ar@{..}[d]\ar@{-}[dr]|{c_1} & \ar@{..}[d] \\
    \ar@{..}[r] &
}
\end{gathered}
=
\begin{gathered}
\xymatrix@=1cm{
    IZ\ar@{..}[r]\ar@{..}[d]\ar@{-}[dr] & \ar@{..}[d] \\
    \ar@{..}[r] & IZ
}
\end{gathered},
\end{equation}
In this way, our one-step gauging process (with the systematical construction of $A_{v,g}$ using the trivial choice $C_v=\{c\mid v\in \supp c\}$ and reference state $|00\rangle$) gives exactly the $A_{v,g}$ presented in Ref.~\cite{tantivasadakarn2021hybrid} in the unitary gauge. Moreover, the fluxes constructed using our gauging process will be the same as theirs except for the additional zero-flux projectors.
This implies that in the exactly solvable limit, we reproduce the ground state and excitations pattern exactly.

\end{appendix}


\bibliography{references}

\end{document}